\documentclass[conference]{IEEEtran}
\usepackage{color,graphicx}
\usepackage{url}
\usepackage{subfig}
\usepackage[ruled, linesnumbered]{algorithm2e}
\usepackage{cite}
\usepackage{amsmath,amssymb,amsfonts}
\usepackage{xspace}
\usepackage{xifthen}
\usepackage{balance}
\usepackage{booktabs} 

\definecolor{dkgreen}{rgb}{0,0.6,0}
\definecolor{gray}{rgb}{0.5,0.5,0.5}
\definecolor{mauve}{rgb}{0.58,0,0.82}

\newcommand{\myparagraph}[2][]{\ifthenelse{\isempty{#1}}{\vspace{1ex} \noindent {\bf #2}}{\noindent {\bf #2}}}

\newcommand{\guangyannote}[1]{{\textcolor{red}{[Guangyan: #1]}}}

\newcommand{\futurenote}[1]{
}

\newcommand{\reconsider}[1]{
}

\begin{document}
\title{Approximation with Error Bounds in Spark}
\author{\IEEEauthorblockN{Guangyan Hu}
	\IEEEauthorblockA{\textit{Rutgers University} \\
		New Brunswick, NJ \\
		gh279@cs.rutgers.edu}
	\and
		\IEEEauthorblockN{Sandro Rigo}
	\IEEEauthorblockA{\textit{University of Campinas} \\
		Campinas - SP, Brazil \\
		srigo@unicamp.br}
	\and
	\IEEEauthorblockN{Desheng Zhang}
	\IEEEauthorblockA{\textit{Rutgers University} \\
		New Brunswick, NJ \\
		d.z@rutgers.edu}
		\and
\IEEEauthorblockN{Thu D. Nguyen}
\IEEEauthorblockA{\textit{Rutgers University} \\
	New Brunswick, NJ \\
	tdnguyen@cs.rutgers.edu}
	}
\maketitle

\begin{abstract}
Many decision-making queries are based on aggregating massive amounts of data, where sampling is an important approximation technique for reducing execution times. It is important to estimate error bounds when sampling to help users balance between accuracy and performance. However, error bound estimation is challenging because data processing pipelines often transform the input dataset in complex ways before computing the final aggregated values. In this paper, we introduce a sampling framework to support approximate computing with estimated error bounds in Spark. Our framework allows sampling to be performed at multiple arbitrary points within a sequence of transformations preceding an aggregation operation. The framework constructs a data provenance tree to maintain information about how transformations are clustering output data items to be aggregated. It then uses the tree and multi-stage sampling theories to compute the approximate aggregate values and corresponding error bounds. When information about output keys are available early, the framework can also use adaptive stratified reservoir sampling to avoid (or reduce) key losses in the final output and to achieve more consistent error bounds across popular and rare keys. Finally, the framework includes an algorithm to dynamically choose sampling rates to meet user-specified constraints on the CDF of error bounds in the outputs. We have implemented a prototype of our framework called ApproxSpark and used it to implement five approximate applications from different domains. Evaluation results show that ApproxSpark can (a) significantly reduce execution time if users can tolerate small amounts of uncertainties and, in many cases, loss of rare keys, and (b) automatically find sampling rates to meet user-specified constraints on error bounds. We also explore and discuss extensively tradeoffs between sampling rates, execution time, accuracy and key loss.
\end{abstract}

\section{Introduction}
\label{sec:intro}

Data-driven discovery and decision support have become critical to the missions of many businesses, scientific and government enterprises. At the same time, the rate of data production and collection is outpacing technology scaling, implying that significant future investment, time, and energy will be needed for data processing~\cite{Chong2014,dai2018cloud}. Approximate computing is a powerful tool to reduce these processing needs. Many data analytic applications such as data mining, log processing, and data visualization are amenable to approximation~\cite{approxSurvey}. As a concrete example, suppose a company wants to know the age distribution of its customers for a particular product. In such an application, estimated counts derived from data samples may be sufficient, allowing tradeoffs between precision and processing time, energy consumption and/or cost.

In this paper, we propose a framework for creating and running approximate Spark programs that use online sampling to efficiently aggregate massive amounts of data. The framework computes error bounds (i.e., confidence intervals) along with the approximate aggregate values. We focus on aggregation because many decision support tasks require aggregation queries: e.g., a study of a Microsoft SCOPE~\cite{chaiken2008scope} data processing cluster reveals that 90\% of 2,000 data mining jobs were aggregations~\cite{yan2014error}. Aggregation is also an important component in online analytical (OLAP) systems for summarizing data patterns in business intelligence~\cite{gray1997data, xie2019olap}.

Spark is a popular data processing system that has been widely adopted in different domains~\cite{shanahan2015large, armbrust2015spark, yu2015geospark, wiewiorka2014sparkseq}. Thus, embedding a general approximation framework in Spark will make approximation easily accessible to application developers in many different fields. In addition, while our work is specific to Spark, it should also be portable to other similar data processing systems.

\begin{figure}
	\begin{center}
		\includegraphics[width=6cm]{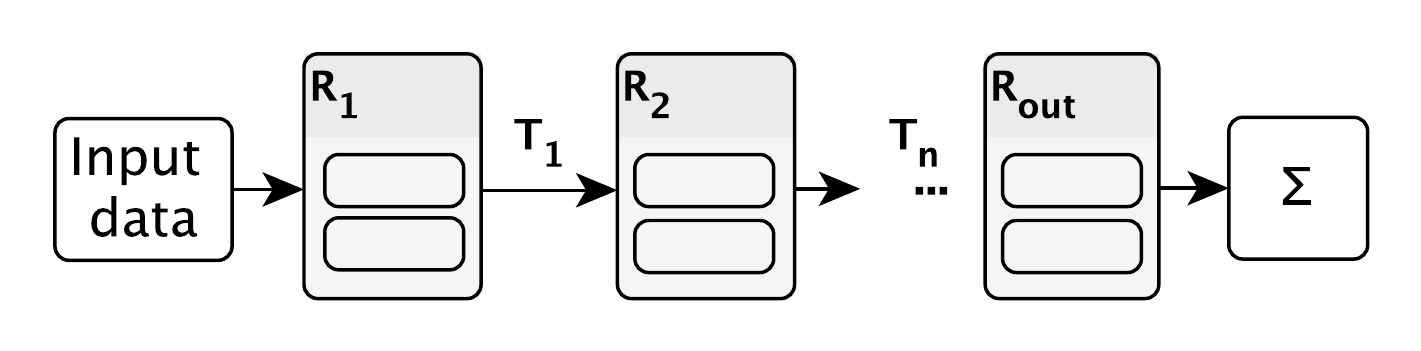} \\
	\end{center}
	\vspace{-0.2cm}
	\caption{A Spark computation having a chain of transformations~($\{T\}$), where each box in an RDD is a partition.}
	\label{fig:example-spark}
	\vspace{-0.5cm}
\end{figure}

Estimating error bounds is important, especially for decision support queries, because it allows users to intelligently balance precision and performance. However, Spark programs (and data processing pipelines in general) often include multiple complex transformations of the input data before the final aggregation~\cite{hashem2015rise, dataflow}, making it challenging to compute error bounds. Consider a Spark computation comprising of a chain of transformations ending with a summation as shown in Figure~\ref{fig:example-spark}. If we sample data items in the resilient distributed dataset (RDD) $R_{\textit{out}}$ immediately before the aggregation, then it is straightforward to use simple random sampling~(SRS) theories to estimate the sums with error bounds~\cite{lohr2009sampling}. However, this sampling is unlikely to reduce execution time by much since the additions saved are relatively inexpensive.

Alternatively, we can view each partition of $R_{\textit{out}}$ as a cluster and apply cluster sampling. We can then use two-stage cluster sampling theories for estimating sums and error bounds~\cite{lohr2009sampling}, although we would need to estimate populations in multi-key computations (see the discussion on multi-stage sampling in multi-key computations below). This can lead to much greater execution time savings since we can avoid performing {\em all} of the transformations on the dropped partitions. Unfortunately, this locks the computation into a very coarse-grained sampling process that may not be tunable to achieve the desired tradeoff between precision and performance.

A natural solution is to sample earlier, e.g., sample when creating $R_1$ from the input data, where we can use a combination of dropping partitions and data items to achieve the right balance between precision and performance. As we discuss in Section~\ref{sec:sampling}, a key insight behind our work is that it is possible to map such a sampling process to a multi-stage sampling process on $R_{\textit{out}}$, and use the accompanying theories to compute the estimated aggregate values and error bounds.

As a concrete example, consider a  program to count word occurrences in a text dataset, where a {\tt map} parses each sentence and produces a list of {\tt (word, 1)} pairs, and a subsequent {\tt flatMap} breaks the lists to produce the final set of {\tt (word, 1)} pairs, followed by summing the count of each unique word. In this computation, there are two levels of clustering, with each partition of the input being a cluster of sentences, and each sentence a cluster of words. Therefore, when sampling at the creation of $R_1$ by selecting a random subset of partitions and a random subset of sentences from each selected partition, the sampling errors need to be estimated using three-stage cluster sampling theories since the end populations are actually words rather than the sentences. The population size of each word also has to be estimated from its sample size given the sampling rate over the sentences, because if a sentence is not chosen for the sample, then it is unknown whether that dropped sentence would have produced counts for a particular word.

In Section~\ref{sec:sampling}, we first explain how sampling at multiple arbitrary points within a sequence of transformations can be mapped to a multi-stage sampling process on the output RDD. We then propose an algorithm to build a {\em data provenance tree} to maintain information about this mapping. Finally, we propose another algorithm to extract information from the tree, estimate populations where needed, and compute the final approximate aggregate values and their error bounds. Critically, we show how to account for the imprecision introduced by population estimation. If the final keys are known early in the transformation sequence, we show how adaptive stratified reservoir sampling (ASRS)~\cite{al2014adaptive} can be integrated with multi-stage sampling to avoid losing rare keys, as well as balancing the sampling errors between popular and rare keys (Section~\ref{sec:reservoir-sampling}).

We have implemented the proposed framework in a prototype system called ApproxSpark (Section~\ref{sec:implementation}).  Our framework supports a subset of common Spark transformations, including {\tt map}, {\tt flatMap}, {\tt mapValues}, {\tt sample} and {\tt filter}, and aggregation operations {\tt mean} and {\tt sum}. When running an approximate computation, users have the flexibility to specify sampling rates or constraints on the CDF of relative error bounds for values associated with output keys---if the computation produces a single value or key-value pair, then the latter reduces to just the maximum allowable relative error bound.  When the user specifies constraints for the error bound CDF, ApproxSpark will run pilot executions of several partitions and use the results to select appropriate sampling rates. 

We have used ApproxSpark to implement five approximate applications from different domains in text mining, graph analysis, and log analysis. We use the applications to evaluate ApproxSpark and explore the tradeoffs between performance and accuracy/precision. Among other findings, our results show that (i) ApproxSpark can significantly reduce execution time if users can tolerate small amounts of uncertainties and, in many cases, loss of rare keys; (ii) it is possible to automatically find sampling rates to meet user-specified constraints on the CDF of error bounds in the output; (iii) partition sampling can lead to greater reduction in execution time than data item sampling, but lead to more key loss and significantly larger error bounds, especially for the rarer keys; and (iv) ASRS with multi-stage sampling avoids or reduces key loss and leads to more consistent error bounds across keys.

In summary, our contributions include: (i) to our knowledge, our work is the first to apply multi-stage sampling theories to estimate aggregate values and error bounds when sampling within arbitrarily long sequences of transformations; (ii) we introduce algorithms for maintaining provenance information during the execution of the transformations and computing the approximate aggregate values and error bounds; (iii) we show how ASRS can be combined with multi-stage sampling for some applications to reduce key loss and equalize error bounds across popular and rare keys; (iv) we explore extensively the tradeoffs between sampling rates, execution time, key loss, and error bounds; and (v) we present an algorithm for automatically choosing sampling rates to meet user-specified constraints on the CDF of error bounds for output values.

\section{background and related work}
\label{sec:related}

\myparagraph[top]{Spark.} Spark has emerged as a popular distributed data processing engine. Spark introduces RDDs, which are fault-tolerant collections of data partitioned across server clusters that can be processed in parallel~\cite{Zaharia2010}. Spark has two types of operations: transformations and actions. A transformation is a lazy operation that produces an output RDD from an input RDD, where as an action computes non-RDD values from an input RDD, and triggers preceding transformations needed to produce the input RDD. Data items in RDDs can be key/value pairs, such that a Spark program may be computing a number of different aggregations in parallel. The word counting program in Section~\ref{sec:intro} is a good example. It is counting potentially many different unique words at once, computing an aggregation for each word.

Spark already contains random and stratified sampling transformations with several important limitations. First, there is no support for computing error bounds, especially across a sequence of multiple transformations. Second,  stratified sampling can still lose some keys, because it adopts Bernoulli Sampling. Third, sampling is only implemented on existing RDDs, so that the entire input dataset has to be loaded before sampling can be applied.

\myparagraph{Approximate query processing (AQP).} A variety of approximation techniques have been employed by query processing systems to reduce execution time. These techniques include using random or stratified sampling to construct samples to provide bounded errors~\cite{chaudhuri2007optimized,ssdbm10, purdueStratified, peng2018aqp++, yan2014error, Agarwal2013, Sapprox} or online aggregation to sample data and produce a result within a time-bound~\cite{hellerstein1997online,Kumar:2016:HEF:2901318.2901351}. BlinkDB~\cite{Agarwal2013} maintains a set of offline-generated stratified samples by using an error-latency profile based on past queries. Sapprox~\cite{Sapprox} collects the occurrences of sub-datasets in offline preprocessing and uses it to drive online sampling. Many AQP systems use offline processing under the assumption that data will be used repeatedly. Online sampling is an efficient approximation method when the large dataset~(e.g., logs) will be used only once or a few times~\cite{ApproxHadoop}. 

\myparagraph{Online sampling.} ApproxHadoop~\cite{ApproxHadoop} introduces approximation to the MapReduce paradigm~\cite{mapreduce}. It uses multi-stage sampling to trade off precision and performance similar to ApproxSpark (we discuss differences below). Users can specify sampling rates or a target maximum relative error. StreamApprox~\cite{StreamApprox} approximates stream processing workloads based on Spark Streaming~\cite{SparkStreaming}. MaRSOS \cite{purdueStratified} is related to ApproxHadoop but proposes a stratified sampling algorithm to avoid losing keys and balance error bounds for popular and rare keys. Compared to MaRSOS, ApproxSpark's implementation of stratified sampling using ASRS avoids the overheads of coordination between parallel tasks while still being able to balance error bounds.

\myparagraph{Comparison with ApproxHadoop.} While ApproxSpark and ApproxHadoop both use multi-stage sampling, there are important differences. First, ApproxSpark generalizes multi-stage sampling to handle sequences of transformations with arbitrary lengths, allowing sampling anywhere within the sequences, whereas ApproxHadoop is limited to using two- and three-stage sampling to handle a single map phase in MapReduce computations. Second, ApproxHadoop also relies on population estimation but does not account for the added uncertainties; ApproxSpark does. ApproxSpark implements ASRS to avoid losing keys and balance error bounds when output keys are known early in the computation. Finally, in this paper, we explore the rich space of tradeoffs between sampling rates, execution times, error bound distributions across all output keys, and loss of rare keys far beyond what was considered in the ApproxHadoop study~\cite{ApproxHadoop}.

\section{Multi-stage sampling in spark}
\label{sec:sampling}

Suppose we have a simple Spark program that reads a set of values into an RDD $R_{\textit{in}}$ and sums the values.  We can reduce the execution time of this computation by (1) reading only a randomly selected subset of input partitions, (2) load a randomly selected subset of data items from each selected partition into $R_{\textit{in}}$, and (3) compute an estimated sum $\hat{\tau}$ and its variance $\hat{\textit{V}}$, which is needed for computing confidence intervals around $\hat{\tau}$, using two-stage cluster sampling theories as follows~\cite{lohr2009sampling}:


\begin{gather}
\hat{\tau} = \frac{N}{n}
\sum_{i=1}^{n}(\frac{M_i}{m_i}\sum_{j=1}^{m_i}v_{ij})
\label{eq:twostage-sum}
\\
\hspace{-0.1in}\hat{\textit{V}}(\hat{\tau}) = N(N-n) \frac{S_u^2}{n} + \frac{N}{n}
\sum_{i=1}^{n} {M_i (M_i-m_i) \frac{S_i^2}{m_i}}
\label{eq:twostage-variance}
\end{gather}
	
\noindent
where $N$ is the total number of partitions in the input data set, $n$ is the number of selected partitions, $M_i$ is the total number of values in partition $i$ of the input data set, $m_i$ is the number of values selected from partition $i$ and loaded into $R_{\textit{in}}$, $v_{ij}$ is the $j^{\textit{th}}$ value from partition $i$ in $R_{\textit{in}}$, $S_i^2$ is the intra-cluster variance for partition $i$, and $S_u^2$ is the inter-cluster variance. Note that $N$ and $M_i$'s are attributes of the input data set, while $n$ and $m_i$'s are attributes of the sample. $S_u^2$ and $S_i^2$ are both computed using the sample.



Now consider a program where $R_{\textit{in}}$ is transformed by a sequence of transformation $T_0, T_1, ..., T_n$ to produce $R_{\textit{out}}$, which is then summed. If each transformation $T_i$ is a one-to-one mapping of an input value to a single output value (e.g., a Spark {\tt map} operation), such that $R_{\textit{in}}$, $R_{\textit{out}}$ and all intermediate RDDs contain the same number of data items, then it is possible to sample the input data when creating $R_{\textit{in}}$ in the same manner as above and still use the estimators given in Equations~\ref{eq:twostage-sum} and~\ref{eq:twostage-variance}. Sampling the input data is exactly equivalent to sampling the $R_{\textit{out}}$ that would have been produced by processing the entire input dataset.

Spark, however, includes transformations that map input items to output items in more complex ways than one-to-one. As already mentioned, this complexity makes it much more challenging to compute error bounds when sampling early within a Spark computation. In the remainder of this section, we first show how generalized multi-stage sampling theories can be used when sampling at multiple different points within a Spark program. We then describe two algorithms necessary to track the multi-level clustering of data items in $R_{\textit{out}}$ as the input data is transformed, and to use the tracking information to estimate the aggregate values and error bounds. We discuss summation, but the discussion is equally applicable to average.

\subsection{Multi-stage Sampling}
\label{sec:multstagetheory}

\begin{figure}
	\begin{center}
		\vspace{-0.1cm}
		\includegraphics[width=9cm]{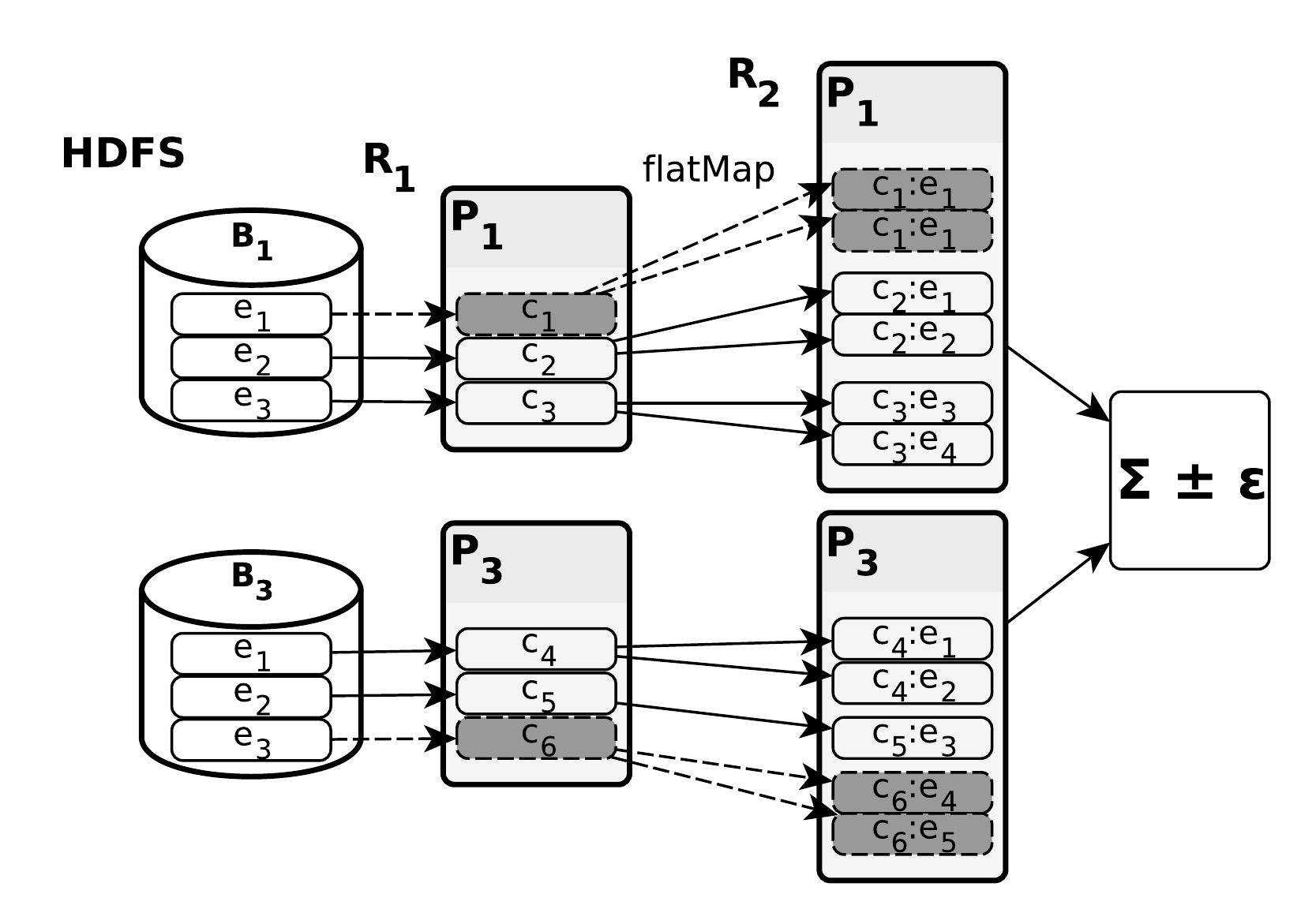}
		\vspace{-0.2cm}
		\caption{HDFS blocks and input data items are sampled when read into RDD $R_1$. Block $B_{2}$ not shown has been dropped. Gray boxes are dropped data items. In $R_{2}$, $c_i:e_j$ means data item $j$ is generated from the data item $i$ in the input partition.}
		\vspace{-0.5cm}
		\label{fig:samplingFramework}
	\end{center}
\end{figure}

Consider the Spark program and its execution as shown in Figure~\ref{fig:samplingFramework}. The {\tt flatMap} transformation can generate multiple output items for each input item, corresponding to a one-to-many mapping. An example is the generation of the two data items $c_2$:$e_1$ and $c_2$:$e_2$ in $R_2$ from the single data item $c_2$ from $R_1$. In this case, when sampling, selecting an input data item to load into $R_1$ is equivalent to selecting a cluster of items from $R_2$, and selecting a partition from the input data set is equivalent to selecting a cluster of clusters from the $R_2$. This corresponds to a three-stage sampling process. In fact, general multi-stage sampling and population estimation can be used to handle Spark programs comprised of a subset of common transformations for both single- and multi-key computations.

\reconsider{
The {\tt filter} and {\tt sample} transformations, on the other hand, can produce zero output items for each input item, corresponding to an one-to-zero mapping. In this case, if we apply a {\tt map} or {\tt flatMap} after sampling or filtering an RDD, we cannot deduce the impact of clusters or items not chosen from the input RDD had {\tt filter} and {\tt sample} not applied on the output RDD; i.e., some cluster/items not chosen from the input RDD would have produced items in the output RDD, while others would not have. This introduces the need to estimate the population of the output RDD from the sample size to apply the cluster sampling theories for estimating the sampling error.}

Below, we generalize the two-stage sampling equations~(Eq~(\ref{eq:twostage-sum}) and (\ref{eq:twostage-variance})) into recurrences for multi-stage sampling with estimated sum and variance. We use $I_k=i_0,i_1,...,i_k$ to denote the index of a specific cluster at level $k$. Note that in a multi-key computation, a sample is chosen for each key, so we will need to estimate the sum and variance for each key.

\reconsider{
Consider an application that computes the distribution of page lengths within a Web site. The input is a set of all Web pages, with the first transformation producing a key/value pair $(length, 1)$ for each page, while the action is to count the number of pages for each distinct length. In this case, the input dataset really contains a mix of populations (pages of different lengths) that are being counted. The first transformation and the data shuffle that occurs between the transformation and action effectively sort the input into a set of different populations, each population corresponding to the pages of a particular length. Each population is then counted by the summation on each different length. This implies that the number of items in the input does not correspond to the size of any of the subpopulations. We will need to maintain sufficient information to estimate the size of each subpopulation, and compute its impact on \textit{V($\hat{\tau}$)}. In this case, we will still need to estimate the subpopulations of each key.
}

\myparagraph{Sum estimation.} We estimate the sum of a multi-stage sample with $d$ sampling stages using the following recurrence:
\begin{equation}
\hat{\tau}_{I_k} = 
\begin{cases}
\frac{N_{I_{k}}}{n_{I_{k}}}\sum _{j = 1}^{n_{I_{k}}}\hat{\tau}_{I_{k},j} & 0\leq k< d~,\\
v_{I_k} & k = d
\end{cases} 
\label{eq:sumEstimate}
\end{equation}
where $\hat{\tau}_{I_k}$ is the estimated sum of cluster $I_k$ (at level $k$), $N_{I_{k}}$ is the total number of sub-clusters of cluster $I_k$, $n_{I_{k}}$ is the number of sub-clusters chosen from cluster $I_k$, $I_k,j$ is the index $i_0,i_1,...,i_k,j$ such that $\hat{\tau}_{I_k,j}$ is the estimated sum of a sub-cluster of cluster $I_k$, and $v_{I_k}$ is the value in the sample (at the last level $k=d$) with index $I_k$. The $0^{\textit{th}}$ stage contains just one cluster comprising the entire population, so $\hat{\tau}_0$ is then the overall estimated sum. 

\myparagraph{Variance estimation.} Similarly, we estimate the variance using the recurrence:
\begin{equation}
\hat{V} (\hat{{\tau}}_{I_{k}}) = 
\begin{cases}
{N_{I_{k}}(N_{I_{k}}-{n_{I_{k}}}})\frac{S_{u,I_k}^{2}}{n_{I_{k}}} \\+ \frac{N_{I_{k}}}{n_{I_{k}}}\sum _{j = 1}^{n_{I_{k}}}\hat{V}(\hat{\tau}_{{{I_{k}},j}}) & 0\leq k< d - 1~,\\\
M_{I_{k}}({M_{I_{k}}}-{m_{I_{k}}})\frac{S_{i,I_k}^{2}}{m_{I_{k}}} &k = d - 1
\end{cases} 
\label{eq:multi-stage}
\end{equation}
where $\hat{V}(\hat{\tau}_{{I_k}})$ is the variance of $\hat{\tau}_{{I_k}}$, $S_{u,I_k}^{2}$ is the inter-cluster variance of the sub-clusters of cluster $I_k$, $M_{I_k}$ is the total number of values in cluster $I_k$, $m_{I_k}$ is the number of values from cluster $I_k$ in the sample, and  $S^2_{i,I_k}$ is the intra-cluster variance of cluster $I_k$. $V(\hat{\tau}_0)$ is then the overall estimated variance.

\myparagraph{Confidence interval.} Given the above estimated sum and variance, we can compute the confidence interval as: $\hat{\tau}_0 \pm \epsilon$, where $\epsilon = t_{n-1, 1-\alpha/2} \sqrt{\hat{V} (\hat{\tau}_0)}$,  $t_{n-1, 1-\alpha/2}$ is the critical value under the Student's t distribution at the desired level of confidence $\alpha$, and $n$ is the degree of freedom (i.e., the number of chosen clusters at level 1)~\cite{lohr2009sampling}.

\begin{figure}
	\begin{center}
		\includegraphics[width=8cm]{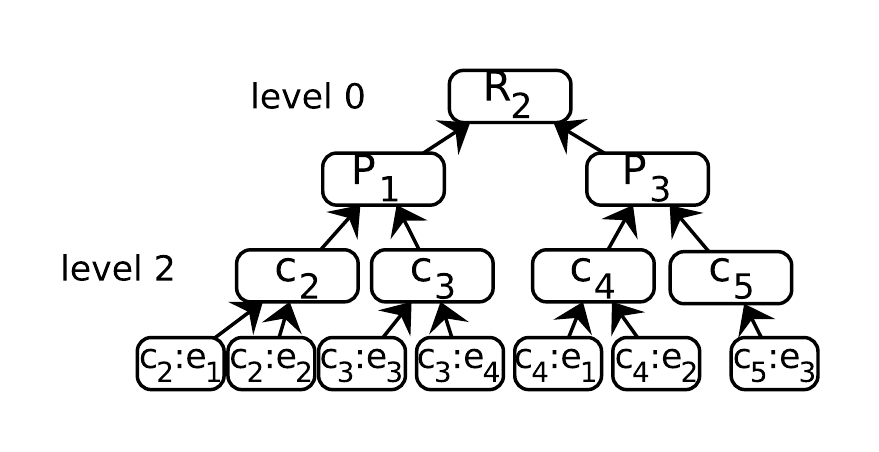}
		\caption{An equivalent tree to the transformation chain shown in Figure~\ref{fig:samplingFramework}, where $R_2$ is the final sample at level 0.}
		\label{fig:provenanceTree}
	\end{center}
\end{figure}

\begin{table}
	{
		\begin{center}
			 {\begin{tabular}{lp{5cm}}
					\hline
					{\bf Transformation} & {\bf {Semantics}} \\ \hline
					{\tt map(func)} & Applies $func$ to each data item~(di). \\ \hline
					{\tt flatMap(func)} & Applies $func$ to each di and flatten. \\ \hline
					{\tt mapValues(func)} & Applies $func$ only to the value of each di.
				 \\ \hline
					{\tt filter(func)} &  
					Selects di's that satisfy a predicate~$func$.
				 \\ \hline
					{\tt sample(r)} &  
			        samples di's using sampling rate $r$.
				\\ \hline
			\end{tabular}}
		\end{center}
	}
	\caption{Spark transformations that can be approximated by ApproxSpark and their semantics. A transformation generates a new RDD from a source RDD.}
	\label{tbl:trans}
\end{table}

\begin{table}
	{
		\begin{center}
			\scalebox{1.0} {\begin{tabular}{lp{4.5cm}}
					\hline
					{\bf Subroutine} & {\bf {Semantics}} \\ \hline
					{\tt sampleInputPar(rate)} & Samples input partitions with $rate$ and returns selected partitions.\\ \hline
					{\tt sampleInputDI(rate)} & Samples input data items in the selected partitions with $rate$ and returns selected data items.\\ \hline
					{\tt createRoot()} & Creates root for the tree. \\ \hline
					{\tt createNodes(\{DI\})} & Creates new nodes using data items $\{DI\}$, where each node corresponds to one item, where an item is a data item in an RDD.\\ \hline
					{\tt addLevel(\{node\})} & Adds a new level to the tree using $\{node\}$, where each node's parent is the parent data item that has generated the data item that this node represents. \\ \hline
					{\tt replaceLast(\{node\})} & Replaces last level nodes with $\{node\}$. A new node~($e$) shares the parent of the node that has generated $e$ in this transformation. Then the nodes that were originally in the last level are deleted.\\ \hline
			\end{tabular}}
		\end{center}
	}
	
	\caption{Description of subroutines used in Algorithm~\ref{algo: buildTree}.}
		\vspace{-0.5cm}
	\label{tbl:provenanceSupp}
\end{table}

\subsection{Data Provenance Tree}

In this subsection, we propose to model the multi-stage sampling clusters resulted from a transformation chain as a {\em data provenance tree}, in order to compute Eq~(\ref{eq:sumEstimate}) and (\ref{eq:multi-stage}). The tree is in essence distributed since each partition builds and maintains the subtree that represents the multi-level clusters contained in it. Figure~\ref{fig:samplingFramework} can be seen as a tree shown in Figure~\ref{fig:provenanceTree}, which maps the input to output items by each transformation. A node in the tree is used as a container for necessary parameters such as variance of each cluster at every level~(sampling stage) in order to recurrently compute Eq~(\ref{eq:sumEstimate}) and (\ref{eq:multi-stage}) eventually at the root. The tree mirrors the translation of sampling during the execution of an RDD transformation sequence, to an equivalent multi-stage sample being taken from the final output RDD. Therefore a level in the tree corresponds to a level of sampling clusters. Table \ref{tbl:trans} shows the subset of Spark transformations that our framework handles.

\myparagraph{Overview.}~In the provenance tree, a $node$ represents a sampling unit at a cluster level, which can be a partition or data item. We define two types of nodes, one is internal node, the other is leaf node. An internal node represents a sampling cluster, such as partition or data item that generates a cluster of data items, and leaf node represents a final output data item. A provenance tree will be incrementally built as each transformation executes in parallel. The computation of Eq~(\ref{eq:sumEstimate}) and (\ref{eq:multi-stage}) occurs after the tree has been built, which is accomplished by traversing the tree level by level from bottom to the top. The computation does not depend on the values of intermediate data items themselves, so an internal node stores its cluster members~(children nodes), estimated sum/variance of the cluster it represents. Note that the estimated sum/variance of clusters for an internal node is not computed until the entire tree has been built, whereas a leaf node always keeps the value of the data item it represents.

\begin{algorithm}
	\SetAlgoLined
	\SetKwInOut{Input}{input}
	\SetKwInOut{Output}{output}
	\SetKwFunction{algo}{DataProvenance}\SetKwFunction{proc}{buildSubtree}
	\SetKwProg{myalg}{Algorithm}{}{}
	\myalg{\algo{$\{T\}$, $pRate$, $iRate$}}{\label{ln:mainStart}
	{\tt createRoot()}\tcp*{level 0}\label{ln:seq1}
	$\{P\}$ = {\tt sampleInputPar}($pRate$)\;\label{ln:parSampling}
	$\{node\}_P$ = {\tt createNodes($\{P\}$)}\;\label{ln:parNodes}
	$rate_1$ = $pRate$\;
	{\tt addLevel}($\{node\}_P$)\;\label{ln:seq2}
	\ForPar{$P_i$ $\in$ $\{P\}$}
	{
		{\tt buildSubtree}($\{T\}$, $iRate$)\;\label{ln:buildTreePara}
	}\label{ln:mainEnd}
	}
	\SetKwProg{myproc}{subroutine}{}{}
	\myproc{\proc{$\{T\}$, $iRate$}}{\label{ln:subStart}
	$\{DI\}$ = {\tt sampleInputDI}($iRate$)\;\label{ln:diSampling}
	$\{node\}_{DI}$ = {\tt createNodes($\{DI\}$)}\;
	{\tt addLevel}($\{node\}_{DI}$)\;\label{ln:idiNodes}
	$rate_2$ = $iRate$\;
	$k$ = 3\tcp*{tracks tree level}
	$rate_{k} = 1.0$\tcp*{initializing $rate_{3}$}
	\For{$T_i$ $\in$ $\{T\}$}{
		$\{DI\}_k$ = {\tt exec}($T_i$)\;
		$\{node\}_k$ = {\tt createNodes($\{DI\}_k$)}\;
		\uIf{$T_i$ is \tt sample}{ 
			{\tt replaceLast}($\{node\}_k$)\;
			$rate_{k}$ *= {\tt sample}.$rate$\;
		}
		\uElseIf{$rate_{k} < 1.0$ and $T_i$ is {\tt flatMap}}{
			{\tt addLevel}($\{node\}_k$)\;
			$rate_{k}$ = $1.0$, $k$ ++\;
		}
		\uElseIf{$T_i$ is {\tt map} or {\tt flatMap} or {\tt mapValues} or {\tt filter}}{
			{\tt replaceLast}($\{node\}_k$)\;
		}
		}
	}\label{ln:subEnd}
	\caption{Building data provenance tree}
	\label{algo: buildTree}
\end{algorithm}

\myparagraph{Tree building.}~We introduce Algorithm~\ref{algo: buildTree} for building a {\em data provenance tree} mapped from an RDD transformation chain. Table~\ref{tbl:provenanceSupp} shows the semantics of subroutines in the algorithm that are not explicitly defined. We assume that input partitions and data items are sampled when input data is being loaded, so the tree's total number of levels would be at least three: the final sample at the root~(level $0$), chosen input RDD partitions and input data items in the chosen partitions. The tree will have more levels if the transformation chain is mapped to more than two cluster levels, depending on each transformation's semantics. Lines~\ref{ln:mainStart} to \ref{ln:mainEnd} contain the main algorithm, which takes as input a transformation chain~$\{T\}$, partition sampling rate~$pRate$ and input data item sampling rate~$iRate$. Line~\ref{ln:seq1} to \ref{ln:seq2} are executed sequentially, where it first creates a root node, then samples the input partitions with rate $pRate$~(line~\ref{ln:parSampling}) and adds the sampled partitions~($\{P\}$) to the tree as a children to the root. We use $\{rate\}$ to keep track of each level's sampling rate. Line~\ref{ln:buildTreePara} is the parallel execution of {\tt buildSubtree}~(line~\ref{ln:subStart} to \ref{ln:subEnd}) for each partition, which builds a subtree rooted at each partition node~(sans the partition node itself) as each transformation executes. Lines~\ref{ln:diSampling} to \ref{ln:idiNodes} add the sampled input data items in each chosen partition as a new level. Then the rest of the algorithm will update the tree based on each $T_i$'s semantics, using $\{node\}_k$ created from data items generated from $T_i$. If $T_i$ is {\tt sample}, it replaces the nodes in the last tree level with $\{node\}_k$ generated by sampling the previous level, then updates $rate_k$ using {\tt sample}'s sampling rate. If $T_i$ is {\tt flatMap} and there is sampling operation before it, a new level is added because sampling data items before applying {\tt flatMap} is equivalent of dropping groups of data items generated from this {\tt flatMap}, thus adding a new level of clusters. In other cases, the last level's nodes will be replaced by $\{node\}_k$ without adding a new level.

\myparagraph{Multi-key computation.}~A transformation can produce multiple keys, and a transformation chain finally leads to multiple final output key spaces. However, only each final output key, instead of an intermediate key, defines an independent Spark computation. Since we are only interested in the estimator and error bounds of the final output RDD, the multi-level clustering in the final sample would only be determined by the leaf nodes in the same key space. Therefore in the provenance tree building process, the intermediate key spaces need not to be explicitly reflected in the internal nodes. The presence of multiple keys also introduces the need of population estimation which can be handled by the theory introduced earlier.

\myparagraph{Limitations.}~We assume {\tt sample} and {\tt filter} will not eliminate all the data items from a particular partition, so that number of partitions stay the same after loading the input data. We does not consider {\tt filter}'s effect over the sampling error, specifically we are not sure about its impact over the variance and how it is propagated through clusters to the final error bound. It is because {\tt filter} deterministically eliminates some data items based on its predicate, instead of randomly selecting data items where the sample sum and variance would follow a certain distribution. We leave exploring the impact of {\tt filter} over error bounds as a future work.

\subsection{Tree traversal-based statistics computation}
Eq~(\ref{eq:sumEstimate}) and (\ref{eq:multi-stage}) can be computed by traversing the provenance tree built using Algorithm~\ref{algo: buildTree}. The tree is traversed level by level starting from the leaf nodes, from which the estimated sum/variance of internal nodes at each level can be incrementally computed, and the desired final output is estimated sum/variance at the root. We introduce Algorithm~\ref{algo: computeTree} for this computation process. Lines~\ref{ln:parCompBegin} to \ref{ln:parCompEnd} compute in parallel each partition's statistics by calling the subroutine {\tt ComputeNodeI}, which computes Eq~(\ref{eq:sumEstimate}) and (\ref{eq:multi-stage}) of a given node at level $k$. Line~\ref{ln:rootComp} computes the root's estimated sum/variance for the final confidence interval output.
\begin{algorithm} 
\SetKwInOut{Input}{input}
\SetKwInOut{Output}{output}
\SetKwFunction{algo}{ComputeTree}\SetKwFunction{proc}{ComputeNodeI}
\SetKwProg{myalg}{Algorithm}{}{}
\SetKwProg{myproc}{subroutine}{}{}
\myalg{\algo{$tree$}}{
	$d=tree.numLevels - 1$\;
		\ForPar{$k \gets d $ to $1$}{\label{ln:parCompBegin}
			\For{$node_i \in$ all $nodes$ at level k}{
				\proc($node_i$, $k$, $d$)\;
			}
		}\label{ln:parCompEnd}
	\proc($root$, $0$, $d$)\;\label{ln:rootComp}
	\Return $CI(root.\hat{\tau}, root.\hat{V})$\;
	}
 		
	\myproc{\proc{$node$, $k$, $d$}}{
			$\{c\} = node.children$\;
			\uIf{$k$ is $d$}{
				$node$.$\hat{\tau}$ = data item's value\;
			}
			\uElseIf{$k$ is $d -1$}{ 
				$m_{I_k}$ = $\{c\}.size()$,
				$S^2_{i, I_k}$ = $Var(\{c.\hat{{\tau}}\})$\;
				$node$.$\hat{\tau}$ = {\tt Eq\ref{eq:sumEstimate}}$(\{c.\hat\tau\}, m_{I_k}, {M}_{I_k})$\;
				$node$.$\hat{V}$ = {\tt Eq\ref{eq:multi-stage}}$(m_{I_k}, {M}_{I_k}, S^2_{k})$\;
			}
			\Else{
				$n_{I_k}$ = $\{c\}.size()$,
				$S_{u,{I_{k}}}^{2}$ = $Var(\{c.\hat\tau\})$\;
				$node$.$\hat{\tau}$ = {\tt Eq\ref{eq:sumEstimate}}$(\{c.\hat\tau\}, n_{I_k}, {N}_{I_k})$\;
				$node$.$\hat{V}$ = {\tt Eq\ref{eq:multi-stage}}$(n_{I_k}, {N}_{I_k}, S_{u,{I_{k}}}^{2}, \{c.\hat{V}\})$\;
			}
	
}
\caption{Confidence interval computation}
\label{algo: computeTree}
\end{algorithm}

We illustrate the computation process by using the tree in Figure~\ref{fig:provenanceTree} representing three-stage cluster sampling as an example. We begin with the level where $k = 2$, we first compute the intra-variance of $c_2$ formed by $c_2:e_1$ and $c_2:e_2$. After computing other clusters~($c_3$, $c_4$ and $c_5$) in the same level, we decrement $k$ to 1 and moves to second level nodes. Computing statistics~(e.g., intra/inter-cluster variance) for $P_1$ depends on $c_2$ and $c_3$'s statistics~(same for $P_3$), which has already been computed in the previous level. Finally, the variance at the root comprising $P_1$ and $P_3$ can be computed.

\reconsider{
To form a provenance graph, information for computing the error is both logged ``globally" at the driver program and ``locally" at data items in the RDD. Information such as which RDDs have been sampled with their sampling rates, and RDDs that have been filtered is logged ``globally" at the Spark driver program. The ``global" information will be sent to the reducers when errors/cluster sizes are being estimated. Information that is logged ``locally" is that, data items generated from the same parent data items by a {\tt flatMap} will be tagged with the same id to indicate that they should be treated as next level clusters during the error computation process. }

\reconsider{
\begin{algorithm}
	\SetAlgoLined
	\SetKwInOut{Input}{input}
	\SetKwInOut{Output}{output}
	\Input{sampling rates, data item clustering info, sampling stages $d$}
	\Output{Overall estimated variance}
	$k = d$ \;
	Using equation \ref{eq:multi-stage}: \\
\While{k $\geq$ 0}{
	\eIf{k = d}{ Compute each $\hat{V} (\hat{{\tau}}_{I_k})$ from the sample under the case when $k=d$, using the estimated value for $N_{I_k}$\;}{Compute each $\hat{V} (\hat{{\tau}}_{I_k})$ from the sample under the case when $ 0 \leq k <d$ using the estimated value for $M_{I_k}$\;
	}
	$k -= 1$
}
	\caption{Computing equation \ref{eq:multi-stage}}
	\label{algo: multistage}
\end{algorithm}}

\subsection{Per-key population estimation}
A Spark transformation can generate multiple keys, thus sampling before a transformation is equivalent of sampling a mixed-key population where sub-population size of each key is unknown. It is because sampling occurs before the transformation that actually generates a key. However, each sub-population size is needed because variance computation applies to each output key. We model estimated population size of a cluster at each sampling stage as a negative binomial distribution parameterized by sample size and sampling rate i.e. $\hat{N}_{I_k} \sim \mathcal{NB}(n_{I_k}, p)$, where $\hat{N}_{I_k}$ is the population size, $n_{I_k}$ is the sample size and $p$ is the sampling rate applied for the sub-clusters. $n_{I_k}$ at the current stage is equivalent to the estimated population size of the next stage~$(N_{I_{k}})$, $p$, $n_{I_k}$ and $N_{I_k}$ corresponds to the success rate, number of successes and the number of trials in a binomial distribution. The unbiased estimator $\hat{N}_{I_k}$ is $\frac{n_{I_k}}{p}$. The same logic also applies to the last sampling stage where the value of $M_{I_k}$ needs to be estimated. The uncertainty coupled with estimating $N_{I_k}$ and $M_{I_{k}}$, must be included in computing the variances in Eq~(\ref{eq:multi-stage}) since their estimators affects the variance. We have detailed derivation on incorporating it into the variance computation in the Appendix of our technical report~\cite{ApproxSparkTR}.

\futurenote{
The clustering relationship among the data items and partitions is determined by the transformation chain in a forward, however estimating the sums and variance using equations~\ref{eq:sumEstimate} and \ref{eq:multi-stage} are computed backwards from the last stage. In this particular example shown in Figure~\ref{fig:samplingFramework}, the number of stages is 3,}  

\reconsider{
	Logically, an RDD could be viewed as the $zeroth$ level cluster containing partitions, and the sub-clusters within each cluster are its ``data items" containing sub-clusters, with only the final output data items in $R_{out}$ being monolithic.  It is convenient to have this nested view because both sampling the output data items and clusters can be modeled as sampling the immediate higher level clusters' data items. For example, partition sampling can be viewed as sampling the data items of the $zeroth$ cluster, the entire RDD. Estimated statistics such as sums, variance, cluster population sizes can all be computed backward from the output data items in $R_{out}$ to the second cluster level, until the $zeroth$ cluster level in a recurrence fashion.
}

\reconsider{
	\guangyannote{cut?}The strategy we identify cluster sampling stages is through the appearances of {\tt flatMap}, which generates a group of data items from one data item. Thus it establishes one-to-many relationship between an data item in the RDD on which {\tt flatMap} is applied and the resulting group of data items in the following RDD. We treat implicitly the creation process of the first RDD in a transformation as the first {\tt flatMap}. For example, in Figure \ref{fig:samplingFramework}, the raw input is seen as a single data item and the HDFS blocks are modeled as the result of an application of {\tt flatMap} on the singular input which generates a number of blocks. On the other hand, {\tt map} only generates one data item from one data item, which forms one-to-one relationship between them. Thus {\tt map} does not create new cluster sampling stages, which means that RDDs created by consecutive {\tt map}s would be collapsed as the same sampling stage.

	The goal of mapping data items and RDD partitions to multi-stage sampling is to form a confidence interval for the estimated result, which requires estimating the sampling error for the estimator using information on how data items/partitions form multiple levels of clusters.    
}



\section{Stratified Reservoir Sampling}
\label{sec:reservoir-sampling}

An inherent limitation of multi-stage sampling is that some rare output keys may either be lost or have large error bounds. We leverage an one-pass sampling algorithm \textit{Adaptive Stratified Reservoir Sampling} (ASRS)~\cite{al2014adaptive} to address the rare key issues. ASRS combines stratified and reservoir sampling~\cite{lohr2009sampling, vitter1985random}, and uses power allocation\cite{bankier1988power} to divide the total sample size among different strata proportionally to each stratum's running sampling error. ASRS dynamically increases the sampling rates of rare keys to compensate for their larger sampling errors and decreases sampling rates on popular keys~\cite{al2014adaptive}.

\myparagraph{ASRS with partition sampling.}~ASRS has a larger overhead compared to simple random sampling. In order to achieve balance among output key retaining, balanced error bound distributions and the overall execution time, we sample RDD partitions at the input and apply ASRS over the their elements, so that in the chosen RDD partitions, sampling errors among popular and rare keys are more even and rare keys are better retained. Partition sampling at the input will have significant execution time saving since much I/O time is reduced. We can estimate the result and the error bound using standard multi-stage sampling theory using Eq~(\ref{eq:sumEstimate}) and (\ref{eq:multi-stage}), because an ASRS sample is very close to a simple random sample~\cite{al2014adaptive}.

\myparagraph{Limitations.}~ASRS stratifies the sample by output keys, thus it cannot be applied unless the output keys are available. However, sampling right before aggregation will not save execution time since aggregation is relatively cheap. Therefore our solution is to apply ASRS over an intermediate RDD, which would make ASRS suitable for applications where an output key's occurrence is proportional to an intermediate key.

\section{ApproxSpark Implementation} 
\label{sec:implementation}
We have implemented our approximation mechanisms by either modifying/extending the original Spark framework. We extend Spark executor implementation to maintain our data provenance tree. We also extend Spark's {\tt StatCounter} class to store intra and inter-cluster variances, sample sizes, sampling rates, etc. ApproxSpark offers two methods for user to set the degree of approximation, either by specifying the sampling rates or error bound targets. In addition to setting specified sampling rates, user is also able to set target error bounds at different percentiles on the error bound CDF of all keys. For example, a user may specify that the $10^{th}$ percentile of the error bound is at most 0.1, the $50^{th}$ percentile at 0.3, the $90^{th}$ percentile at most 0.6. 

\subsection{User-specified sampling rates}
\myparagraph{Multi-stage sampling.}~We modify the partition loading and computation mechanisms in Spark's {\tt HadoopRDD} class to support partition/input data item sampling when data is being loaded into an RDD. Subsequent RDDs' data items can be sampled using the original {\tt sample} function from Spark API. In order to forward information to the output RDD as in Table \ref{tbl:trans}, we extend the implementations of those transformations in RDD class to support the data provenance building algorithm shown in Algorithm~\ref{algo: buildTree}. For example, {\tt flatMap} not only tags a group of data items generated from the same data item $i$ in the parent RDD with a cluster id $c_i$, it also implements the provenance tree building logic. ApproxSpark provides the user with a new RDD transformation {\tt aggregateByKeyMultiStage}, for both intra and inter partition aggregations when multistage sampling is used. It is similar to RDD's original {\tt aggregateByKey} but has added error bound computation mechanisms. 
\begin{figure}
	\begin{center}
		\includegraphics[width=7cm]{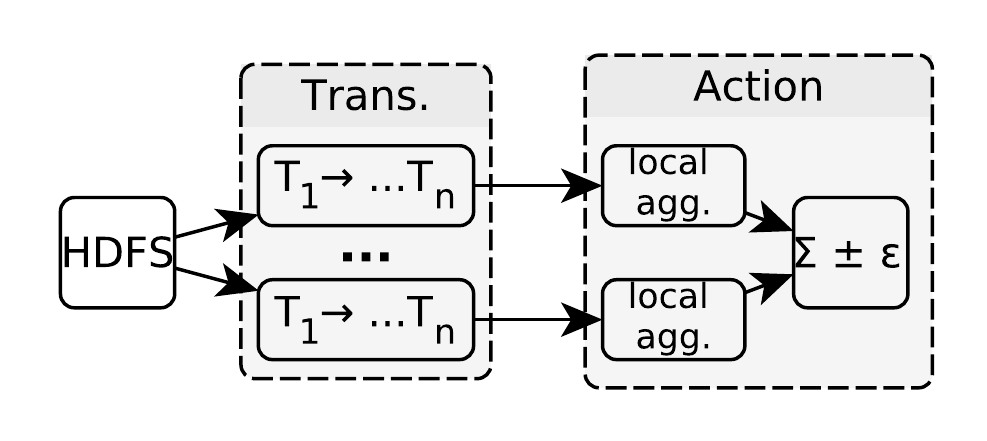} \\
	\end{center}
	\caption{Error bound computation process, divided across the transformation and action phases. The tree building happens in the transformation phase, and the error bound computation happens in the action phase.}
	\label{fig:errorBoundProcess}
\end{figure}

\myparagraph{Error bound estimation.} The error computation process is shown in Figure~\ref{fig:errorBoundProcess}. As introduced in Algorithm~\ref{algo: buildTree}, the first two levels of the provenance tree are sequentially built by the Spark driver program. Then every subtree rooted at each partition node~(sans the partition nodes) are built by each parallel task in the transformation phase, maintained by a coordinator in each Spark executor. In the action phase, an RDD partition is first locally aggregated by each Spark executor before sending them to reducers across the network for final aggregation. In the local aggregation phase, the subtree of each partition is traversed to compute each partition's statistics. Then in the final aggregation, the statistics of each partition are sent to the reduces for computing the inter-cluster variance among the RDD partitions and the final confidence interval. As transformations $T_1 \rightarrow T_n$ execute in parallel on every partition in the transformation phase, the subtree for every partition of the provenance tree is built; in the action phase, partitions are first locally aggregated to compute each partition's statistics, then sent to the reducers across network for the final error bound computation.

\myparagraph{Stratified reservoir sampling.}~We  modify ASRS for Spark's distributed environment by dividing the total reservoir size, taken as a user input, evenly among RDD partitions. Each partition is then sampled using ASRS independently without coordination among them.  We implement ASRS as a transformation that produces another RDD, containing the resulting sample with balanced sampling errors among popular and rare keys. ASRS changes the sampling rate by changing the size of the portion of the reservoir allocated for a particular key. On the other hand, ASRS shrinks the size allocated to each existing key as it discovers more keys in the partition, where the initial reservoir size for the new key is set as the average of the sizes for existing keys. The benefit of implementing ASRS as a transformation is that the resulting RDD can be cached in memory for reuse. We provide {\tt ASRSSample} for the user to sample an RDD using ASRS and {\tt aggregateByKeyStratified} for the aggregation with error bound computation, both implemented as RDD transformations.    

\reconsider{
\begin{lstlisting}[caption={ApproxSpark approximate word count},captionpos=b, style=java]
val sc = new SparkContext(new SparkConf())
.setAppName("WordCount"))
// p1: partition sampling rate
// p2: input data item sampling rate
val tokenized = sc.textFileWithSampling(args(0), p1, p2)
.flatMap(_.split(" ,;."))
// count the occurrence of each word
val wordCounts = tokenized.map((_, 1))
.countByKeyApproxMultiStage() 
// containing statistics of each word
wordCounts.saveAsTextFile()
\end{lstlisting}
}
\subsection{User-specified target error bounds}

\begin{figure}
	\vspace{-0.5cm}
	\begin{center}
		\includegraphics[width=6cm]{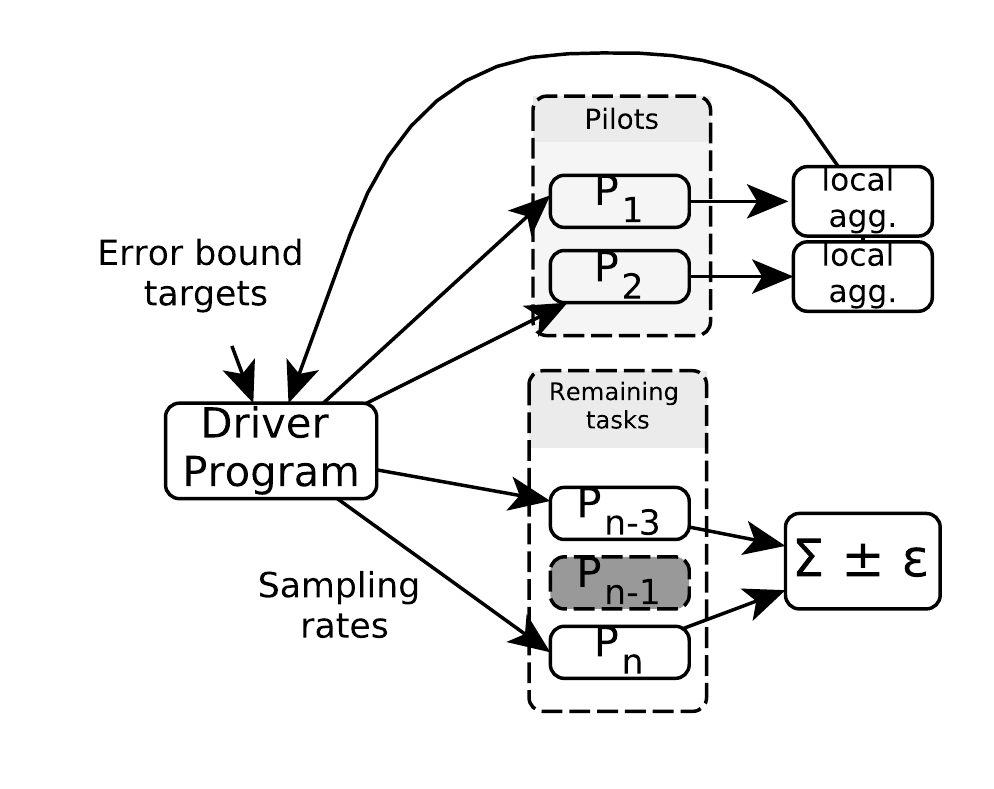} \\
	\end{center}
	\vspace{-0.5cm}
	\caption{User setting error bound targets design, gray shaded box is dropped partition(s).}
	\label{fig:UsersettingErrorBoundDesign}
	\vspace{-0.5cm}
\end{figure}

\reconsider{
\begin{algorithm}
	\SetAlgoLined
	\SetKwInOut{Input}{input}
	\SetKwInOut{Output}{output}
	\Input{$t_{perc1}$,...$t_{percn}$, $pCluster$ = 1.0, $pItem$ = 1.0}
	\Output{$pCluster$ and $pItem$}
	-------------------------------Phase I----------------------------\\
	Execute pilot tasks and return partially aggregated partitions to the driver\;
	Estimate ${M}$, $S^2_{i}$ and $S^2_{u}$

	\While{None of the predicted error exceeds the target error}{
		Use $pCluster$ to  
		compute error bounds in $perc1$, $perc2$...$percn$ \;
		$pCluster$ -= $bstep$\;
	}
	$pCluster$ += $bstep$\;
	fix $pCluster$, find $pItem$ in the same way as $pCluster$\;
	-------------------------------Phase II----------------------------\\
	Continue the remaining Spark tasks setting sampling rates to be $pCluster$ and $pItem$\;
	\caption{Two-stage sampling rates search}
	\label{algo: userSettingErrorBounds}
\end{algorithm}
}

We propose a greedy algorithm to search for a sampling rate combination leading to a potentially tight error bound CDF constrained by the target errors, while aiming to significantly reduce execution time. Initially, partition and data item sampling rates are both initialized as 1.0. The algorithm includes two phases. 1) In the first phase, a wave of pilot tasks are executed and the partially aggregated results from these tasks are sent back to the driver program, where the number of data items $M$ and inter/intra cluster variances for each key are computed. It uses a Spark's job submission mode that returns the partially aggregated partitions to the driver instead of sending them for shuffling. 2) In the second phase, the algorithm uses statistics gathered in the first phase to predict error bounds: it first lowers partition sampling rate for potentially maximum execution time reduction until it would violate any error bound target, then it searches for an appropriate input data item sampling rate, before the predicted error CDF would violate any of the user-specified targets. When predicting errors for keys that are not encountered in the first phase, the algorithm just uses the average of the statistics for the keys obtained in the first phase. When the predicted error distribution meets all the error targets with the lowest possible sampling rates, the algorithm proceeds to the second phase and uses them for the remaining Spark tasks. Figure~\ref{fig:UsersettingErrorBoundDesign} shows the architecture of user setting error bounds.

The algorithm exploits a property that partition sampling may incur more sampling error~\cite{lohr2009sampling}, but reduces more execution time compared with data item sampling. Our algorithm follows a principle in online aggregation - \textit{minimum time to accuracy}~\cite{hellerstein1997online}, i.e. minimizing the time to achieve a useful estimated value. However, online aggregation typically outputs a running confidence interval for a single estimator as data is being aggregated in a random order, whereas ApproxSpark applies multi-stage sampling over the data and outputs the error bounds for multiple keys at the end of execution.
\reconsider{
\myparagraph{Discussions.}~Our proposed searching algorithm only considers partition and data item sampling rate for the input RDD. We think that sampling the input RDD is sufficient to meet the error bound targets without too much added complexity. It is because sampling operations placed far from the input will have diminishing effect over the error bound as well as execution time. In fact, expanding the algorithm to search for more than two sampling rates is straightforward, as we can add more while loops to find sampling rates for following RDDs down the transformation chain. }

\myparagraph{Limitations.}~The algorithm assumes the keys are distributed evenly and the pilot partitions are representative of the entire dataset. However, when keys are not distributed evenly, the pilot wave is not able to accurately estimate the parameters. The error bound computation in our implementation has only considered two-stage sampling, while in theory, user can insert multiple sampling operations along the chain and achieve the target error bounds. We leave this more complicated case as future work.

\section{evaluation}
\label{sec:eval}
We evaluate ApproxSpark using five real world applications from different application domains (see Table \ref{tbl:evalApp}). We begin by briefly describing the applications. We then use them to extensively explore the tradeoff space between sampling and precision. Finally, we explore ApproxSpark's ability to find appropriate sampling rates for user specified target error bound constraints.

\myparagraph{Experimental environment.}
All experiments are run on a cluster of four servers.  Each server is equipped with a 2.5GHZ Intel Xeon CPU  with 12 cores, 256GB of RAM, and a SATA hard disk. The cluster is interconnected with 1Gbps Ethernet.  All servers run Linux 3.10.0.  ApproxSpark is implemented on top of Spark version 1.6.1 and is configured with 16 executors, each of which runs up to 6 tasks, so that each server has 4 executors, running up to 24 tasks.

\subsection{Applications}
\begin{table}
	{\footnotesize
		\begin{center}
			\begin{tabular}{lllr}
				\hline
				& & & \textbf{Size} \\ 
				\textbf{Application} &\textbf{Domain} & \textbf{Input Dataset}& \textbf{(GB)} \\ \hline \hline
				Co-occur  & Text Mining & MEDLINE database  & 7.5 \\ \hline
				Speed  & Smart City & GPS trace &  36.0  \\ \hline
				Twitter  & NLP & Tweets2011 (TREC)  &  2.2 \\ \hline
				PageRank  & Graph Analysis &  Wikipedia snapshot  &  53.0    \\ \hline
				Clickstream & Log Analysis & Wikipedia clickstream & 6.5 \\
				\hline
			\end{tabular}
		\end{center}
	}
	\caption{List of applications, the domains they come from, and the input datasets used in our evaluation.}
		\vspace{-0.5cm}
	\label{tbl:evalApp}
\end{table}
\reconsider{
\begin{figure}
	\vspace{0cm}
	\begin{center}
		\includegraphics[width=7cm]{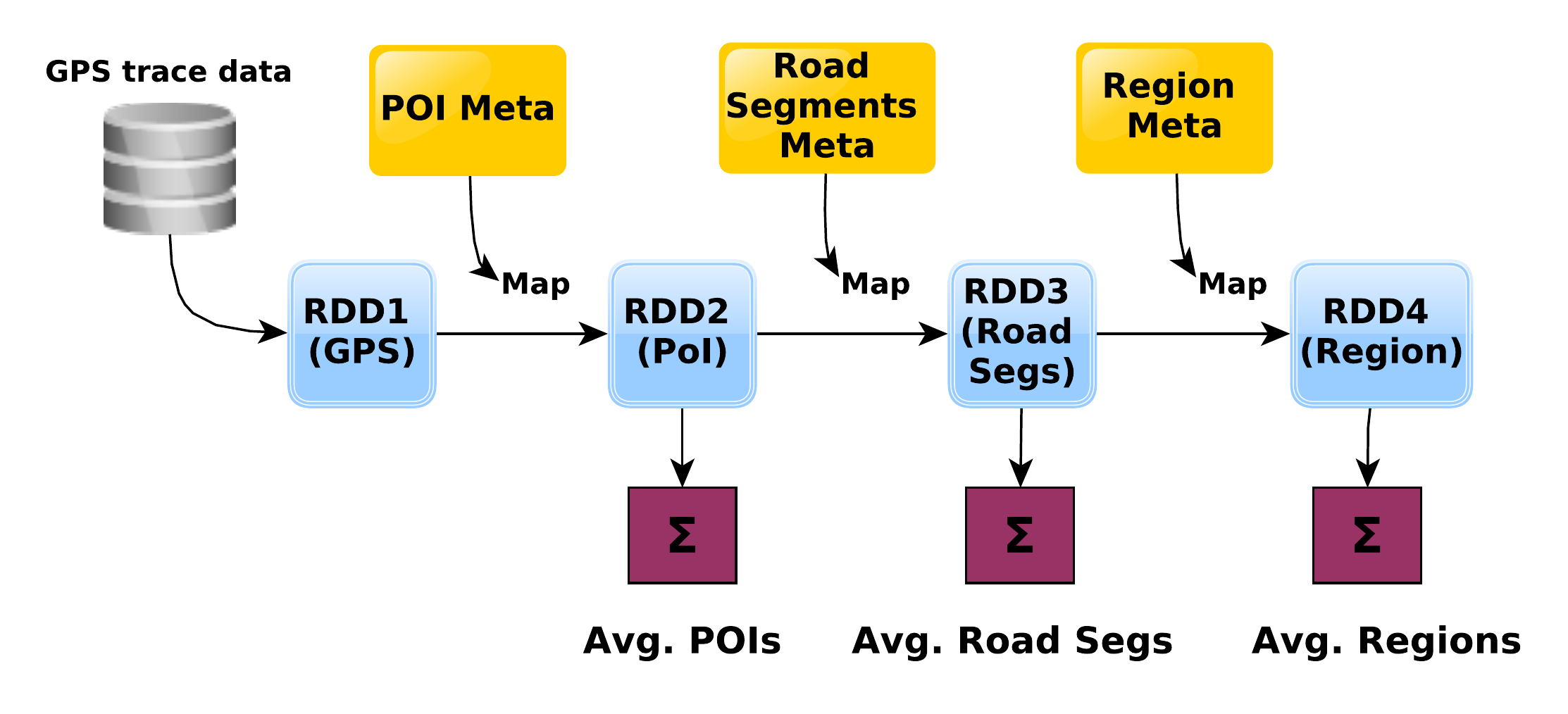} \\
	\end{center}
    \vspace{-0.5cm}
	\caption{Multi-step transformation for Speed}
	\label{fig:SpeedIllustration}
	    \vspace{-0.5cm}
\end{figure}}

\myparagraph[top]{Word Co-occurrence (Co-occur).} Co-occurrence is a common text mining application that computes the frequencies of pairs of words~\cite{berkhin2006survey}. In this study, the application counts co-occurrences of topic tags in the MEDLINE database~\cite{Medline}, containing more than 20M citation records of publications in life sciences. Each citation record contains a set of topic tags, listing the major topics relevant to the publication. The application first reads the input data into an RDD, and then performs a {\tt map} to extract the list of major topic tags from each citation record.  It then performs a {\tt flatMap} to generate key-value pairs ((co-occurring tag pair), 1).  Finally, it sums and outputs the count of each co-occurring tag pairs.

\myparagraph{Vehicular Average Speed Analysis (Speed).}  
This application analyzes the average speed of vehicles moving in a geographical area each hour at three different granularities: 
around a point-of-interest (POI) (e.g., a restaurant), on 
a road segment, and within a region.  
An analysis of vehicular traces is useful for monitoring urban traffic, predicting passenger demand, recommending taxi routes, etc.~\cite{Carpooling}.  
We analyze a taxi GPS dataset containing status records collected every 30 seconds from 14,000 taxis operating in Shenzhen, China, over one week~\cite{Zhang:2014:EHM:2639108.2639116}.  
Each record contains information about a taxi, including a timestamp and the taxi's GPS location and speed. 
The dataset has $\sim$291M records that covers an area of $\sim$790 square miles divided into 491 regions, containing $\sim$569k POIs and $\sim$198k road segments. Each POI is assigned to a road segment and each road segment belongs to a region. The application reads the input data into an RDD, and then performs three transformations using metadata and three actions.  
The three transformations are three {\tt map} operations that: 
(1) transform each GPS entry into a ((POI, hour), speed) key-value pair; 
(2) transform each ((POI, hour), speed) pair into a ((road segment, hour), speed) key-value pair; and,
(3) transform each ((road segment, hour), speed) pair into a ((region, hour), speed) pair.  
The three actions use the three intermediate RDDs to compute the average speed per hour at each POI, each road segment, and each region, respectively.

\myparagraph{Twitter Hashtags Sentiment Analysis (Twitter).} Sentiment analysis computes quantitatively whether a piece of text is positive, negative or neutral using natural language processing (NLP) techniques~\cite{liu2012sentiment}. In this study, the application computes the average sentiment for each unique hashtag in the Tweets2011 Twitter dataset from TREC 2011 \cite{tweets2011}, using the Stanford CoreNLP library \cite{stanfordcorenlp}. This dataset contains $\sim$16M tweets sampled over 17 days in early 2011. The application first reads the input data into an RDD, and then performs a {\tt map} to compute a score in the interval $[0,5]$ with 0 being \textit{very negative}, 3 being \textit{neutral}, and 5 being \textit{very positive} for each tweet. It then performs a {\tt flatMap} to extract all hashtags from each tweet and associates each with the sentiment score for the tweet.  Finally, it computes and outputs the average sentiment for each hashtag.

\myparagraph{WikiPageRank (PageRank).} This application counts the number of articles that link to each article in a set, emulating one of the main processing components of PageRank \cite{PageRank}. We use the Wikipedia data snapshot from 2016 with $\sim$5M articles \cite {Wikipedia}.  The application first loads the data into an RDD, then applies a {\tt map} to parse the XML, generating a list of outbound links for each article.  It next performs a {\tt flatMap} to generate pairs of (destination article, 1).  Finally, it sums and outputs the count for each destination article.

\myparagraph{WikiClickstream (Clickstream).} Clickstream analysis can be used to generate a weighted network of linked articles showing the probability of users navigating from one article to another. We use a Wikipedia clickstream dataset from 2016  \cite{clickstream} containing $\sim$149M tuples of (source, destination, count), where count is the number of times that a user has visited the destination page from the source page.  The application computes the total count for each unique (source, destination) pair. Specifically, it reads the input data into an RDD, performs a {\tt map} to generate a key-value pair for each entry, and then sums and outputs the total count for each unique (source, destination) pair.

\subsection{Results for multi-stage sampling}

We explore the performance and accuracy of multi-stage sampling using four of the above applications: Co-occur, Twitter, WikiPageRank, and WikiClickstream. In most experiments, we sample the input data as it is read into the first RDD because this will lead to the highest speedups. However, we also explore sampling from RDDs later in the applications' transformation chains to explore the trade-off between performance and accuracy of such scenarios.

\begin{figure}
	\begin{center}
		\centering
		\subfloat[Co-occur] {\includegraphics[width=0.5\linewidth]{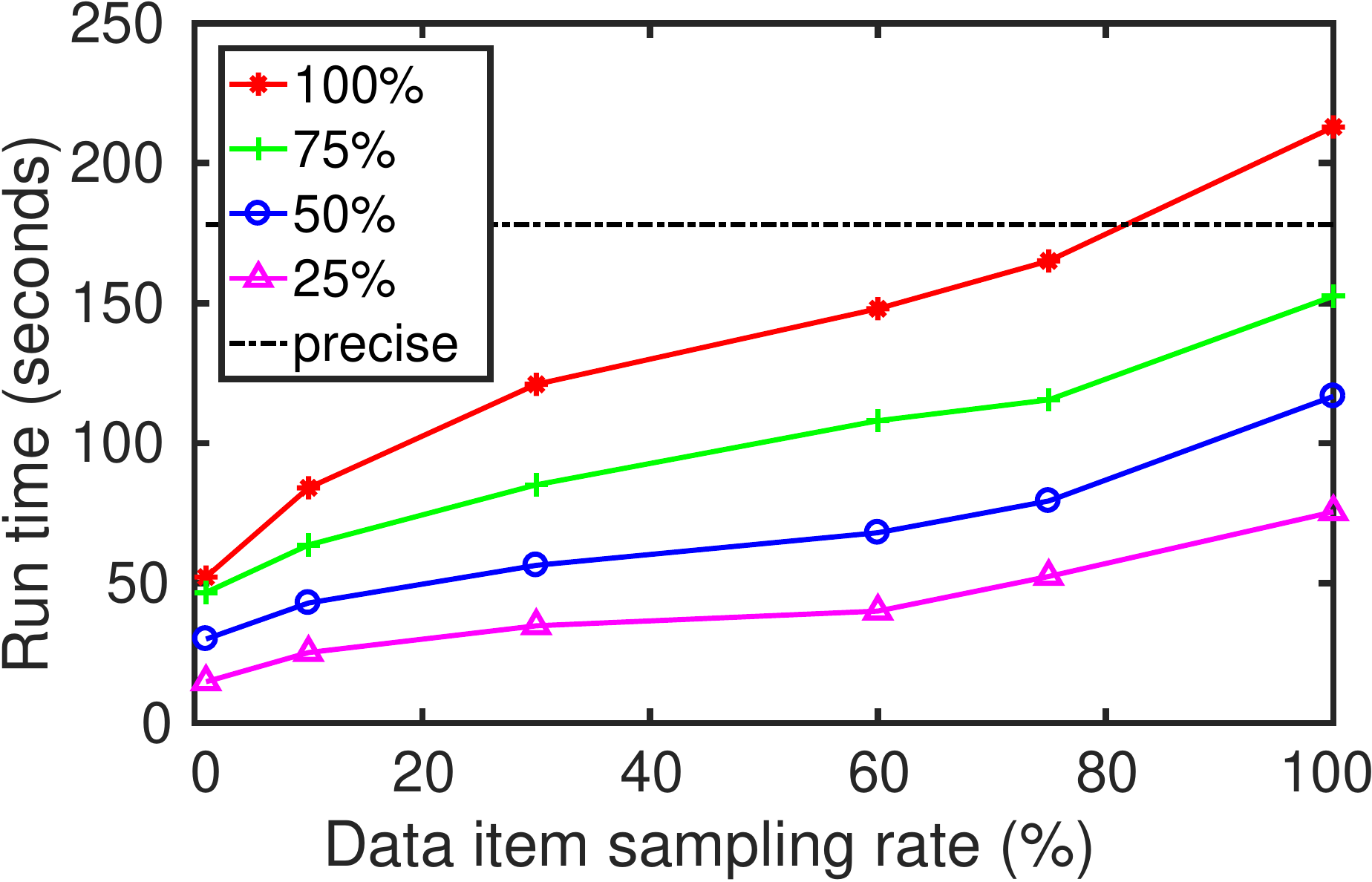}}
		\subfloat[WikiPageRank] {\includegraphics[width=0.5\linewidth]{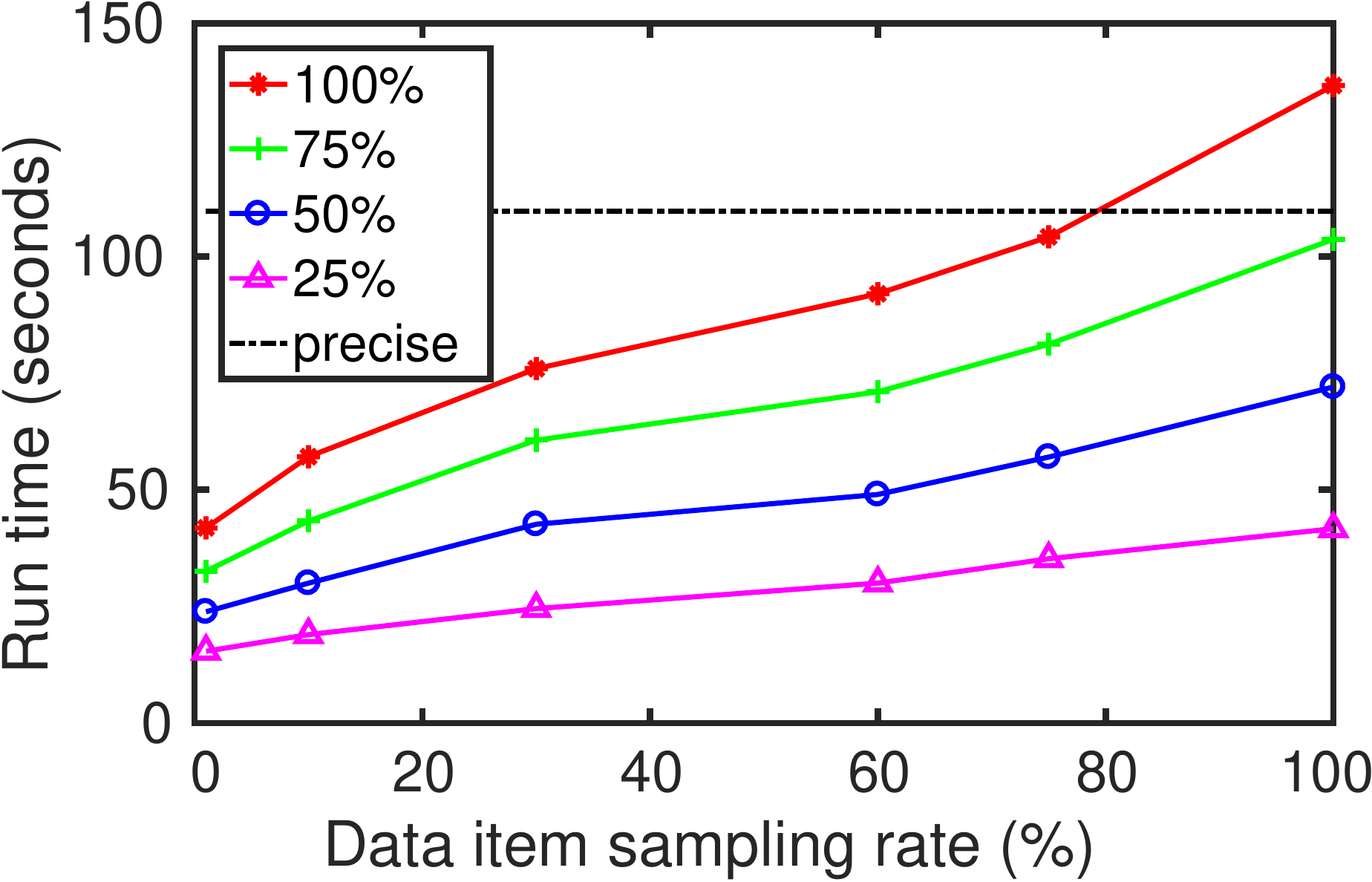}}
	\end{center}
\vspace{-0.2cm}
	\caption{Execution times under different sampling rates. Each line corresponds to a partition sampling rate. The x-axis shows the sampling rate for input data items. The dashed line gives the run time of precise executions.}
	\vspace{-0.5cm}
	\label{fig:clusterRuntime}
\end{figure}

\begin{figure}[!tbp]
	\begin{center}
		\centering
		\subfloat[Co-occur] {\includegraphics[width=0.5\linewidth]{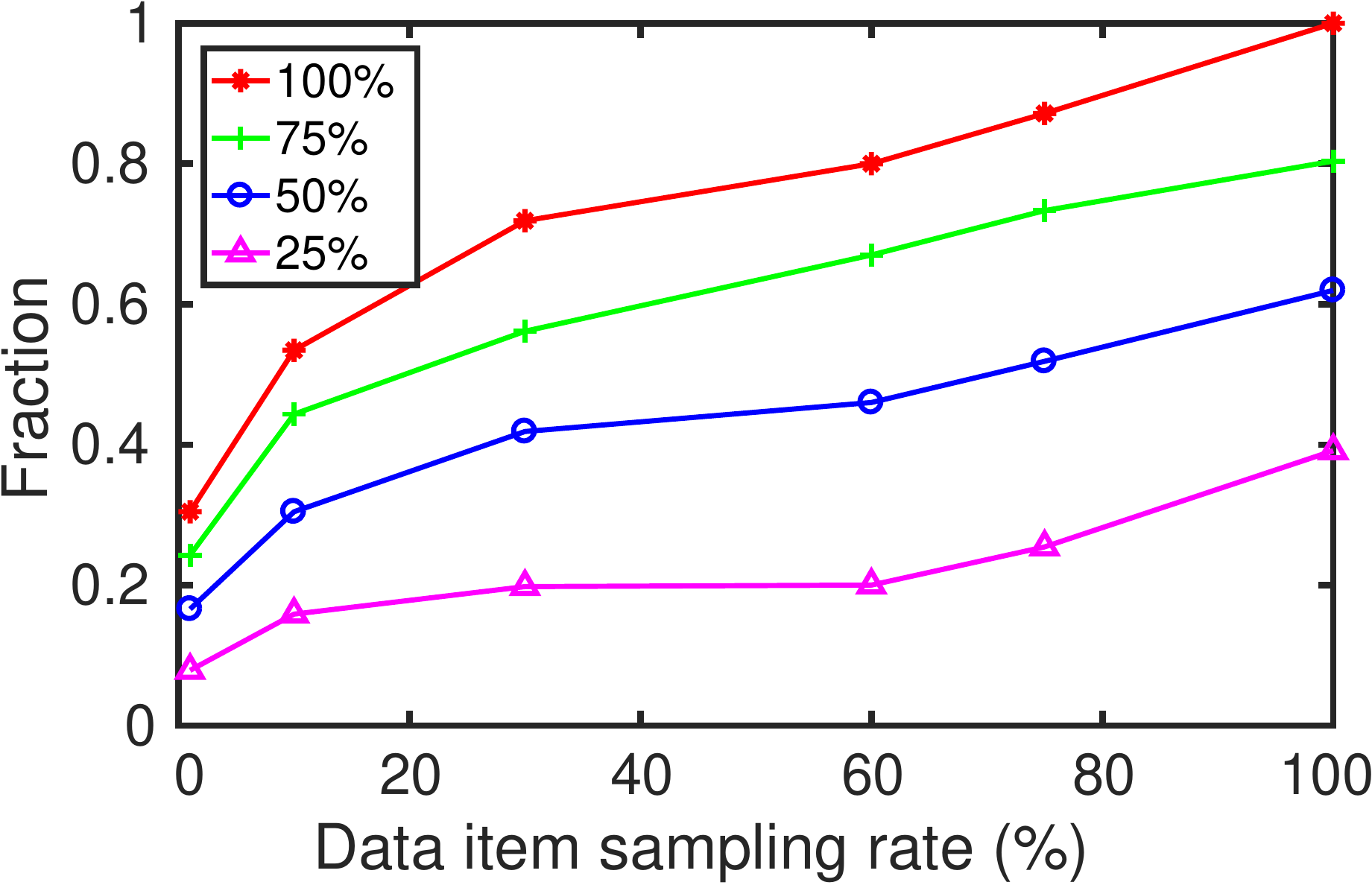}}   
		\subfloat[WikiPageRank] {\includegraphics[width=0.5\linewidth]{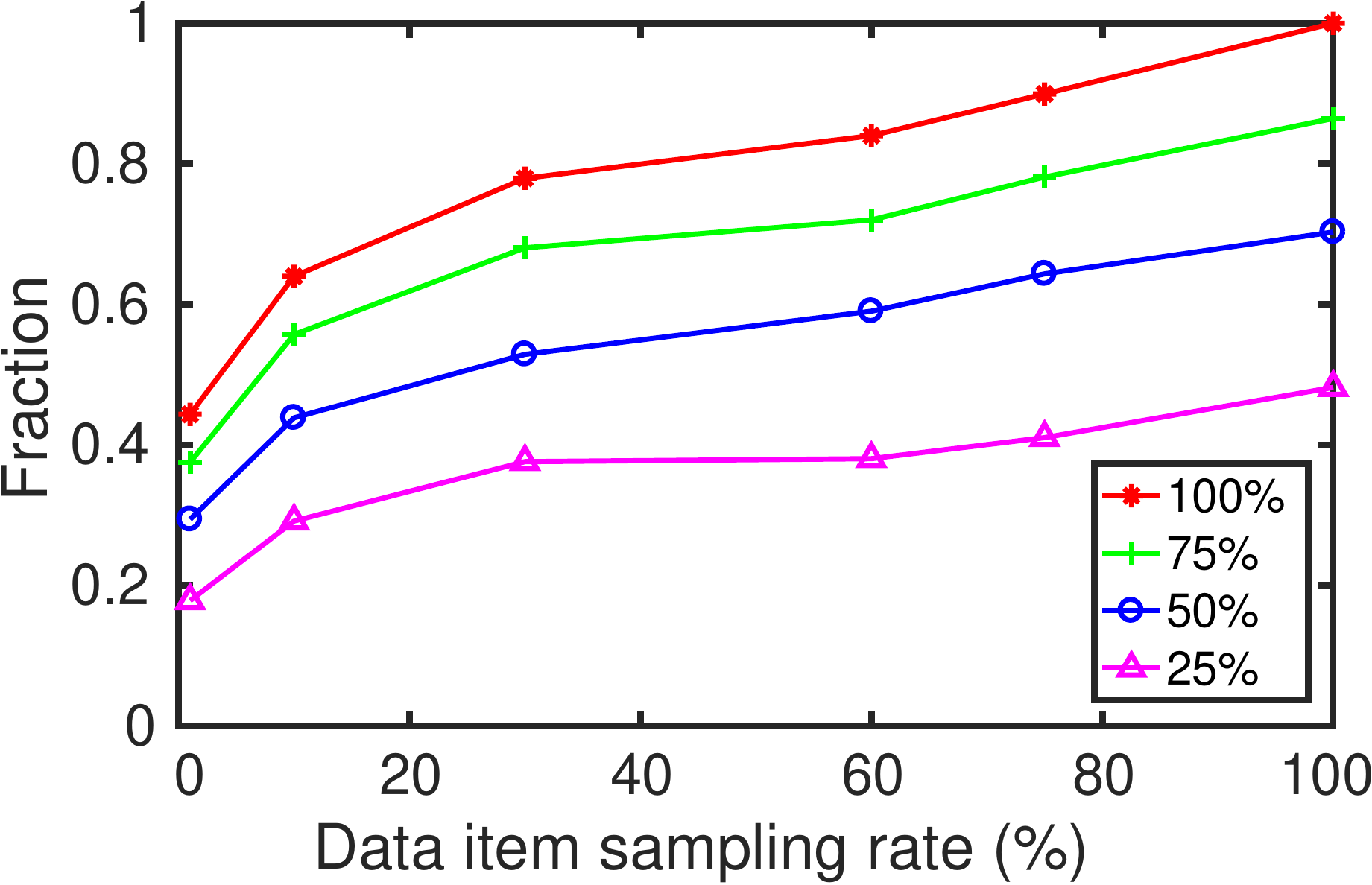}}
	\end{center}
\vspace{-0.2cm}
	\caption{Fraction of unique keys (normalized against number of keys produced under precise execution) outputed under different sampling rates. Each line represents a particular partition sampling rate. 
	}
\vspace{-0.5cm}
	\label{fig:numKeys-ClusterSampling}
\end{figure}

\begin{figure}[!tbp]
	\begin{center}
		\centering
		\subfloat[Co-occur] {\includegraphics[width=0.5\linewidth]{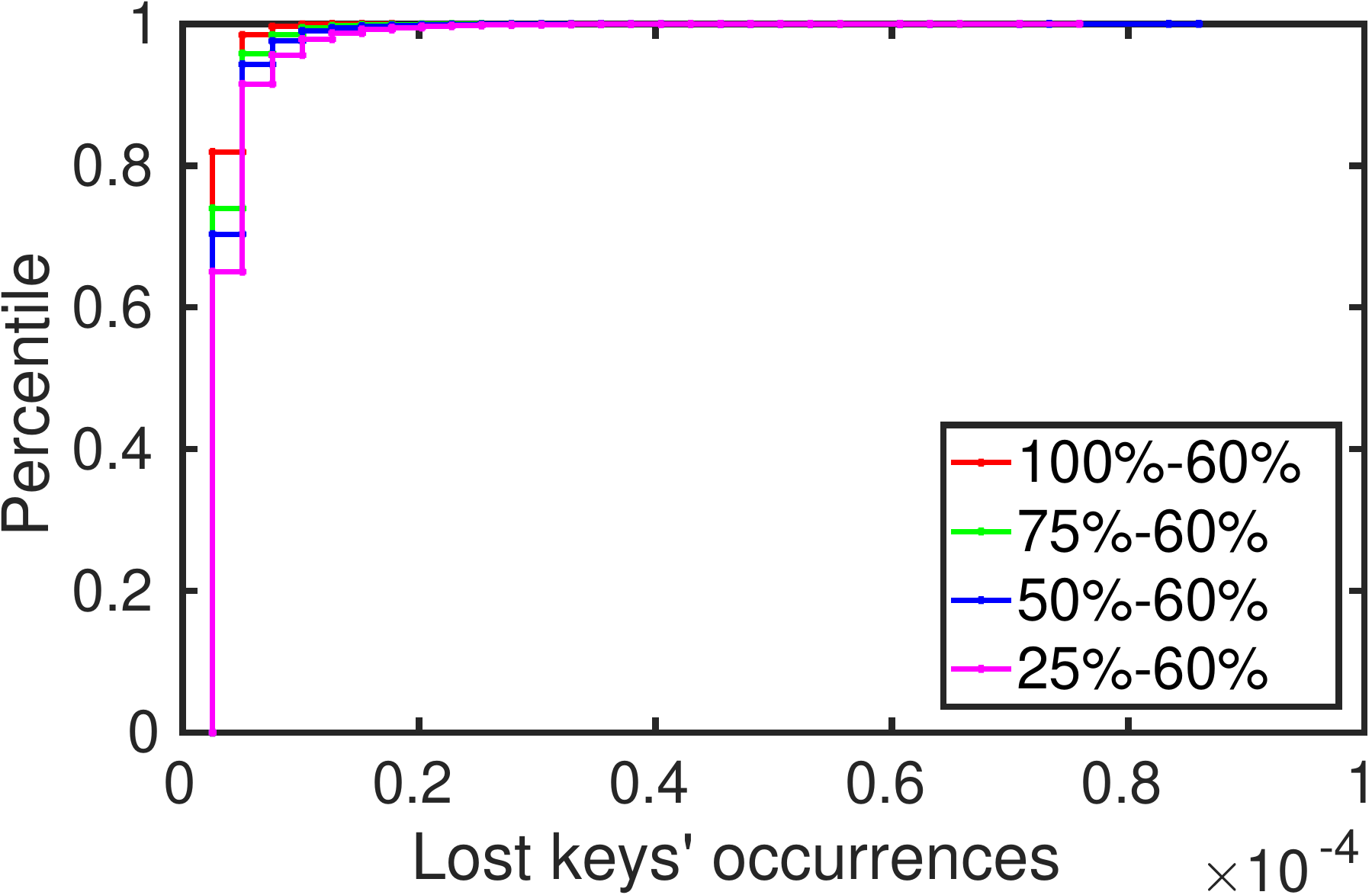}}   
		\subfloat[WikiPageRank] {\includegraphics[width=0.5\linewidth]{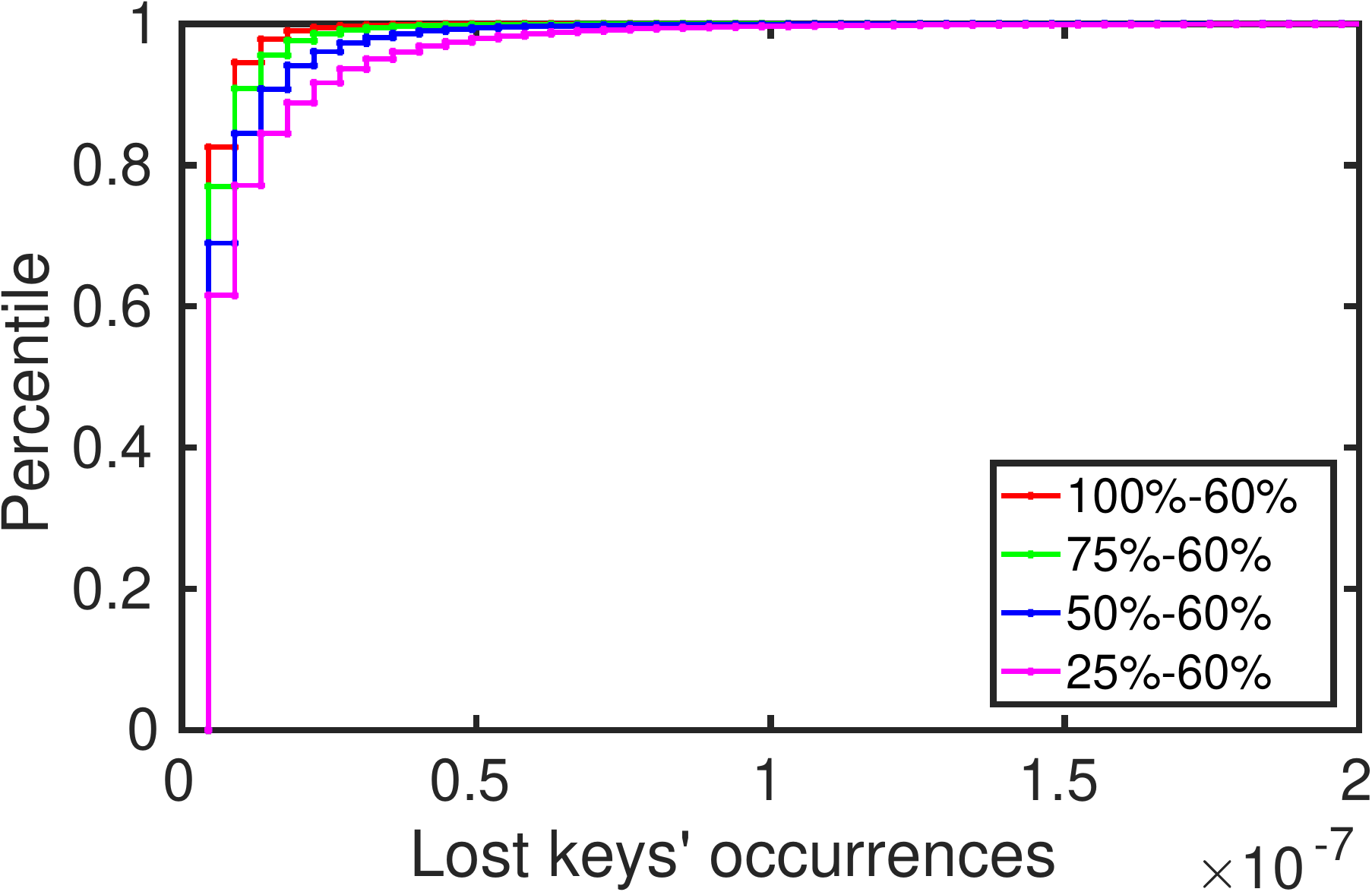}}
	\end{center}
\vspace{-0.2cm}
	\caption{CDFs of occurrences of the lost keys, normalized against the total number of data items across all keys at a data item sampling rate of 60\%.  Each line corresponds to a specific partition sampling rate.}
	\vspace{-0.5cm}
	\label{fig:missedKeys-ClusterSampling}
\end{figure}

\begin{figure}[!tbp]
	\begin{center}

		\centering
		\subfloat[Data item sampling rate - 75\%.] {\includegraphics[width=0.5\linewidth]{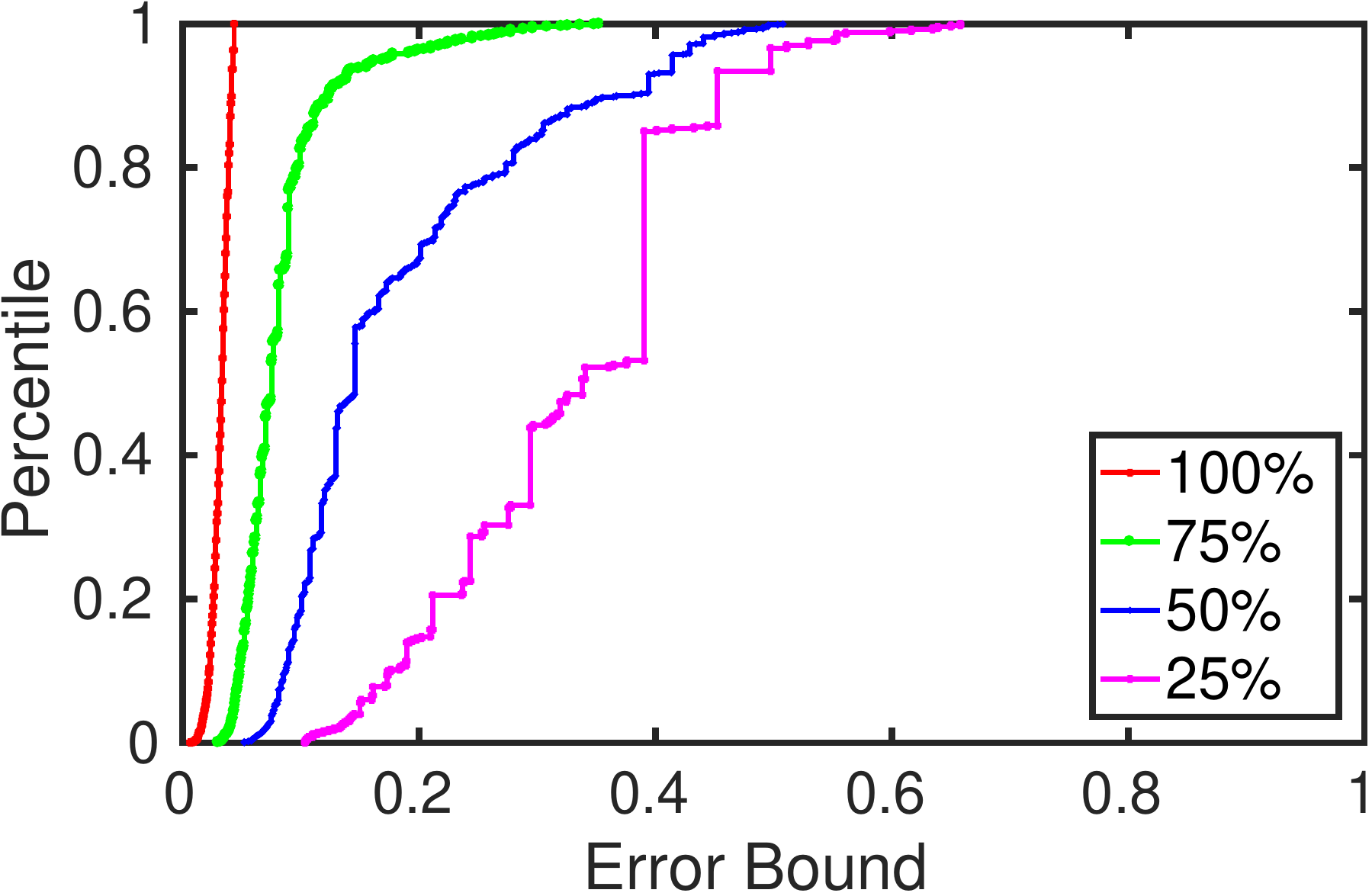}}
		\subfloat[Data item sampling rate - 60\%.] {\includegraphics[width=0.5\linewidth]{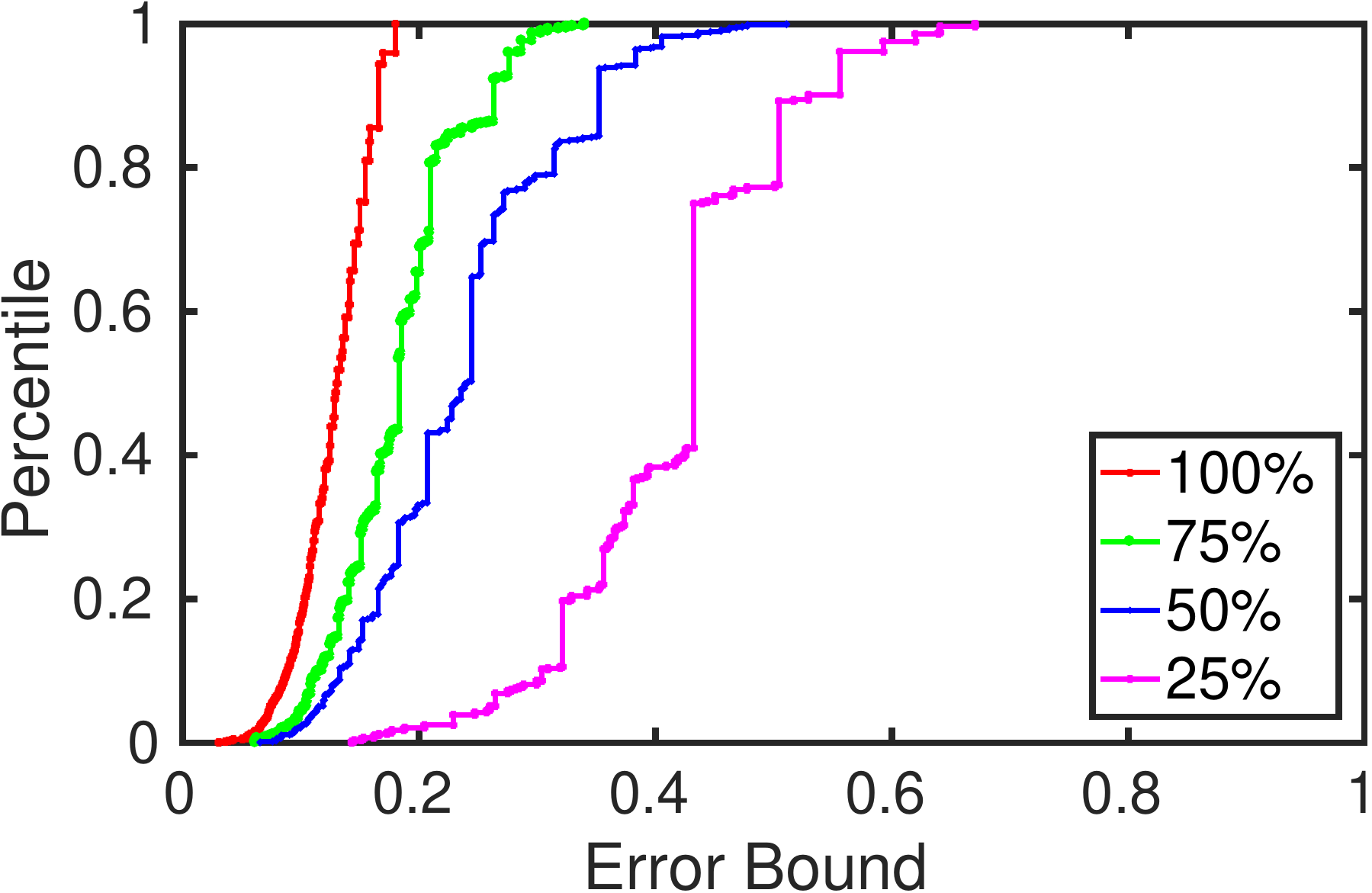}}
	\end{center}
\vspace{-0.2cm}
	\caption{Each graph plots CDFs of errors with 95\% confidence error at a fixed input data item sampling rate for Co-occur application. Each line in a graph plots the error CDF at a particular partition sampling rate. }
		\vspace{-0.5cm}
	\label{fig:medlineErrorpartitionSampling}
\end{figure}

\begin{figure}[!tbp]

	\begin{center}
		\centering
		\subfloat[Partition sampling rate - 50\%.] {\includegraphics[width=0.5\linewidth]{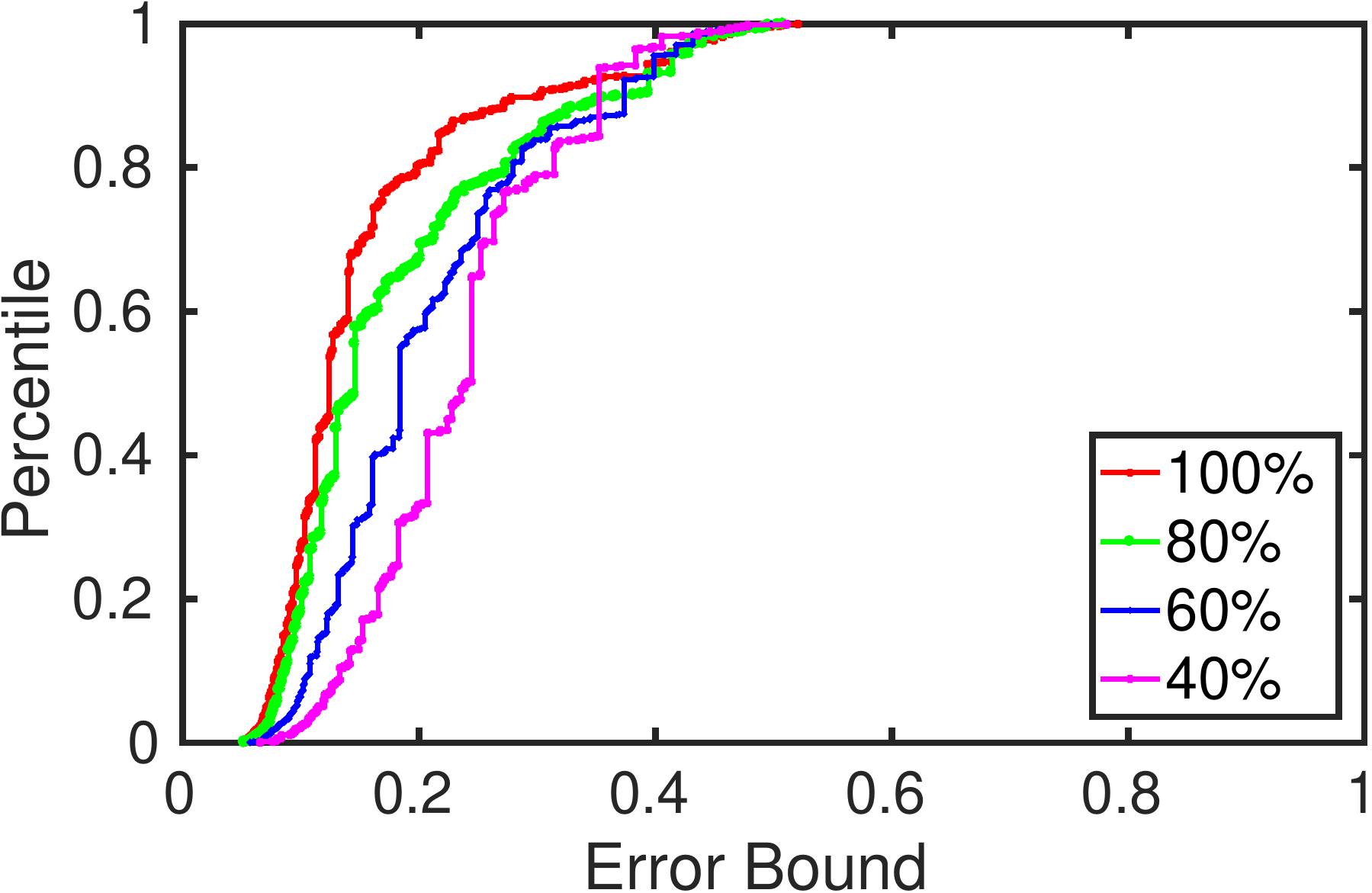}} 
		\subfloat[Partition sampling rate - 25\%.] {\includegraphics[width=0.5\linewidth]{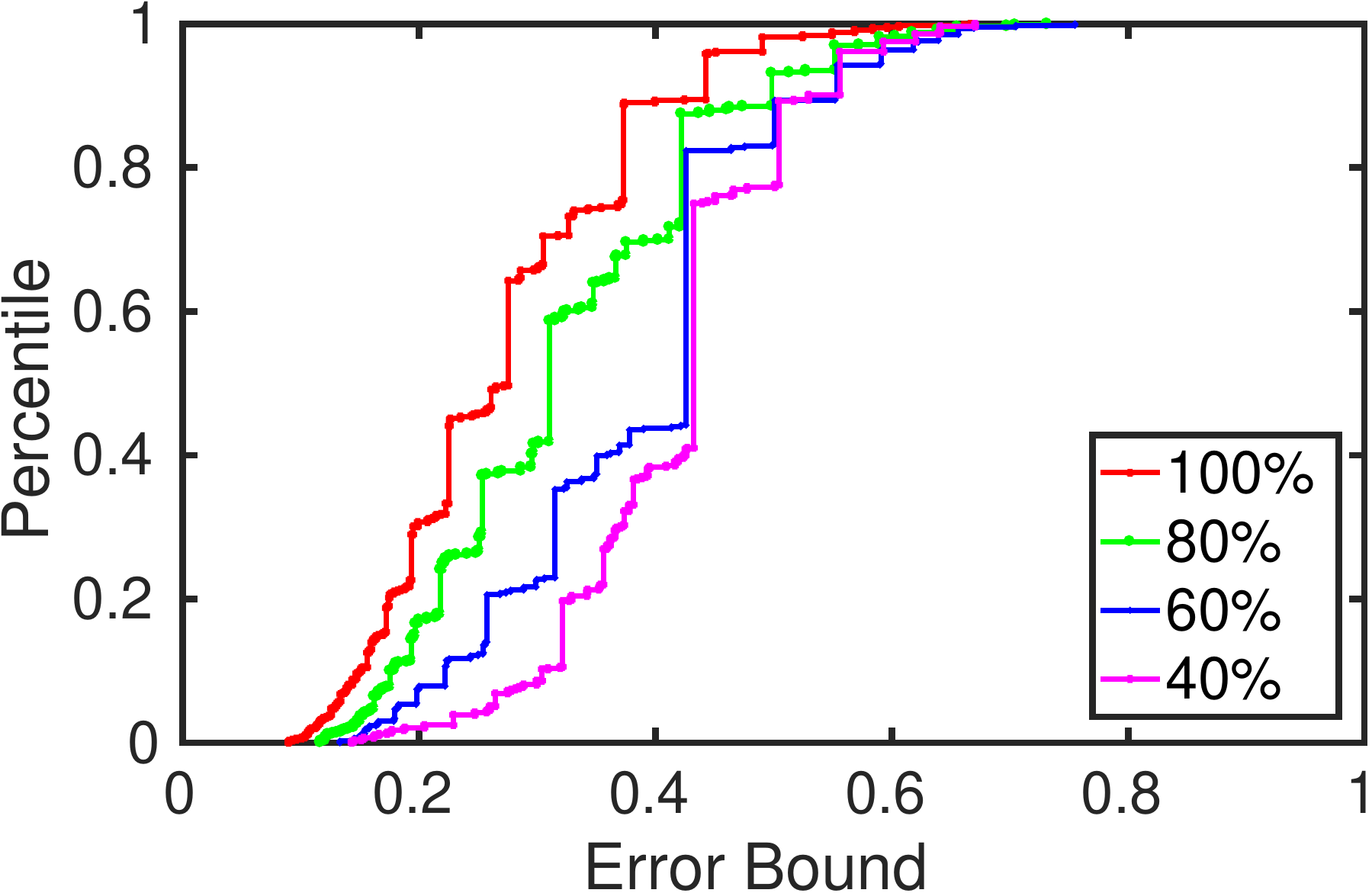}} 
	\end{center}
	\caption{CDFs of errors at a fixed partition sampling rate for Co-occur application. Each line in a graph plots the error CDF at a particular input data item sampling rate. }
	\label{fig:medlineErrorItemSampling}
\end{figure}

\myparagraph{Execution times.} Figure \ref{fig:clusterRuntime} plots the execution times for two of the applications, Co-occur, WikiPageRank at different partition and data item
sampling rates.  Consistent with previous results from
\cite{ApproxHadoop}, we observe that (a) multi-stage sampling significantly reduces execution times, and (b) partition sampling can
lead to larger execution time savings than data item sampling. The
latter is because dropping a partition eliminates overheads such
as I/O time for reading the blocks, the creation of an RDD
partition in memory, etc., whereas data item sampling still requires
some processing for each partition.  The sampling framework imposes some overheads; i.e., execution time for the (100\% (partition sampling), 100\% (data item sampling)) case is somewhat greater than that of the precise version, ran on unmodified Spark.


\begin{figure}[!tbp]
	\begin{center}
		\centering
		\subfloat[] {\includegraphics[width=0.5\linewidth]{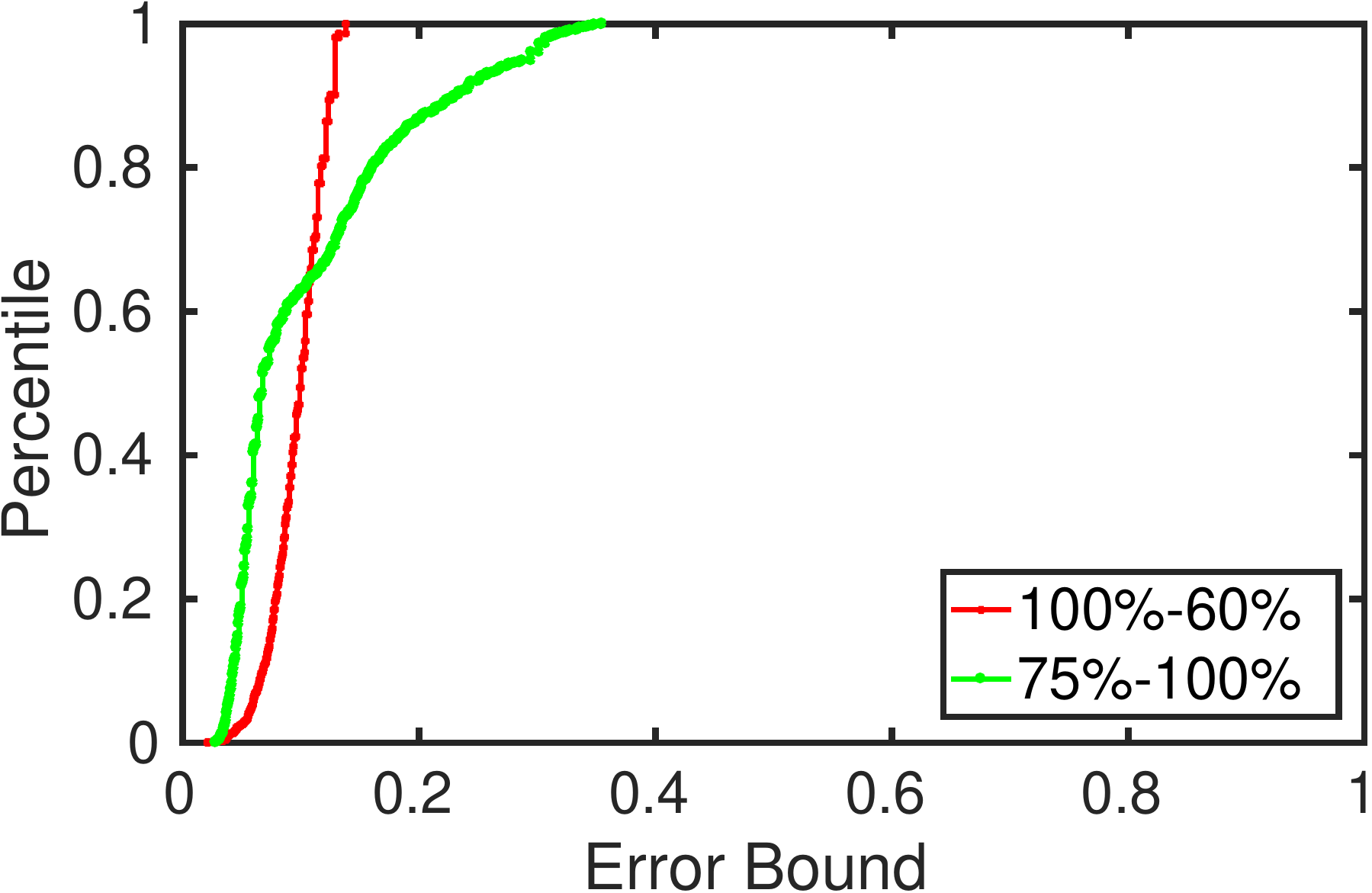}}   
		\subfloat[] {\includegraphics[width=0.5\linewidth]{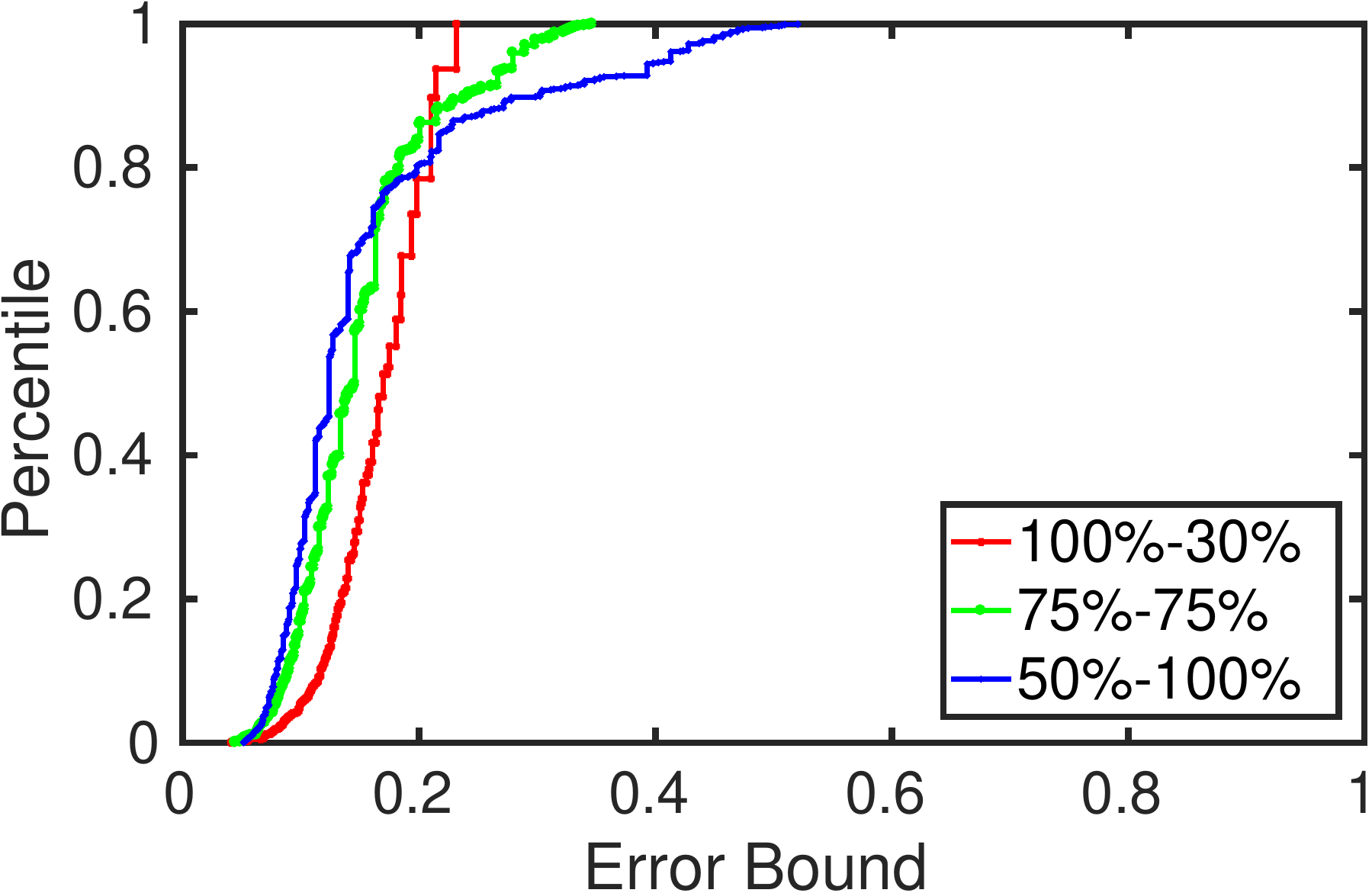}}
	\end{center}
\vspace{-0.2cm}
	\caption{Error distribution trade off under different partition and data item sampling rates combination from the Co-occur application. The legends indicate partition and data item sampling rates respectively.}
		\vspace{-0.5cm}
	\label{fig:medlineErrorTradeoff}
\end{figure}

\begin{figure}[!tbp]
	\begin{center}
		\centering
		\subfloat[WikiPageRank, 75\%-50\%] {\includegraphics[width=0.5\linewidth]{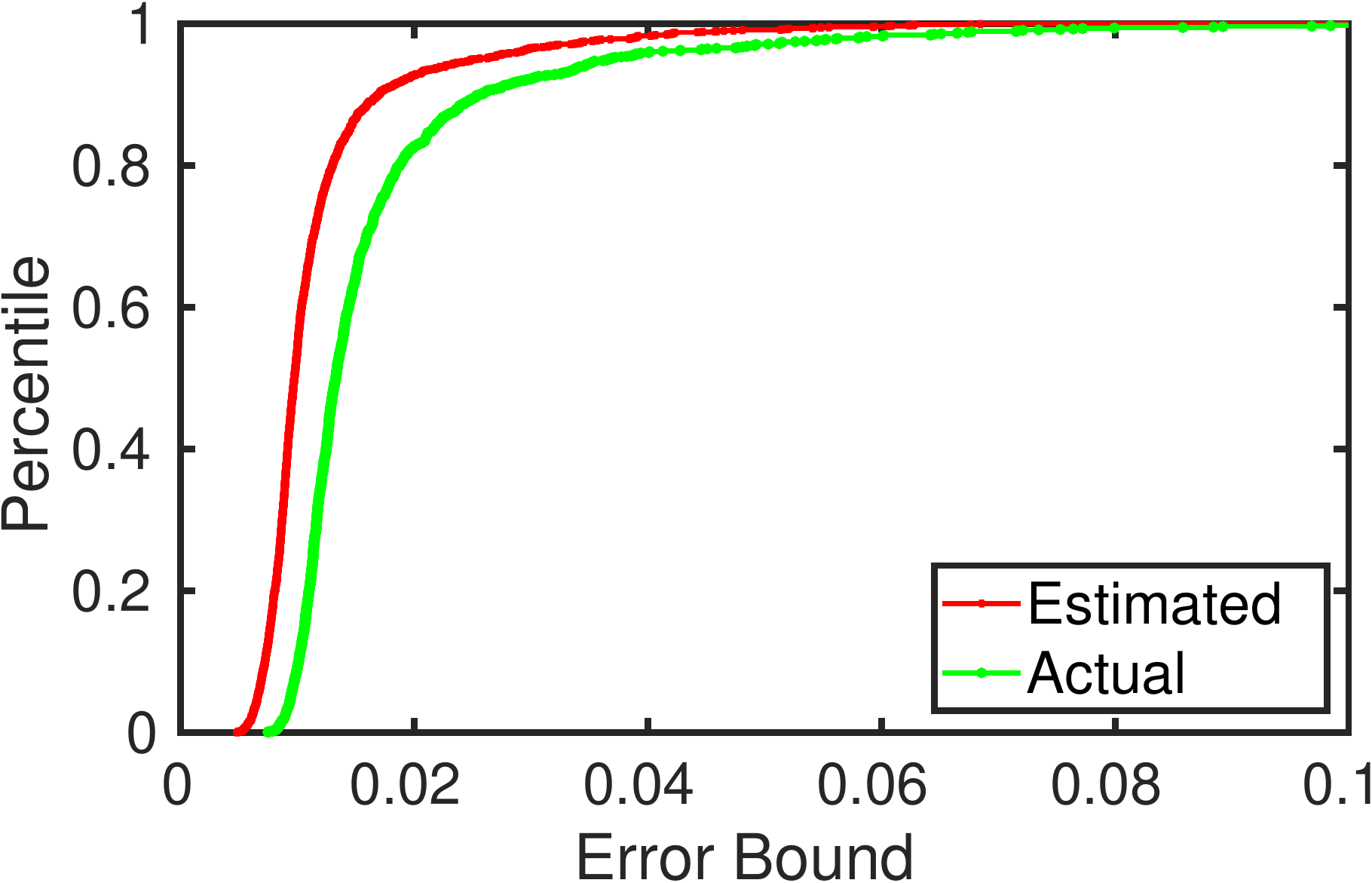}}
		\subfloat[WikiClickstream, 75\%-50\%]
		{\includegraphics[width=0.5\linewidth]{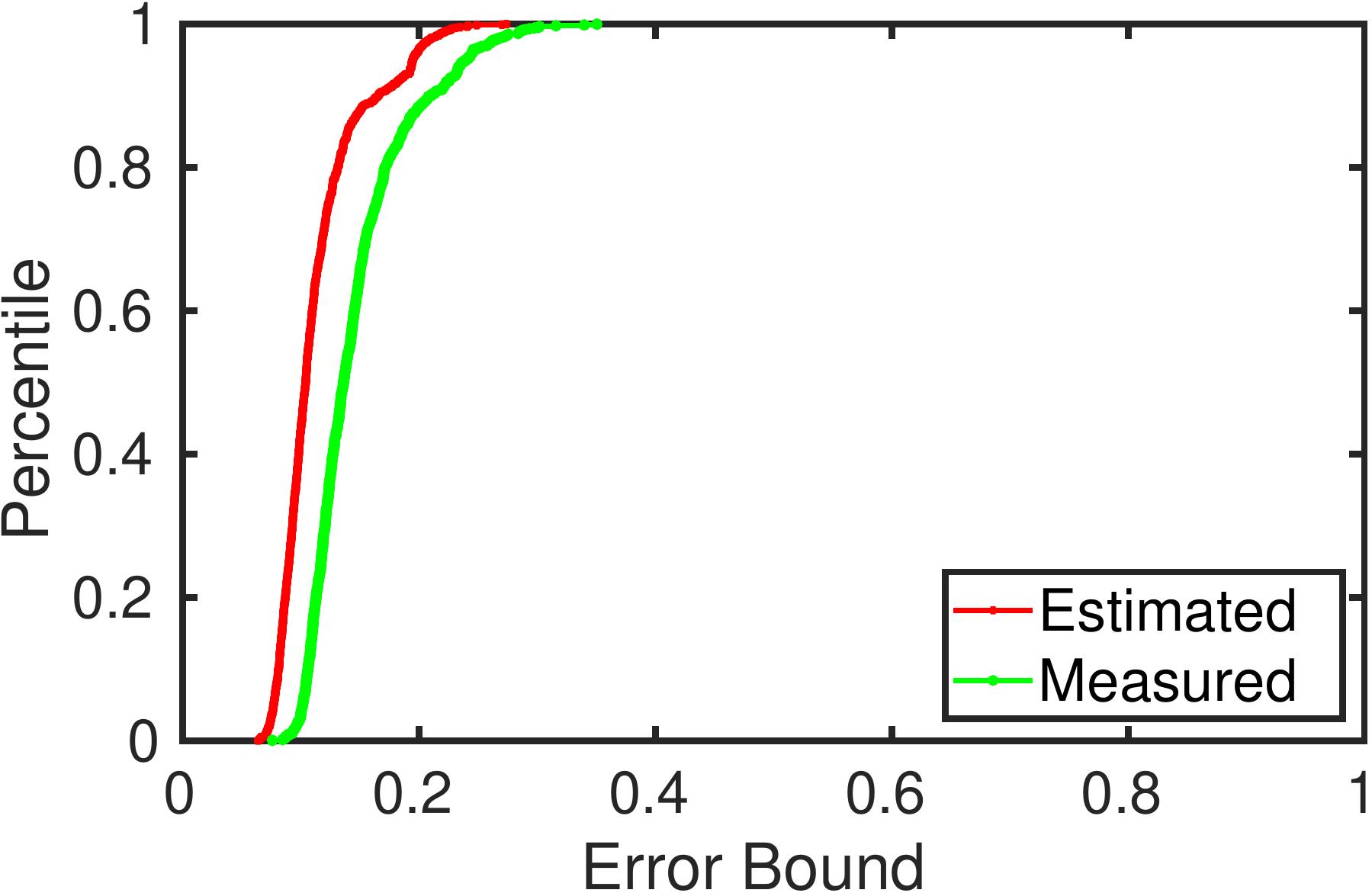}}
	\end{center}
	\caption{Estimated and actual relative error comparison.}
	\label{fig:realVSEstimatedErrors}
	\vspace{-0.5cm}
\end{figure}

\myparagraph{Fraction of keys in output.} As already mentioned, multi-stage sampling can result in loss of keys in the output for jobs that produce more than one key. Figure~\ref{fig:numKeys-ClusterSampling} plots the fractions of keys present in the output at different sampling rates for the same three applications, normalized against the total number of keys produced in the precise executions.  Figure~\ref{fig:missedKeys-ClusterSampling} shows the occurrence frequencies of lost keys in the input RDD of the final aggregation action in a precise execution, normalized against the total number of data items in the RDD. We observe that significant fractions of keys can be lost, especially at higher partitioning sampling rates.  For example, sampling rates of (75\%, 60\%) for Co-occur reduce execution time by 40\% at the expense of losing 25\% of the keys produced by the precise execution.  However, Figure~\ref{fig:missedKeys-ClusterSampling} shows that only {\em rare} keys are lost.  For example, for the same (75\%, 60\%) sampling rates in Co-occur, the most frequently appearing key that was lost accounted for only a very small fraction $0.85 \times 10^{-4}$ of the total number of data items in the input RDD of the final aggregation action, while 90\% of the lost keys each accounted for less than or equal to $0.08 \times 10^{-4}$ of the total number of data items in the RDD.  The lost keys are even more rare in the WikiPageRank and WikiClickstream applications, where the occurrences of each lost key accounting for $10^{-7}$ of the total number of data items.

\myparagraph{Effect of sampling rates on error bounds.}
Figure \ref{fig:medlineErrorpartitionSampling} plots the CDFs of the estimated error bounds computed as $\frac{\epsilon}{v_{approx}}$, which are the ratios of sampling error to estimated value, for all keys with 95\% confidence for Co-occur.  Each graph in the figure plots CDFs for several different partition sampling rates while keeping data item sampling rate fixed. We observe that, as pointed out in \cite{purdueStratified}, multi-stage sampling without considering keys in the final output over-samples popular keys and under-samples rare keys, leading to uneven relative error bounds. This can lead to large relative error bounds in the tails of the relative error bounds CDFs.

We observe that even relatively high partition sampling rates (e.g., 75\% - green curve in Figure~\ref{fig:medlineErrorpartitionSampling}(a)) can significantly impact error bounds for more rare keys (pushing the CDF curve for $>$60\% to the right) while not affecting the frequently appearing keys much (the CDF curve does not change much for $<$60\%).  Interestingly, a 75\% partition sampling rate affects error bounds less or comparable to a 75\% data item sampling rate (red curve in Figure~\ref{fig:medlineErrorpartitionSampling}(b)) for up to 60\% of the keys, but the tail is significantly worse for partition sampling.  We believe this is caused by the clustering of data items with the same keys within partitions. As either or both sampling rates decrease, the entire error bound CDF shifts to the right (larger error bounds).  However, the observation that partition sampling affects the tail of error bounds CDF much more strongly than data item sampling remains consistent throughout. 

Figure~\ref{fig:medlineErrorItemSampling} shows the error bound CDFs when the partition sampling rates are fixed with varying data item sampling rates, the tails of the error bound CDFs are similar under the same partition sampling rates. This points to a fundamental trade-off: partition sampling can reduce execution time over data item sampling, but trades off higher error bounds for the rarer keys to do so. Figure~\ref{fig:medlineErrorTradeoff} shows that the 95\% relative error CDFs can exhibit trade-offs with different combinations of partition and data item sampling rates. In each subgraph, the sampling rates are chosen so that they have similar execution time as in Figure \ref{fig:clusterRuntime}(a). We can see that their error CDFs intersect, with the error CDFs from lower partition sampling rates having worse tails. It shows that different partition and data item sampling rates combinations can achieve similar execution time, but different error bound distributions. For example in Figure~\ref{fig:medlineErrorTradeoff}(a), (100\%-60\%) has better smaller errors after the $62^{th}$ percentile, but performs worse on frequent keys that have smaller errors. It is because (100\%-75\%) processes more data than (100\%-60\%), so the frequent keys result in smaller errors but the rarer keys have worse error due to partition dropping.

\myparagraph{Comparison with relative error.}~Figure \ref{fig:realVSEstimatedErrors} plots the distributions of estimated error bounds, versus the relative error against ground truth - $|1-\frac{\hat{v}}{v}|$. We can see that ApproxSpark's error estimation is constantly lower than the actual relative error. We can also see that the estimation is more accurate at lower percentiles and less so at higher percentiles. It is because the popular keys usually provides more statistical information to the error estimation process than rare keys.


\begin{table}
	{	
		\begin{center}
			\begin{tabular}{|l|r|r|} \hline
				& \multicolumn{2}{c|}{\bf Sampling Rates} \\
				{\bf Source} & {\bf (100\%, 30\%)} & {\bf (75\%, 75\%)} \\ \hline	
				Partition sampling & 0\% & 78\%\\         \hline	
				Data item sampling & 88\% & 12\%  \\           \hline	
				Pop. estimate partitions & 0\% & 5\%  \\          \hline
				Pop. estimate data items  & 12\% & 5\% \\         \hline
			\end{tabular}	
		\end{center}
	}
	\caption{Breakdown of uncertainty on average across all keys for the four sources of errors in multi-stage sampling for Co-occur.}
	\label{tbl:SourcesUncertainty}
\end{table}

\begin{table}
	{
		\begin{center}
			\begin{tabular}{@{}cccccc@{}}
				\hline
				{\bf Sampling} & {\bf Execution} & \multicolumn{3}{c}{\bf Error Bound Percentile} & {\bf \% Keys}  \\	
				{\bf Rates} & {\bf Time (s)} & 	{\bf 100$^{th}$} &  {\bf 90$^{th}$} & {\bf 50$^{th}$}  & {\bf Present} \\ \hline
				\\
				100\%-60\%  & 149.2   & 0.17 &  0.12    &  0.10   &  80.0    \\ \hline
				75\%-100\%  & 150.8   & 0.37 &  0.22      &  0.06   &  80.3 \\ \hline
				\\
				100\%-30\%  & 120.9   & 0.22 &  0.20      &  0.17   &  71.8    \\ \hline
				75\%-75\%  & 121.4   &  0.32 &  0.28      &  0.13   &  72.3 \\ \hline
				50\%-100\%  & 119.7   &  0.51 &  0.31      &  0.11   &  61.5 \\ \hline
			\end{tabular}
		\end{center}
	}
	\caption{Comparison of run times, error bounds at $100^{th}$, $90^{th}$, $50^{th}$ percentiles, and fraction of unique keys for Co-occurrence.}
	\vspace{-0.5cm}
	\label{tbl:runtime-ErrorComp}
\end{table}

\begin{figure}
	\begin{center}  
			\vspace{-0.5cm}
		\centering
		\subfloat[100\% partition sampling rate] {\includegraphics[width=0.5\linewidth]{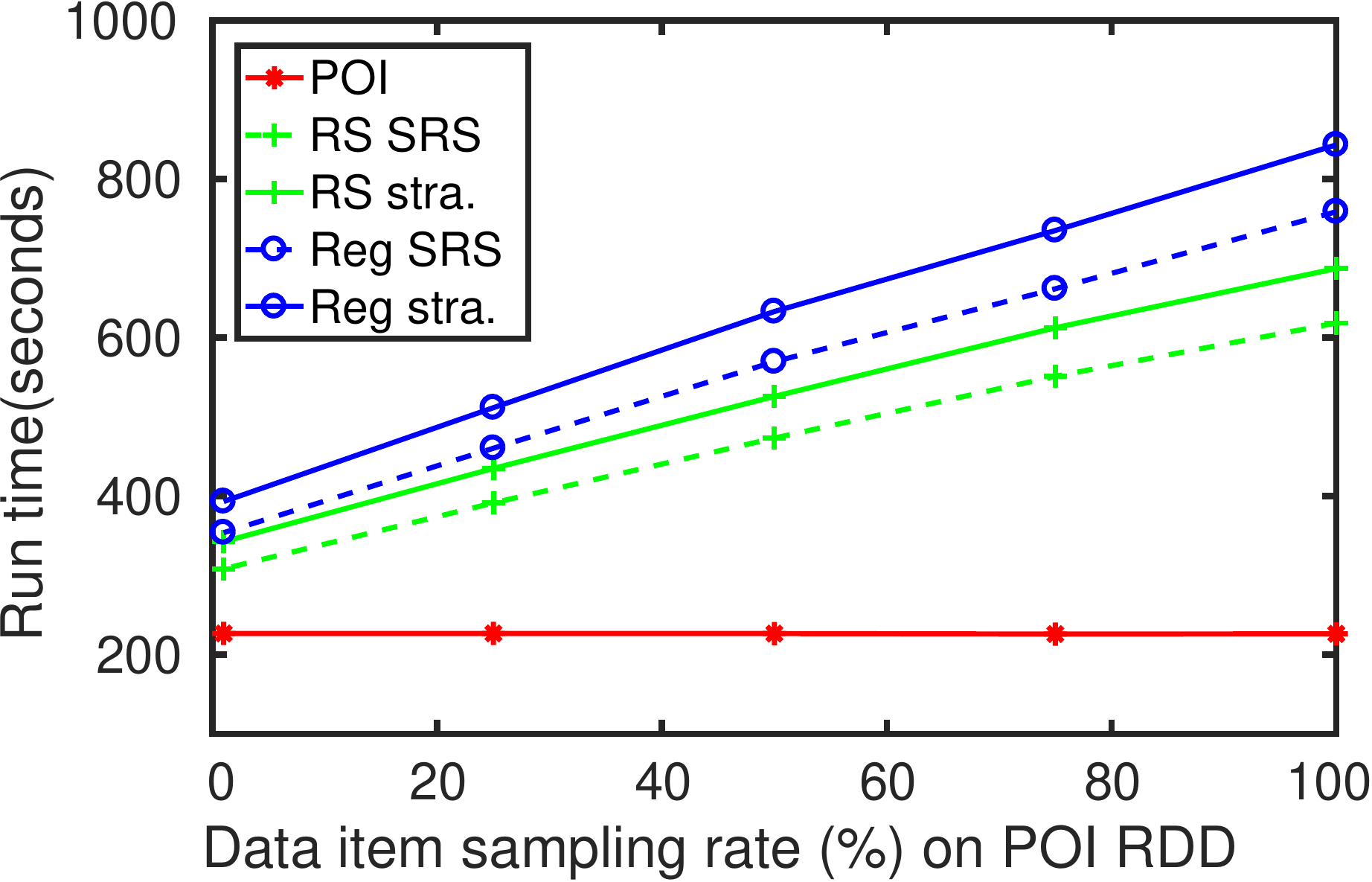}}  
		\subfloat[75\% partition sampling rate] {\includegraphics[width=0.5\linewidth]{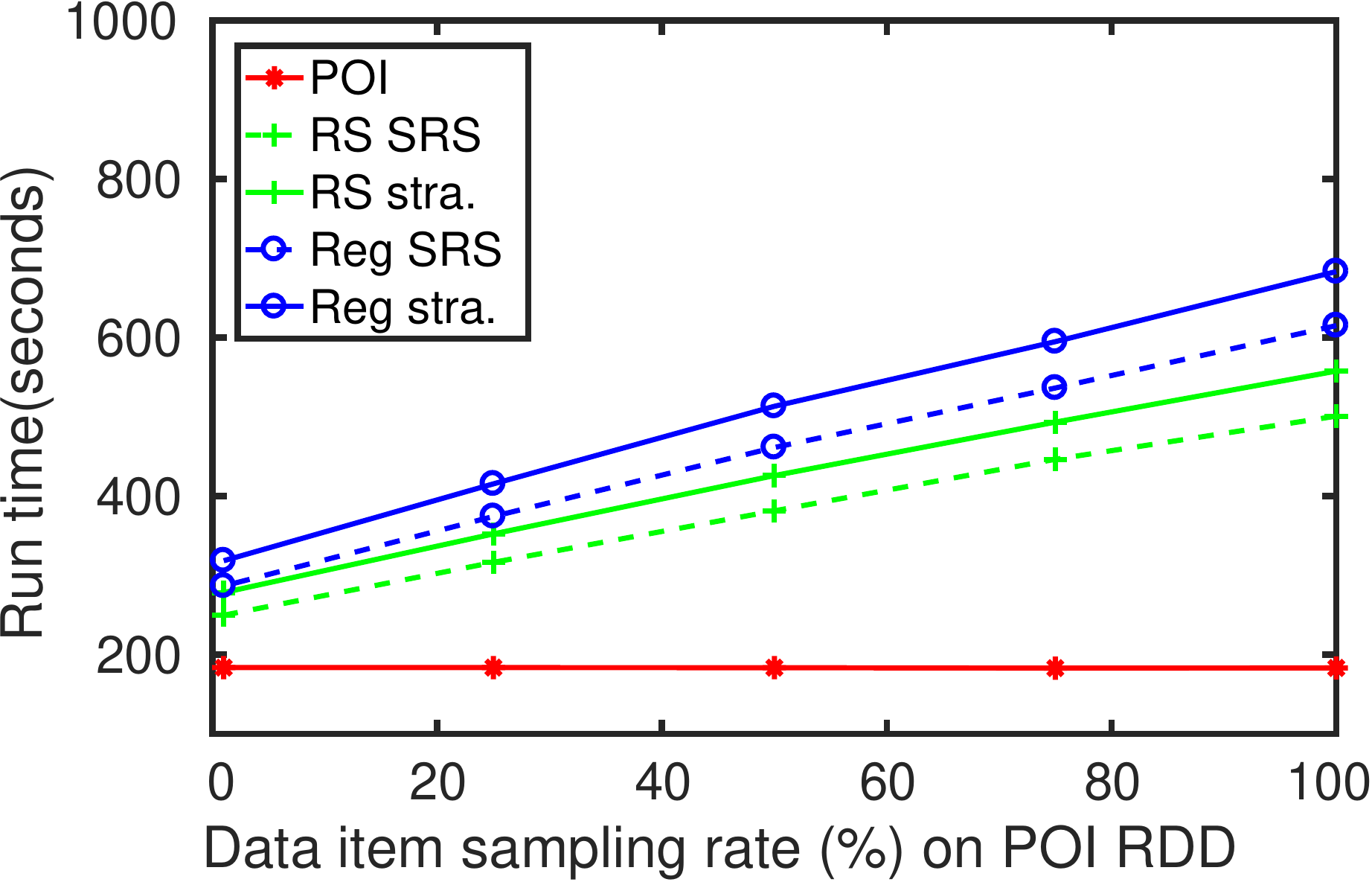}} 
	\end{center}
	
	\caption{Run times for the aggregations at POI, road segment and region RDD. The dashed curves represent run times under SRS, solid curves represent stratified sampling~(stra).}
	\vspace{-0.5cm}
	\label{fig:TaxiHeatMapStratifiedRuntime}
\end{figure}
\begin{figure*}[!tbp]
		\vspace{-0.5cm}
	\subfloat[50\%-25\% SRS] {\includegraphics[width=0.25 \linewidth]{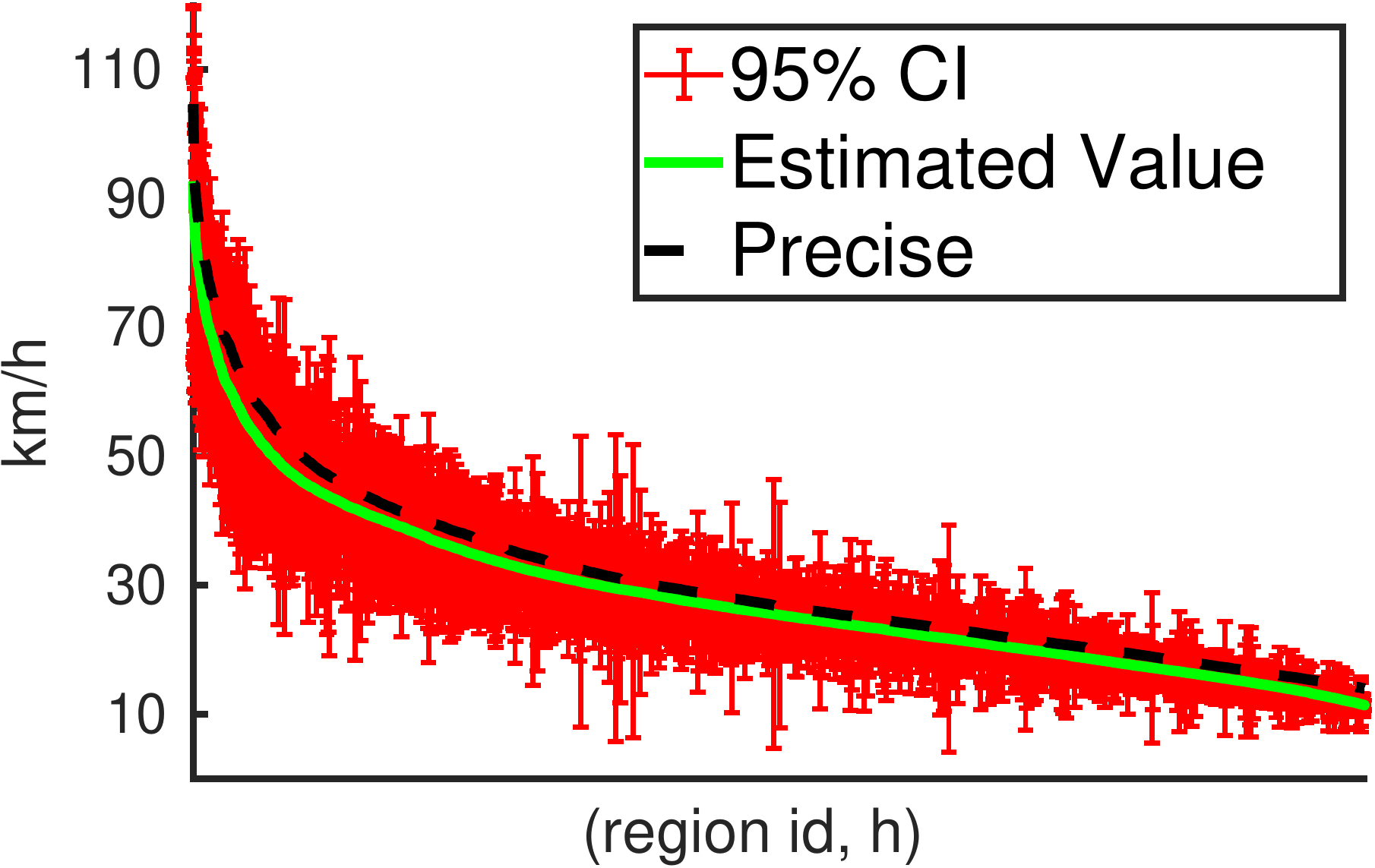}} 
	\subfloat[50\%-25\% Stratified] {\includegraphics[width=0.25 \linewidth]{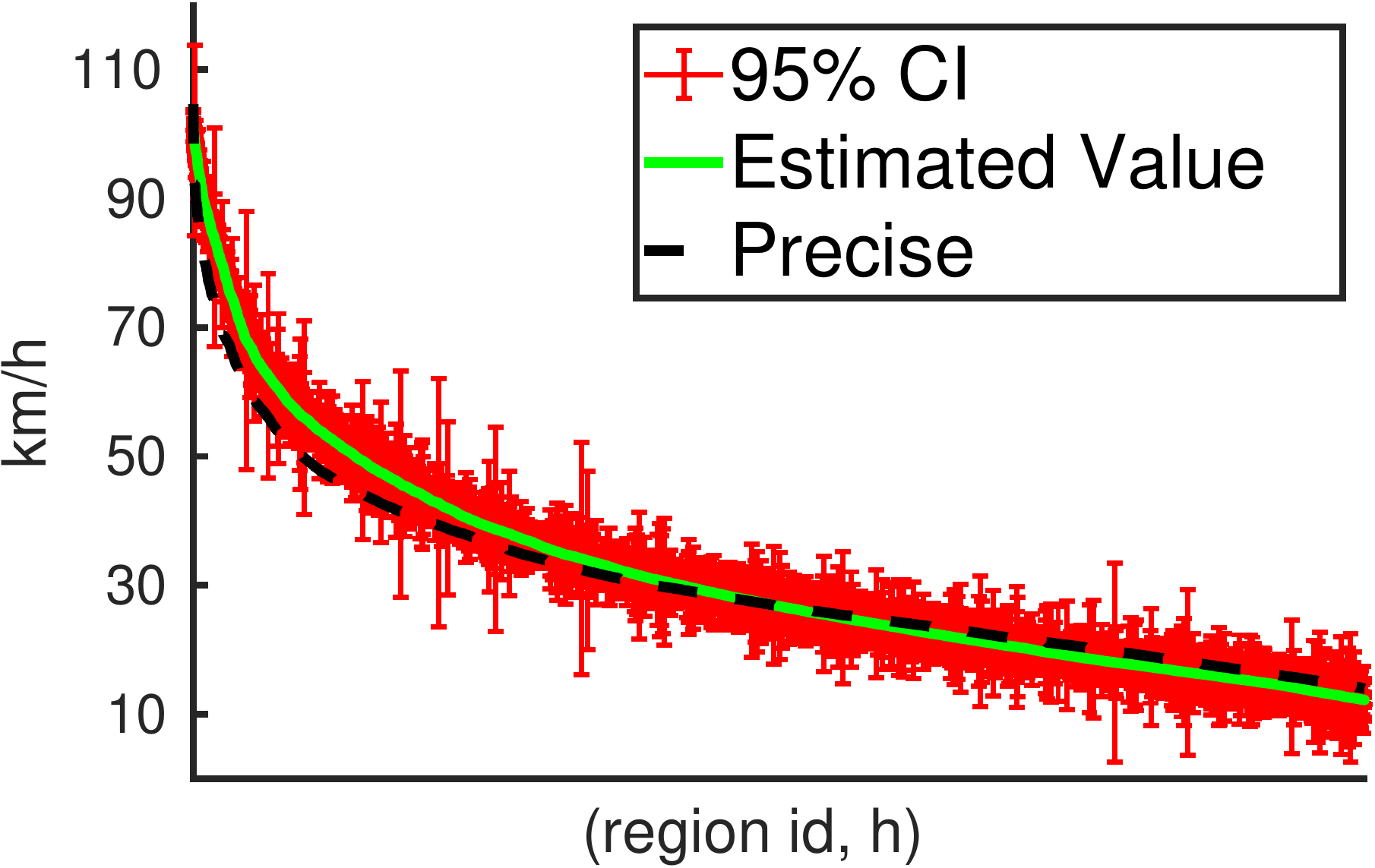}}
	\subfloat[75\%-50\% SRS] {\includegraphics[width=0.25 \linewidth]{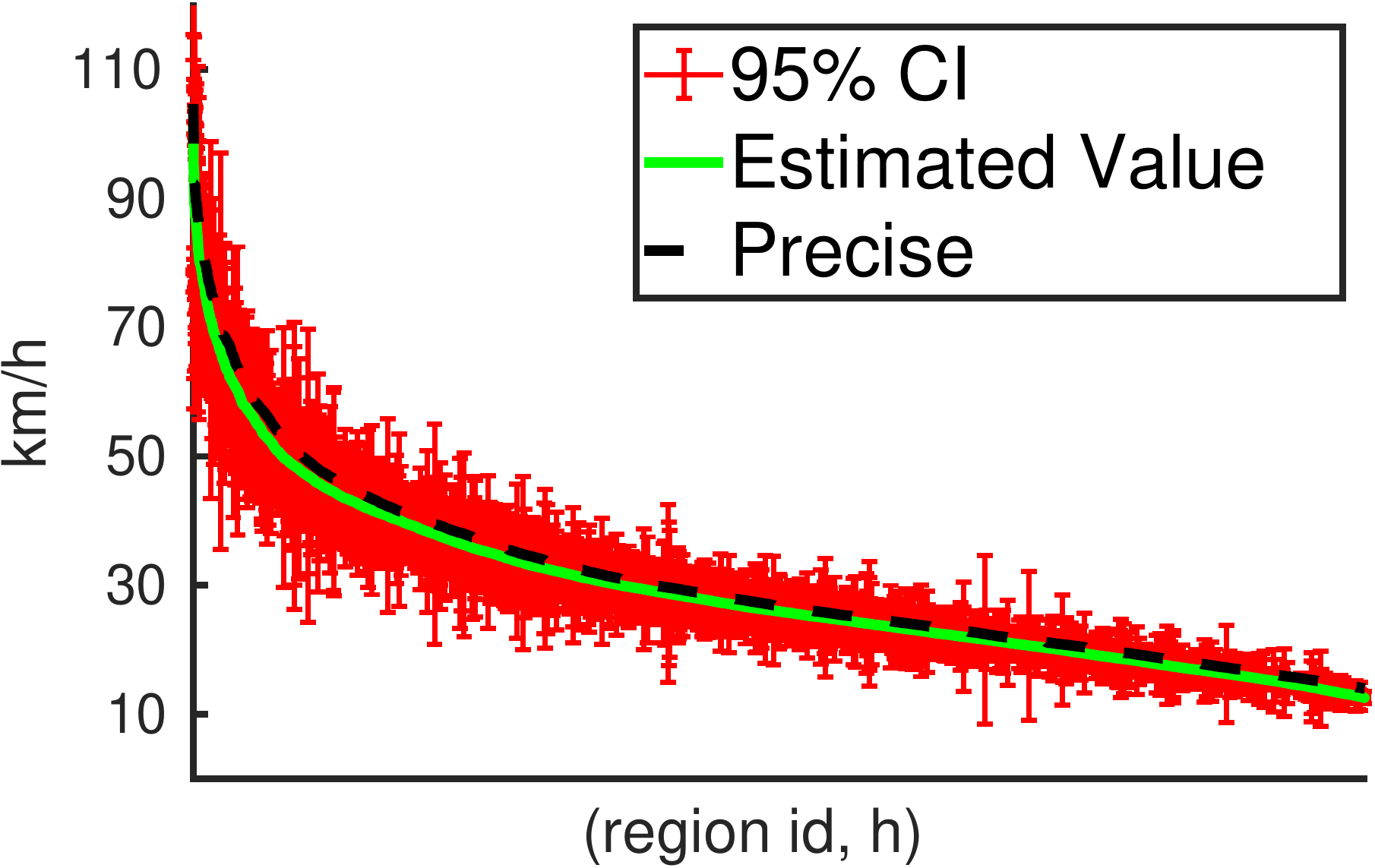}} 
	\subfloat[75\%-50\% Stratified] {\includegraphics[width=0.25 \linewidth]{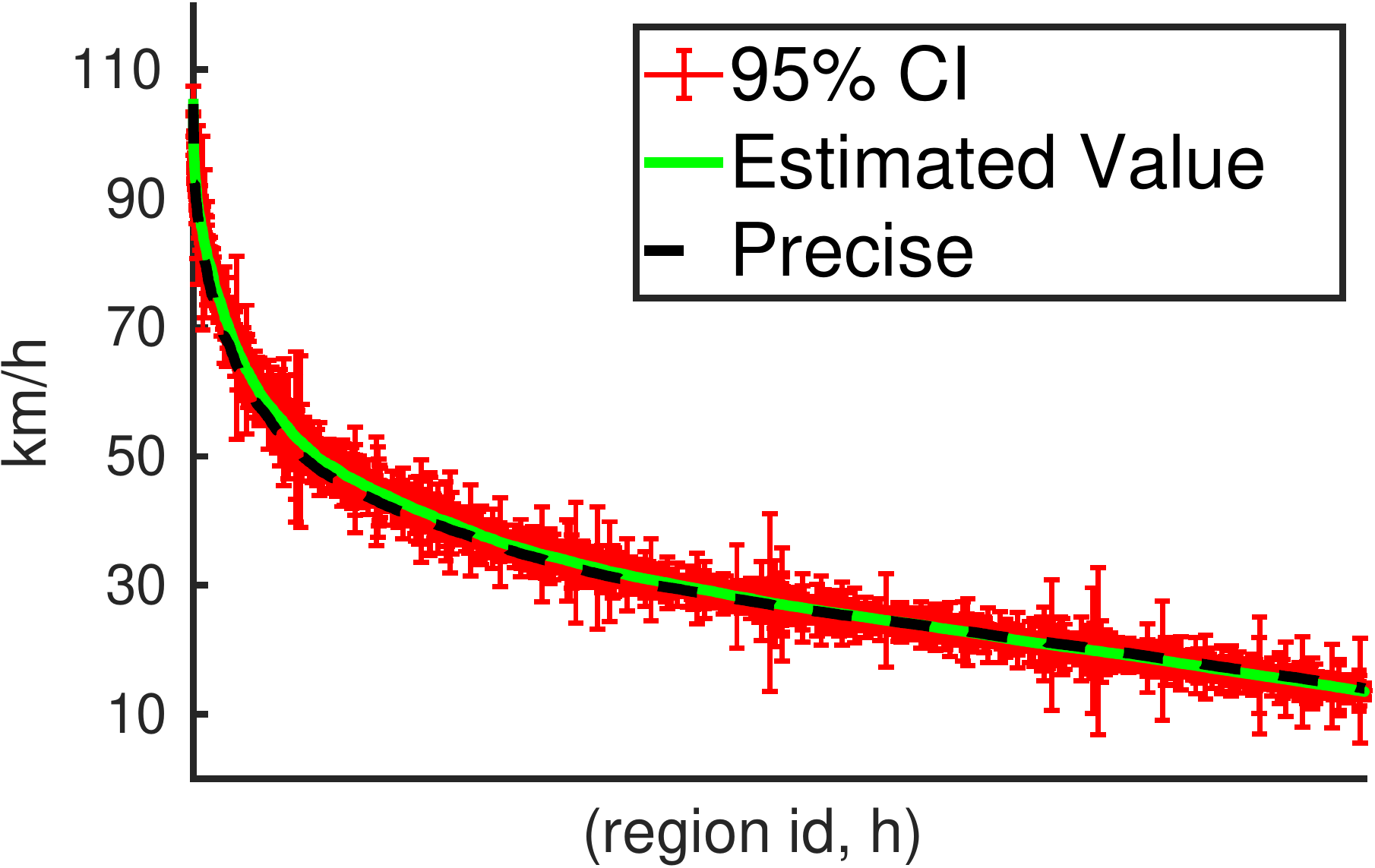}}
	\caption{Region average speed at each hour at different partition and data item sampling rates combinations. Comparisons of 95 \% Confidence Interval width when sample random or stratified sampling is applied at the POI RDD, coupled with partition sampling at the input.}
	\label{fig:Stratified Sampling and multistage error comparison}
	\vspace{-0.5cm}
\end{figure*}
\futurenote{
	\myparagraph{Effect of where to sample.} Sampling at different parts of the RDD transformation chain can produce different error bound distributions. Figure \ref{fig:medlineBeforeAfterFlatMap} shows the error bound distributions when data item sampling occurs at different points in the transformation chain. Since data item sampling before the {\tt flatMap} corresponds to dropping small clusters generated by the {\tt flatMap}, we can see that under the same partition and data item sampling rates, sampling after the {\tt flatMap} results in an error distribution that is smaller than sampling before the {\tt flatMap}.
	
	\begin{figure}[!tbp]
		\vspace{-3cm}
		\begin{center}
			\centering
			\subfloat[Run time] {\includegraphics[width=0.5 \linewidth]{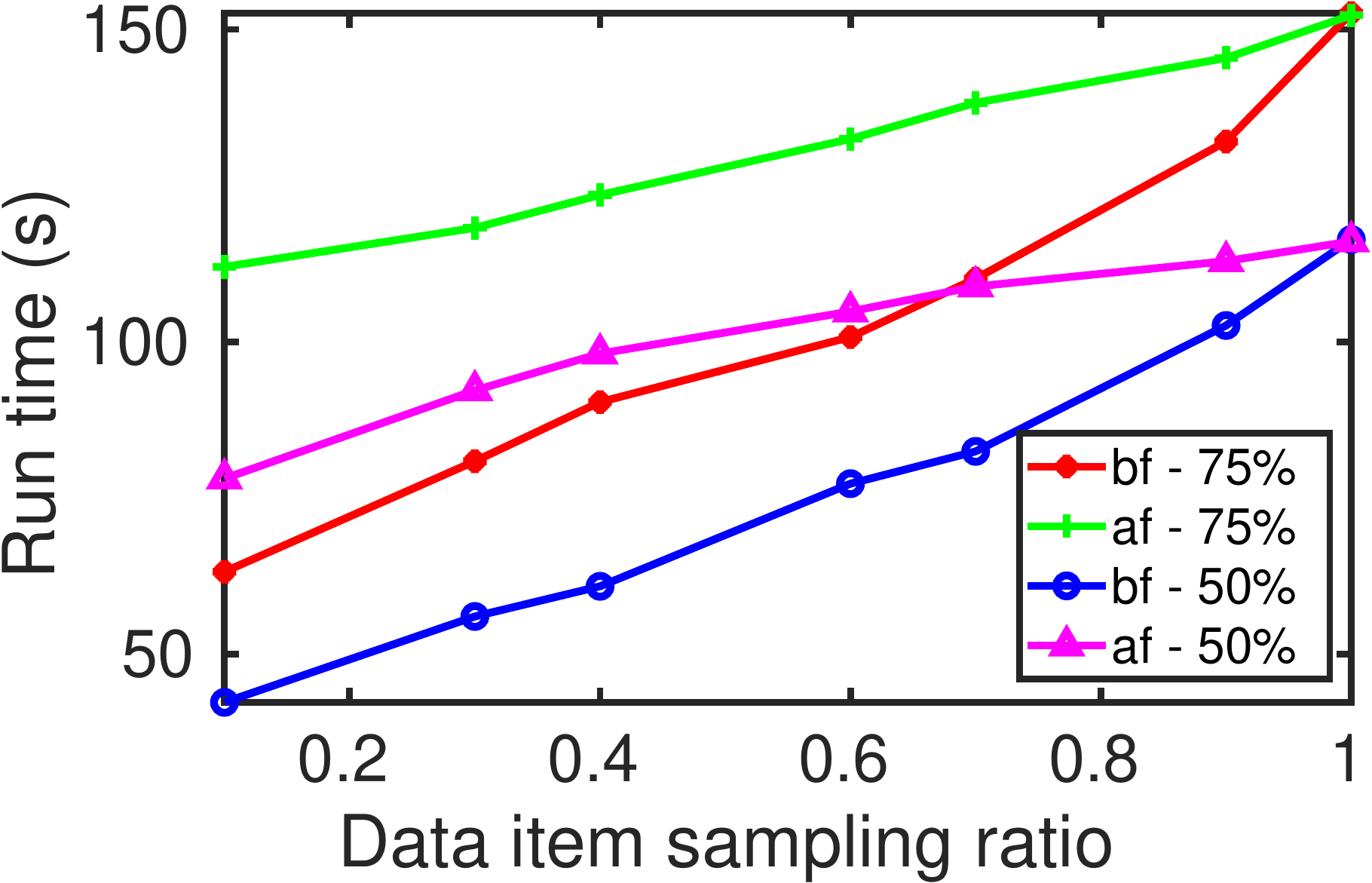}} 
			\hfill
			\subfloat[Error bound CDF] {\includegraphics[width=0.5 \linewidth]{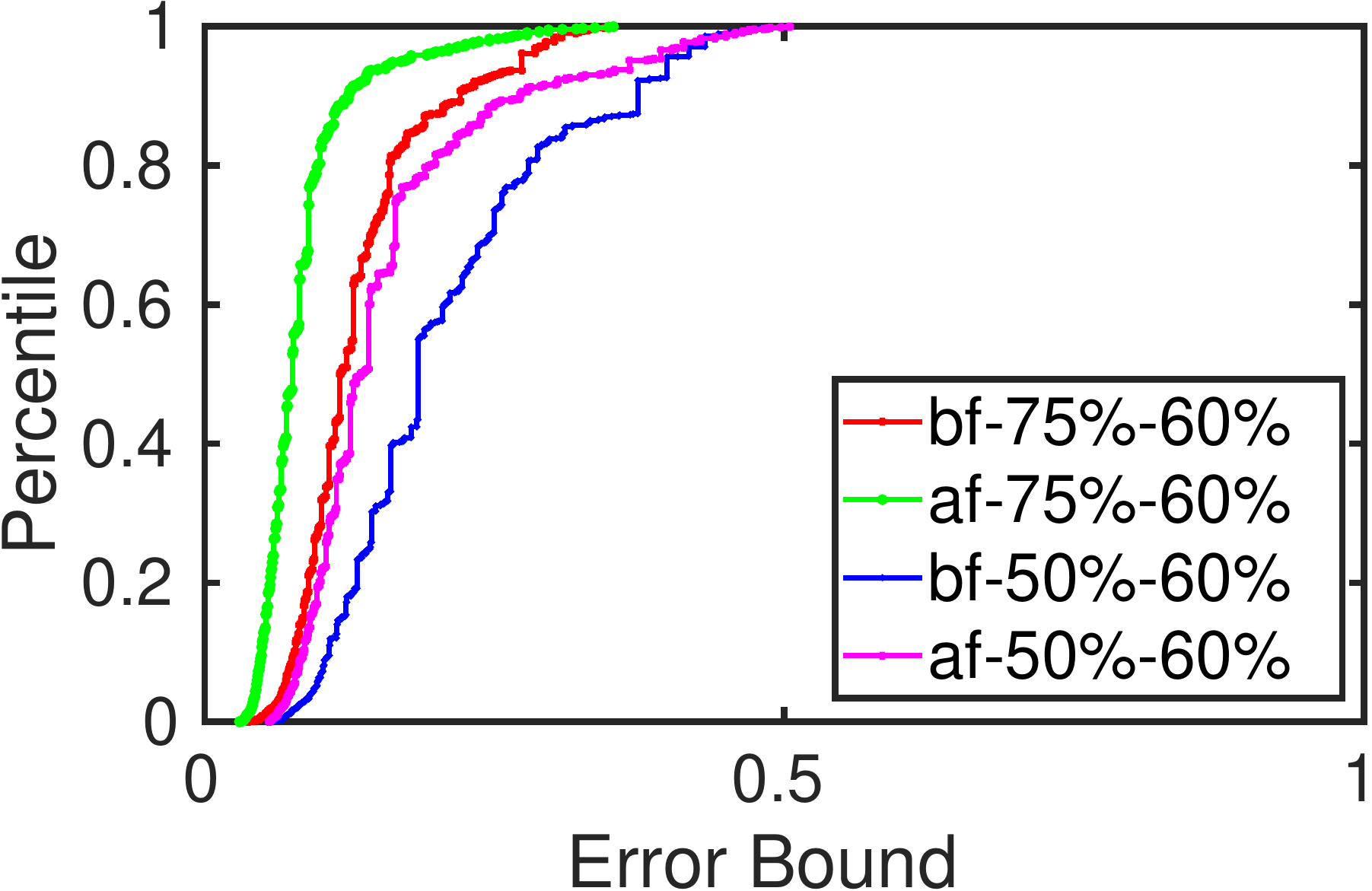}} 
			\hfill
		\end{center}
		\vspace{0cm}
		\caption{Run time comparison between data item sampling at the input data~(before {\tt flatMap}) and after {\tt flatMap}, coupled with different input partition sampling ratios}
		\label{fig:medlineBeforeAfterFlatMap}
	\end{figure}
}

\reconsider{
\myparagraph{Sensitivity to underlying data distribution.}
In the dataset for our Twitter application, the tweets are ordered chronologically which implies that the keys (Hashtag) tend to cluster. We explored shuffling the data items and see what effects it has over the result.

\myparagraph{Sampling error.} Sampling error is smaller in the shuffled data case. It because when data items is shuffled, there is less inter-cluster variance and less uncertainty in population size estimation. Figure \ref{fig:twitterSentiErrorShuffledCompare} shows the comparison of sampling error CDFs between the shuffled and unshuffled data when applying a $50\%$ partition sampling rate. }
\reconsider{
\myparagraph{Fraction of keys shown in the output.} Effect of sampling partitions misses fewer keys when the input data is shuffled, mitigating the clustering effect of the keys.
So if temporal info is not important to the application, shuffling the data offline as preprocessing will improve the result in terms of sampling error and number of missed keys, assuming data will be reused.}

\myparagraph{Sources of uncertainty.} As previously explained, uncertainties (leading to estimated error bounds) can arise from the sampling as well as population estimations.  Table~\ref{tbl:SourcesUncertainty} shows the percentages of the error bounds, averaged across all keys in the output, attributable to each of four sources for Co-occurence. We observe that the inter-cluster variance from partition sampling accounts for by far the largest portion of the estimated error bounds, which is consistent with \cite{lohr2009sampling}.  The intra-cluster variance from data item sampling accounts for the next largest portion, while population estimations for number of partitions, the number of groups of co-occurred words, within each partition for each key, account for only small portions of the error bounds.

\myparagraph{Summary.} Putting together the observations made above, we conclude that multi-stage sampling works well to significantly reduce execution time while introducing small to modest relative errors, as long as the loss of rare keys are acceptable. Further, data item sampling would typically be preferable to partition sampling because it gives more consistent error bounds across keys. To more clearly support this conclusion, Table~\ref{tbl:runtime-ErrorComp} presents data for two sets of sampling rates for Co-occur, $\{(100\%, 60\%), (75\%, 100\%)\}$ and $\{(100\%,30\%), (75\%, 75\%), (50\%, 100\%)\}$, where members within each set have similar execution times.  As the partition sampling rate increases, the tail of the error bounds CDF worsen significantly.  The trend is less clear for lost keys; however, high partition sampling rates (e.g., 50\%) can clearly lead to significantly increased number of lost keys.  Looking at Figure~\ref{fig:clusterRuntime}(a), this implies that partition sampling rates of 50\% and 75\% are not as useful since similar performance is achievable with (100\%, $x$\%) sampling rates.  On the other hand, execution time can be reduced using a partition sampling rate of 25\% if one is willing to tolerate the accompanying key loss and increased error bounds.

\reconsider{
\begin{figure}
	\vspace{-4.2 cm}
	\begin{center}
		\includegraphics[width=2in]{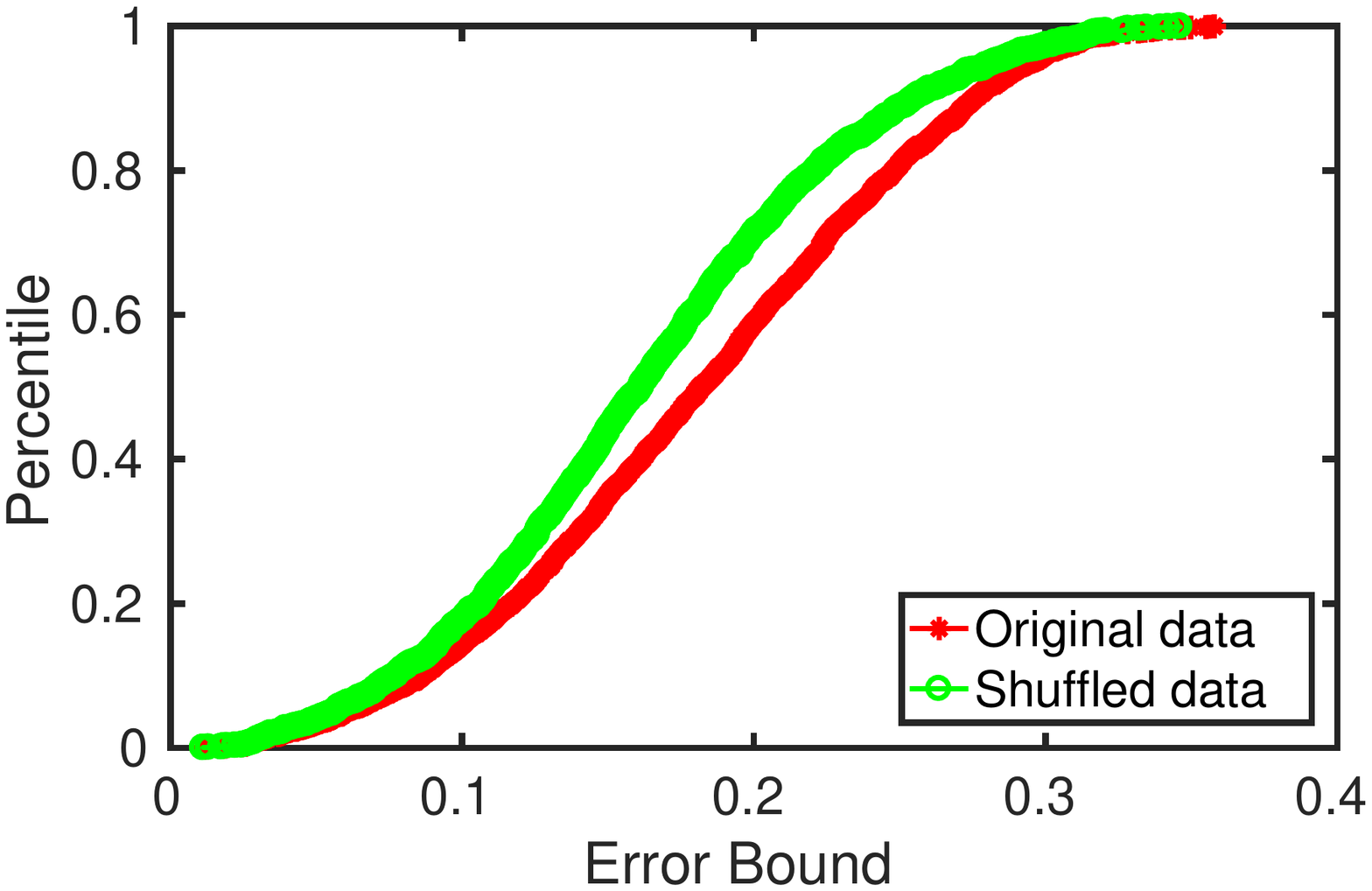} \\
	\end{center}
	\vspace{-0.2in}
	\caption{Comparison of error bound in twitter application between the shuffled data and unshuffled data when applying 50\% partition sampling rate.}
	\label{fig:twitterSentiErrorShuffledCompare}
\end{figure}
}

\subsection{Results for stratified sampling using ASRS}
In the Speed application, we explored both stratified sampling using ASRS with power allocation technique, and simple random sampling~(SRS) on the data items in POI RDD. In addition, partition sampling is also applied when reading the input data. When stratified sampling is performed over the data items in the POI RDD, it also creates stratification effect for both road segment and region RDDs since each POI maps to a road segment, which in turn maps to a region. We use power allocation techniques to balance the sampling errors at each strata in the POI RDD when stratified sampling is applied.

\myparagraph{Execution times.} Figure~\ref{fig:TaxiHeatMapStratifiedRuntime} shows the execution times for aggregating at multiple RDD along the chain. We see that aggregating the street and region RDD both have run time reduction as the sampling rate on the POI RDD lowers, whether stratified sampling or SRS is performed. However, we do see that stratified sampling using ASRS has much higher overhead
than the SRS since it needs to perform stratification and power allocation over the keys.

\myparagraph{Confidence interval.} Figure \ref{fig:Stratified Sampling and multistage error comparison} plots the estimated values with error bars of the average speed at the region level. The precise result is plotted in the dashed black curves, estimated value in green curve and 95 \% confidence intervals as red error bars. Usually higher average speed is observed in regions that are away from city centers and at hours that are early in the morning or late night, as a result these points come with lower taxi densities, i.e., fewer samples which also tend to cluster over a few partitions. These points would have larger error bars without balancing the sample sizes among popular and rare keys, as shown in the Figure \ref{fig:Stratified Sampling and multistage error comparison}(a) and (c), whereas stratified reservoir sampling coupled with power allocation technique increases the sampling rates of these rare keys, resulting in smaller error bars for them. However, we do observe that the popular keys have shorter error bars under SRS compared to stratified, it is because the popular keys have a much larger representation in the sample than the rare keys.

\myparagraph{Fraction of keys shown in the output.}
\begin{figure}[!tbp]
	\begin{center}
		\centering
		\subfloat[SRS] {\includegraphics[width=0.5 \linewidth]{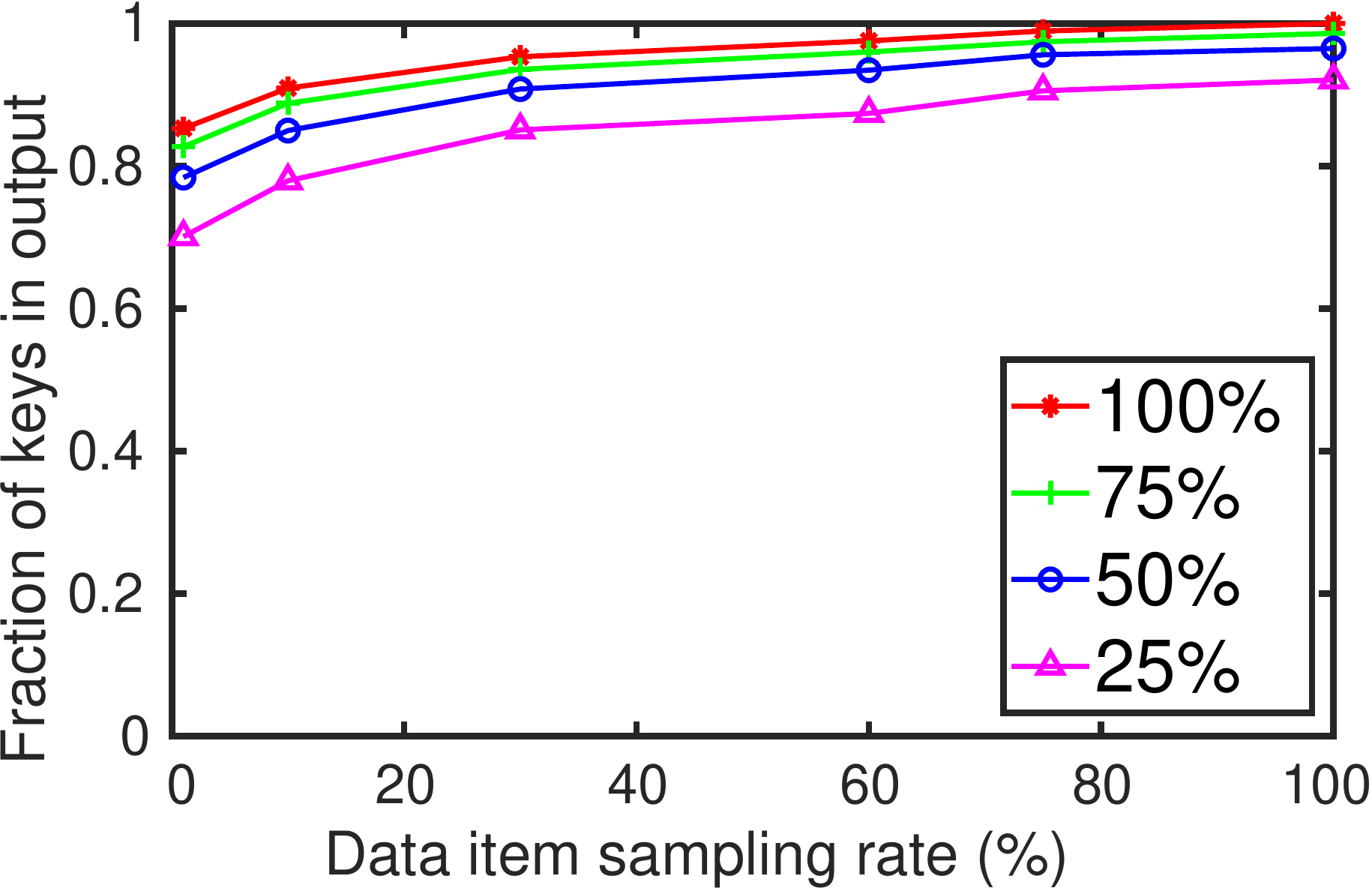}} 
		\hfill
		\subfloat[Stratified] {\includegraphics[width=0.5 \linewidth]{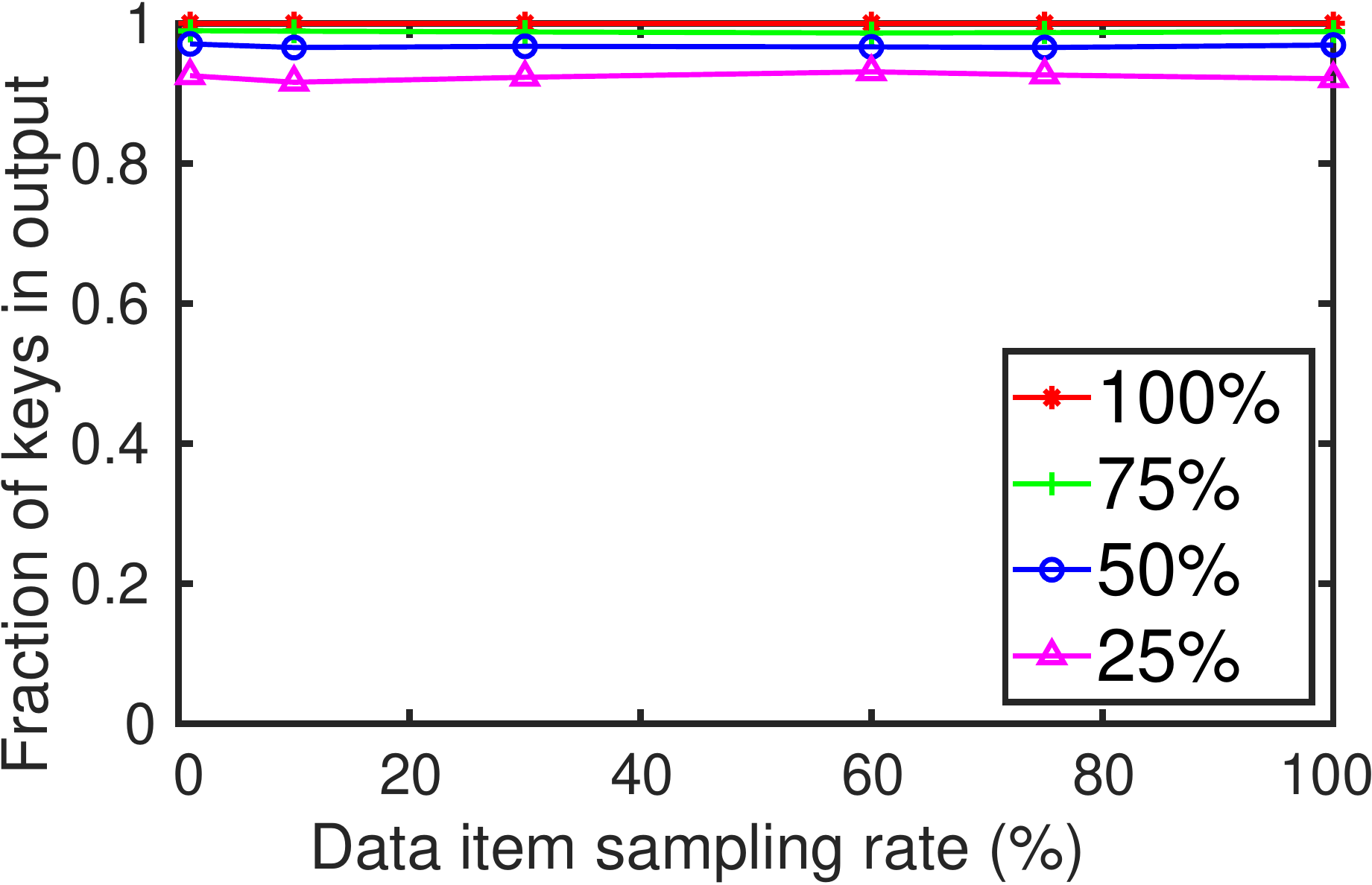}}
		\hfill
	\end{center}
	\vspace{-0.2cm}
	\caption{Number of output keys (normalized) occurred in the output, SRS or stratified sampling performed at the road segment RDD. Each line represents a partition sampling rate at the initial RDD.}
		\vspace{-0.5cm}
	\label{fig:NumKeysTaxiStreet}
\end{figure}
In Figure \ref{fig:NumKeysTaxiStreet}, we see that the stratified sampling constantly loses less keys than SRS at same sampling rate. In Figure \ref{fig:NumKeysTaxiStreet}(b), we see that when partition sampling rate is set, stratified sampling can preserve the number of output keys without being affected by the sampling rate over the data item shown in (a). This shows that stratified reservoir sampling is much better at preserving output keys in the result.

\myparagraph{Summary.}~ASRS can not only achieves a balanced error distribution among popular and rare keys, it also loses much fewer output keys, consistent throughout varying sampling rates over the data items. However, ASRS has a higher sampling overhead than SRS which is reflected in the execution times.

\begin{figure}[!tbp]
	\begin{center}
		\centering
		\subfloat[Speed] {\includegraphics[width=0.5 \linewidth]{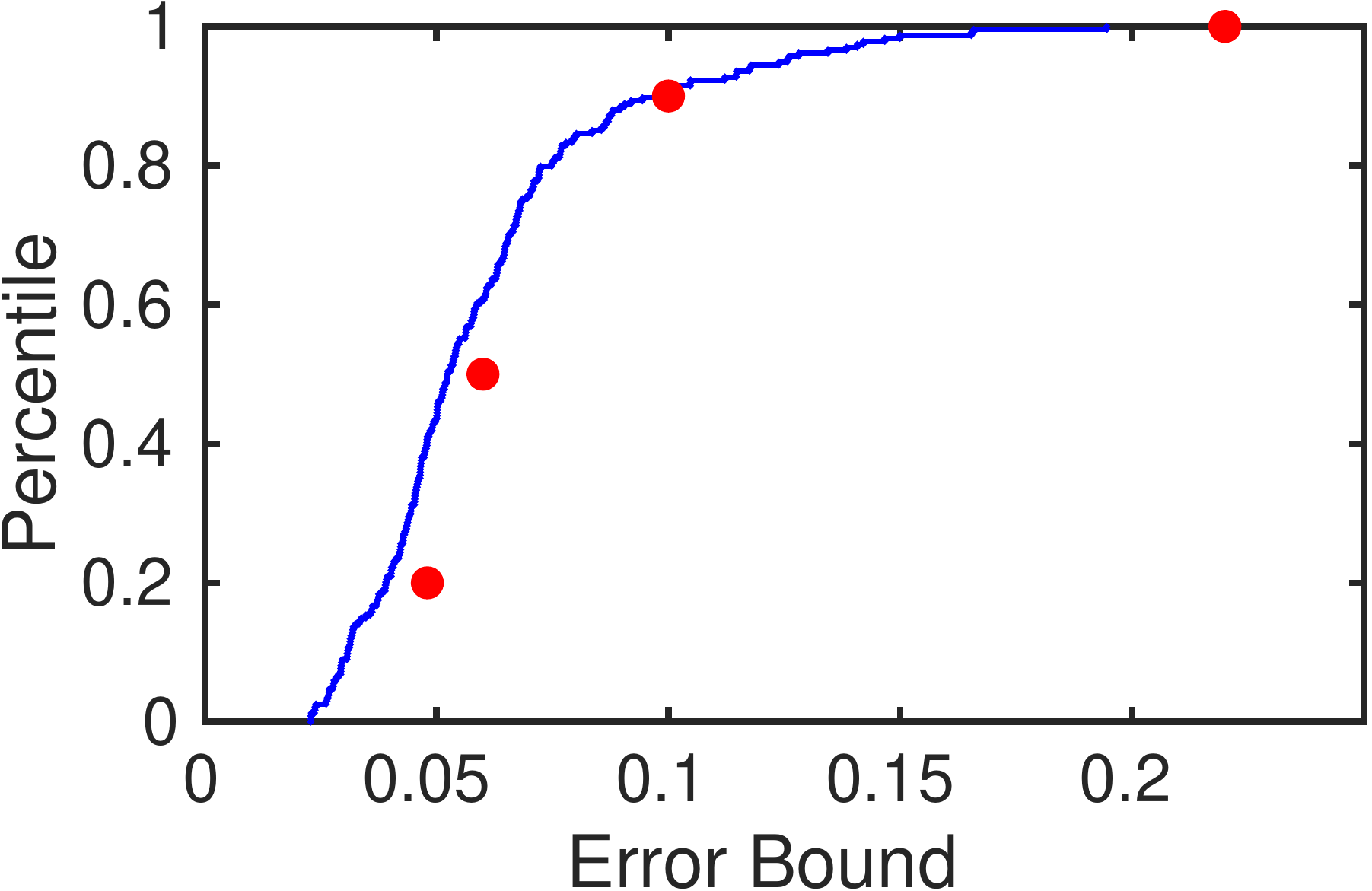}} 
		\subfloat[WikiClickstream] {\includegraphics[width=0.5 \linewidth]{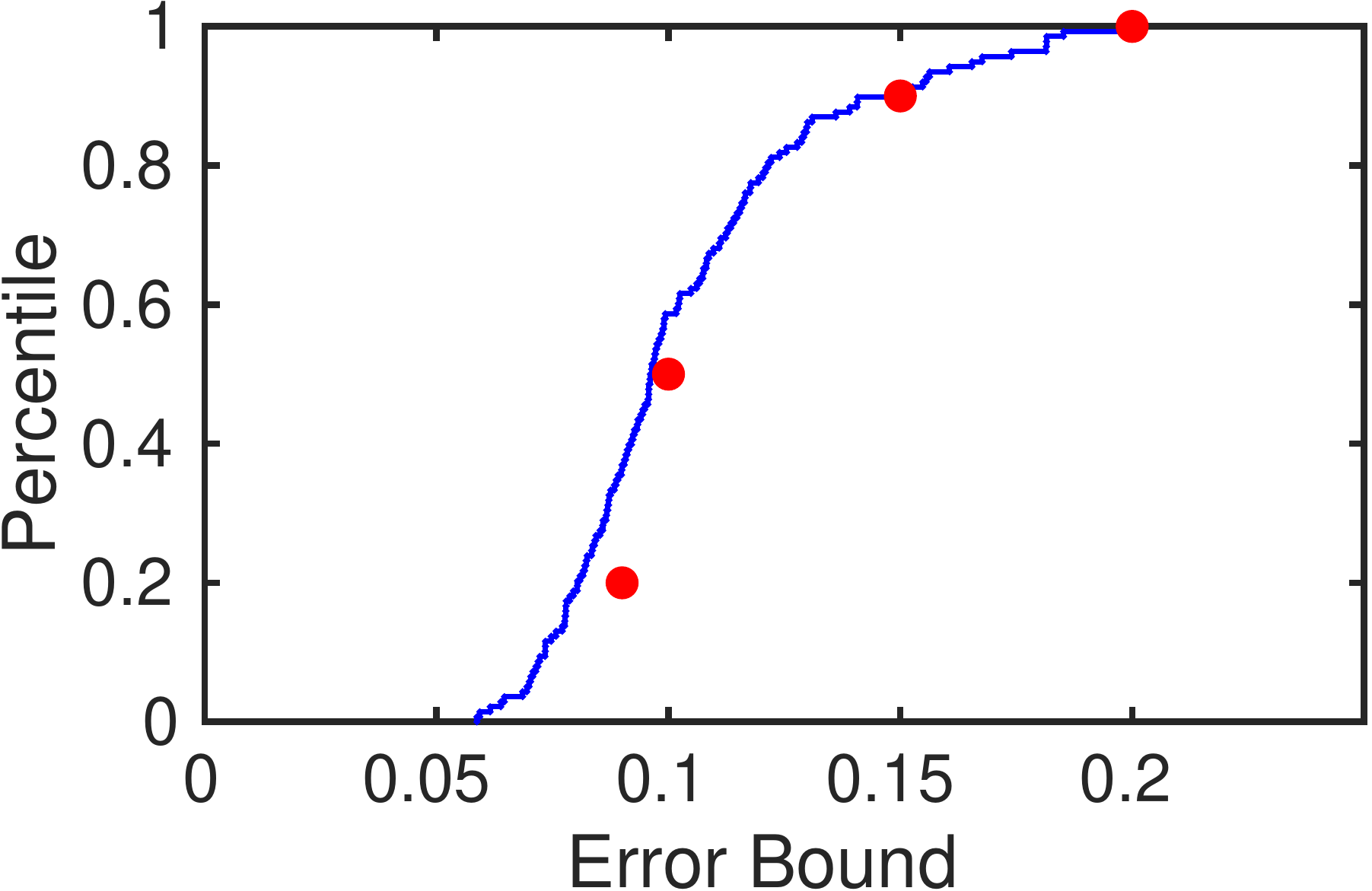}}
	\end{center}
	\vspace{-0.2cm}
	\caption{User-specified error bounds targets shown on the error CDF achieved by ApproxSpark.}
	\vspace{-0.5cm}
	\label{fig:UserSettingErrorBoundTargets}
\end{figure}

\subsection{Results for user-specified error targets}
We now demonstrate ApproxSpark's capability in allowing users to set error bounds target at different percentiles over the relative error distribution. The red dots in Figure \ref{fig:UserSettingErrorBoundTargets} show the target error bounds at $20^{th}$, $50^{th}$, $90^{th}$, $100^{th}$ percentiles; the blue curves show the resulting CDF achieved by ApproxSpark, setting both partition and data item sampling step sizes to be 0.1\%. We can see that the CDFs are bounded by the targets set by the user. Figure \ref{fig:UserSettingErrorBoundTargets}(a) is the error distribution for average Taxi speed with $50\%$ partition, and $80\%$ data item sampling rates at the POI RDD. Figure \ref{fig:UserSettingErrorBoundTargets}(b) is the error distribution for WikiClickstream aggregation result with $40\%$ partition and $60\%$ data item sampling rates. We randomly select 10\% of the RDD partitions to be executed in the pilot wave. This approximation mode incurs more overhead compared with setting the sampling rates that satisfy the user-specified error targets. We have observed that the pilot wave causes about 20\% and 25\% extra execution time respectively in the two applications compared with setting the sampling rates directly.


\section{Conclusion}
In this paper, we motivate, design, and implement a system called ApproxSpark, which features a set of approximation mechanisms leveraging sampling theories to adapt to Spark's computing paradigm. Our proposed multi-stage sampling theories together with a data provenance tree allows for a general approximate computing framework under Spark's parallel computing model with multi-step RDD transformations. We utilize a set of metrics to rigorously evaluate ApproxSpark including run times, the error bound distribution for all keys, and the number of missed keys using applications from different domains. We have also gained important insights on an interesting yet complicated trade-off space in terms of the error bounds, run times and number of keys, when choosing different sampling schemes, such as specific sampling rates, partition vs data item sampling, whether to use stratified sampling, different combinations of partition/data item sampling rates. We also have found that input data that contains less rare keys for the application is more amenable to approximation. In a real-world setting, a user would choose the most appropriate sampling setup catering to her approximation goal. Based on our experience and results, we conclude that our framework and system can make efficient and customized approximation to big data practioners using Spark.

\balance
\appendix


\section{Cluster sampling variance with population estimation}
\label{sec:popEsti}
Estimated sum for all clusters is:
\begin{equation}
\hat{\tau}=\frac{\hat{N}}{n}\sum _{i\in S}v _{i}=\hat{N}\bar{\tau}
\end{equation}
Sample mean among the cluster totals is:
\begin{equation}
\bar{\tau}=\frac{1}{n}\sum _{i\in S}v _{i}
\end{equation}
Estimated total number of clusters $N$ is:
\begin{equation}
\hat{N} = \frac{n}{p_{1}}  
\end{equation}
Since $\hat{N} \sim \mathcal{NB}(n, p_{1})$, the variance of $\hat{N}$ is:
\begin{equation}
Var(\hat{N})=\frac{n(1-p_{1})}{p_{1}^2}
\end{equation}
If we treat it as simple random sampling, the variance of mean of cluster total is:
\begin{equation}
Var(\bar{\tau})=(1-p_{1})\frac{s_{t}^{2}}{n}
\end{equation}
Variance of cluster totals $Var(\hat{\tau})_{srs} =$
\begin{equation}
\begin{split}
 &Var(\hat{N}\bar{\tau}) \\ &= \hat{N} ^{2}Var(\bar{\tau}) +\bar{\tau}^{2} Var(\hat{N})+Var(\hat{N})Var(\bar{\tau})
\\&= (\frac{n}{p_{1}})^2(1-p_{1})\frac{s_{t}^{2}}{n}  +  \frac{n(1-p_{1})}{p_{1}^2}(\bar{\tau}^2 + (1-p_{1})\frac{s_{t}^{2}}{n})
\end{split}
\label{eq:sumVarInter}
\end{equation}
thus:
\begin{equation}
Var_{inter} = (1-\frac{1}{p_{1}})Var_{srs}(\hat{\tau}) 
\end{equation} 
Estimated sum of cluster $i$ is:
\begin{equation}
\hat{\tau} _{i} = \hat{M_{i}}\bar{\tau_{i}}
\end{equation}
where sample mean $\bar{\tau_{i}}$ in cluster $i$ is:
\begin{equation}
\bar{\tau_{i}}=\frac{1}{m_{i}}\sum _{j\in S_{i}}v_{ij}
\end{equation}
where $m_{i}$ is the number of sampled items in cluster $i$, $M_{i}$ is the population total in cluster $i$ and $p_{2}$ is the data item sampling rate. \newline \newline Since $\hat{M_{i}} \sim \mathcal{NB}(m_{i}, p_{2}$), estimated $\hat{M_{i}}$ is:
\begin{equation}
\hat{M_{i}}=\frac{m_{i}}{p_{2}}
\end{equation}
with variance:
\begin{equation}
Var(\hat{M_{i}}) = \frac{(m_{i})(1-p_{1})}{p_{2}^2} 
\end{equation}
The variance of sample mean in cluster $i$ is:
\begin{equation}
Var(\bar{\tau _{i}}) = (1-p_{2})\frac{s_{i}^{2}}{m_{i}}
\end{equation}
The variance of estimated sum in cluster $i$ is:
\begin{equation}
\begin{split}
Var(\hat{\tau} _{i})  &= \hat{M_{i}}^{2}\hat{Var}(\bar{\tau _{i}})+\bar{\tau _{i}}^{2}Var(M_{i})+Var(\hat{M_{i}})Var(\bar{\tau _{i}}) 
\\ &= (\frac{m_{i}}{p_{2}}) ^2(1-p_{2})\frac{s_{i}^{2}}{m_{i}} 
\\ &+  \frac{(m_{i})(1-p_{1})}{p_{1}^2} (\bar{\tau_{i}} + \frac{(m_{i})(1-p_{1})}{p_{1}^2})
\end{split}
\end{equation}
Intra-cluster variance is:
\begin{equation}
Var_{intra} = \frac{1}{p_{1}}\sum _{i\in S}V(\hat{\tau} _{i})
\end{equation}
The total variance is:
\begin{equation}
\begin{split}
Var(\hat{\tau}) &= Var_{inter} + Var_{intra} \\ &= Var(\hat{\tau})_{srs} +\frac{1}{p_{1}}\sum _{i\in S}Var(\hat{\tau} _{i}) 
\end{split}
\end{equation}


\bibliographystyle{IEEEtran}
\bibliography{ApproxSpark}

\begin{thebibliography}{10}
\providecommand{\url}[1]{#1}
\csname url@samestyle\endcsname
\providecommand{\newblock}{\relax}
\providecommand{\bibinfo}[2]{#2}
\providecommand{\BIBentrySTDinterwordspacing}{\spaceskip=0pt\relax}
\providecommand{\BIBentryALTinterwordstretchfactor}{4}
\providecommand{\BIBentryALTinterwordspacing}{\spaceskip=\fontdimen2\font plus
\BIBentryALTinterwordstretchfactor\fontdimen3\font minus
  \fontdimen4\font\relax}
\providecommand{\BIBforeignlanguage}[2]{{%
\expandafter\ifx\csname l@#1\endcsname\relax
\typeout{** WARNING: IEEEtran.bst: No hyphenation pattern has been}%
\typeout{** loaded for the language `#1'. Using the pattern for}%
\typeout{** the default language instead.}%
\else
\language=\csname l@#1\endcsname
\fi
#2}}
\providecommand{\BIBdecl}{\relax}
\BIBdecl

\bibitem{Chong2014}
F.~T. Chong, M.~J.~R. Heck, P.~Ranganathan, A.~A.~M. Saleh, and H.~M.~G.
  Wassel, ``{Data Center Energy Efficiency:Improving Energy Efficiency in Data
  Centers Beyond Technology Scaling},'' \emph{IEEE Design {\&} Test}, vol.~31,
  no.~1, 2014.

\bibitem{dai2018cloud}
W.~Dai, L.~Qiu, A.~Wu, and M.~Qiu, ``Cloud infrastructure resource allocation
  for big data applications,'' \emph{IEEE Transactions on Big Data}, vol.~4,
  no.~3, pp. 313--324, 2018.

\bibitem{approxSurvey}
\BIBentryALTinterwordspacing
S.~Mittal, ``A survey of techniques for approximate computing,'' \emph{ACM
  Comput. Surv.}, vol.~48, no.~4, pp. 62:1--62:33, Mar. 2016. [Online].
  Available: \url{http://doi.acm.org/10.1145/2893356}
\BIBentrySTDinterwordspacing

\bibitem{chaiken2008scope}
R.~Chaiken, B.~Jenkins, P.-{\AA}. Larson, B.~Ramsey, D.~Shakib, S.~Weaver, and
  J.~Zhou, ``Scope: easy and efficient parallel processing of massive data
  sets,'' \emph{Proceedings of the VLDB Endowment}, vol.~1, no.~2, pp.
  1265--1276, 2008.

\bibitem{yan2014error}
Y.~Yan, L.~J. Chen, and Z.~Zhang, ``Error-bounded sampling for analytics on big
  sparse data,'' \emph{Proceedings of the VLDB Endowment}, vol.~7, no.~13, pp.
  1508--1519, 2014.

\bibitem{gray1997data}
J.~Gray, S.~Chaudhuri, A.~Bosworth, A.~Layman, D.~Reichart, M.~Venkatrao,
  F.~Pellow, and H.~Pirahesh, ``Data cube: A relational aggregation operator
  generalizing group-by, cross-tab, and sub-totals,'' \emph{Data mining and
  knowledge discovery}, vol.~1, no.~1, pp. 29--53, 1997.

\bibitem{xie2019olap}
X.~Xie, K.~Zou, X.~Hao, T.~B. Pedersen, P.~Jin, and W.~Yang, ``Olap over
  probabilistic data cubes ii: Parallel materialization and extended
  aggregates,'' \emph{IEEE Transactions on Knowledge and Data Engineering},
  2019.

\bibitem{shanahan2015large}
J.~G. Shanahan and L.~Dai, ``Large scale distributed data science using apache
  spark,'' in \emph{Proceedings of the 21th ACM SIGKDD international conference
  on knowledge discovery and data mining}.\hskip 1em plus 0.5em minus
  0.4em\relax ACM, 2015, pp. 2323--2324.

\bibitem{armbrust2015spark}
M.~Armbrust, R.~S. Xin, C.~Lian, Y.~Huai, D.~Liu, J.~K. Bradley, X.~Meng,
  T.~Kaftan, M.~J. Franklin, A.~Ghodsi \emph{et~al.}, ``Spark sql: Relational
  data processing in spark,'' in \emph{Proceedings of the 2015 ACM SIGMOD
  International Conference on Management of Data}.\hskip 1em plus 0.5em minus
  0.4em\relax ACM, 2015, pp. 1383--1394.

\bibitem{yu2015geospark}
J.~Yu, J.~Wu, and M.~Sarwat, ``Geospark: A cluster computing framework for
  processing large-scale spatial data,'' in \emph{Proceedings of the 23rd
  SIGSPATIAL International Conference on Advances in Geographic Information
  Systems}.\hskip 1em plus 0.5em minus 0.4em\relax ACM, 2015, p.~70.

\bibitem{wiewiorka2014sparkseq}
M.~S. Wiewi{\'o}rka, A.~Messina, A.~Pacholewska, S.~Maffioletti, P.~Gawrysiak,
  and M.~J. Okoniewski, ``Sparkseq: fast, scalable and cloud-ready tool for the
  interactive genomic data analysis with nucleotide precision,''
  \emph{Bioinformatics}, vol.~30, no.~18, pp. 2652--2653, 2014.

\bibitem{hashem2015rise}
I.~A.~T. Hashem, I.~Yaqoob, N.~B. Anuar, S.~Mokhtar, A.~Gani, and S.~U. Khan,
  ``The rise of “big data” on cloud computing: Review and open research
  issues,'' \emph{Information Systems}, vol.~47, pp. 98--115, 2015.

\bibitem{dataflow}
\BIBentryALTinterwordspacing
T.~Akidau, R.~Bradshaw, C.~Chambers, S.~Chernyak, R.~J.
  Fern\'{a}ndez-Moctezuma, R.~Lax, S.~McVeety, D.~Mills, F.~Perry, E.~Schmidt,
  and S.~Whittle, ``The dataflow model: A practical approach to balancing
  correctness, latency, and cost in massive-scale, unbounded, out-of-order data
  processing,'' \emph{Proc. VLDB Endow.}, vol.~8, no.~12, pp. 1792--1803, Aug.
  2015. [Online]. Available: \url{http://dx.doi.org/10.14778/2824032.2824076}
\BIBentrySTDinterwordspacing

\bibitem{lohr2009sampling}
S.~Lohr, \emph{{Sampling: Design and Analysis}}.\hskip 1em plus 0.5em minus
  0.4em\relax Cengage Learning, 2009.

\bibitem{al2014adaptive}
M.~Al-Kateb and B.~S. Lee, ``Adaptive stratified reservoir sampling over
  heterogeneous data streams,'' \emph{Information Systems}, vol.~39, pp.
  199--216, 2014.

\bibitem{Zaharia2010}
\BIBentryALTinterwordspacing
M.~Zaharia, M.~Chowdhury, T.~Das, A.~Dave, J.~Ma, M.~McCauly, M.~J. Franklin,
  S.~Shenker, and I.~Stoica, ``Resilient distributed datasets: A fault-tolerant
  abstraction for in-memory cluster computing,'' in \emph{Presented as part of
  the 9th USENIX Symposium on Networked Systems Design and Implementation (NSDI
  12)}.\hskip 1em plus 0.5em minus 0.4em\relax San Jose, CA: USENIX, 2012, pp.
  15--28. [Online]. Available:
  \url{https://www.usenix.org/conference/nsdi12/technical-sessions/presentation/zaharia}
\BIBentrySTDinterwordspacing

\bibitem{chaudhuri2007optimized}
S.~Chaudhuri, G.~Das, and V.~Narasayya, ``{Optimized Stratified Sampling for
  Approximate Query Processing},'' \emph{ACM Transactions on Database Systems
  (TODS)}, vol.~32, no.~2, 2007.

\bibitem{ssdbm10}
M.~Al-Kateb and B.~S. Lee, ``Stratified reservoir sampling over heterogeneous
  data streams,'' in \emph{Proceedings of the 22nd International Conference on
  Scientific and Statistical Database Management (SSDBM)}.\hskip 1em plus 0.5em
  minus 0.4em\relax Springer Berlin Heidelberg, 2010, pp. 621--639.

\bibitem{purdueStratified}
M.~Thottethodi, T.~Vijaykumar, M.~Kulkarni \emph{et~al.}, ``Stratified online
  sampling for sound approximation in mapreduce,'' 2015.

\bibitem{peng2018aqp++}
J.~Peng, D.~Zhang, J.~Wang, and J.~Pei, ``Aqp++: connecting approximate query
  processing with aggregate precomputation for interactive analytics,'' in
  \emph{Proceedings of the 2018 International Conference on Management of
  Data}.\hskip 1em plus 0.5em minus 0.4em\relax ACM, 2018, pp. 1477--1492.

\bibitem{Agarwal2013}
S.~Agarwal, B.~Mozafari, A.~Panda, H.~Milner, S.~Madden, and I.~Stoica,
  ``{BlinkDB: Queries with Bounded Errors and Bounded Response Times on Very
  Large Data},'' in \emph{Proceedings of the Eurosys Conference}, 2013.

\bibitem{Sapprox}
\BIBentryALTinterwordspacing
X.~Zhang, J.~Wang, and J.~Yin, ``Sapprox: Enabling efficient and accurate
  approximations on sub-datasets with distribution-aware online sampling,''
  \emph{Proc. VLDB Endow.}, vol.~10, no.~3, pp. 109--120, Nov. 2016. [Online].
  Available: \url{https://doi.org/10.14778/3021924.3021928}
\BIBentrySTDinterwordspacing

\bibitem{hellerstein1997online}
J.~M. Hellerstein, P.~J. Haas, and H.~J. Wang, ``{Online Aggregation},'' in
  \emph{Proceedings of the ACM SIGMOD International Conference on Management of
  Data (SIGMOD)}, 1997.

\bibitem{Kumar:2016:HEF:2901318.2901351}
\BIBentryALTinterwordspacing
G.~Kumar, G.~Ananthanarayanan, S.~Ratnasamy, and I.~Stoica, ``Hold 'em or fold
  'em?: Aggregation queries under performance variations,'' in
  \emph{Proceedings of the Eleventh European Conference on Computer Systems},
  ser. EuroSys '16.\hskip 1em plus 0.5em minus 0.4em\relax New York, NY, USA:
  ACM, 2016, pp. 7:1--7:14. [Online]. Available:
  \url{http://doi.acm.org/10.1145/2901318.2901351}
\BIBentrySTDinterwordspacing

\bibitem{ApproxHadoop}
\BIBentryALTinterwordspacing
I.~Goiri, R.~Bianchini, S.~Nagarakatte, and T.~D. Nguyen, ``Approxhadoop:
  Bringing approximations to mapreduce frameworks,'' in \emph{Proceedings of
  the Twentieth International Conference on Architectural Support for
  Programming Languages and Operating Systems}, ser. ASPLOS '15.\hskip 1em plus
  0.5em minus 0.4em\relax New York, NY, USA: ACM, 2015, pp. 383--397. [Online].
  Available: \url{http://doi.acm.org/10.1145/2694344.2694351}
\BIBentrySTDinterwordspacing

\bibitem{mapreduce}
\BIBentryALTinterwordspacing
J.~Dean and S.~Ghemawat, ``Mapreduce: Simplified data processing on large
  clusters,'' in \emph{Proceedings of the 6th Conference on Symposium on
  Opearting Systems Design \& Implementation - Volume 6}, ser. OSDI'04.\hskip
  1em plus 0.5em minus 0.4em\relax Berkeley, CA, USA: USENIX Association, 2004,
  pp. 10--10. [Online]. Available:
  \url{http://dl.acm.org/citation.cfm?id=1251254.1251264}
\BIBentrySTDinterwordspacing

\bibitem{StreamApprox}
\BIBentryALTinterwordspacing
D.~L. Quoc, R.~Chen, P.~Bhatotia, C.~Fetzer, V.~Hilt, and T.~Strufe,
  ``Streamapprox: Approximate computing for stream analytics,'' in
  \emph{Proceedings of the 18th ACM/IFIP/USENIX Middleware Conference}, ser.
  Middleware '17.\hskip 1em plus 0.5em minus 0.4em\relax New York, NY, USA:
  ACM, 2017, pp. 185--197. [Online]. Available:
  \url{http://doi.acm.org/10.1145/3135974.3135989}
\BIBentrySTDinterwordspacing

\bibitem{SparkStreaming}
M.~Zaharia, T.~Das, H.~Li, S.~Shenker, and I.~Stoica, ``Discretized streams: An
  efficient and fault-tolerant model for stream processing on large clusters.''
  \emph{HotCloud}, vol.~12, pp. 10--10, 2012.

\bibitem{ApproxSparkTR}
G.~Hu, D.~Zhang, S.~Rigo, and T.~D. Nguyen, ``Approximation with error bounds
  in spark,'' \emph{arXiv preprint arXiv:1812.01823}, 2018.

\bibitem{vitter1985random}
J.~S. Vitter, ``Random sampling with a reservoir,'' \emph{ACM Transactions on
  Mathematical Software (TOMS)}, vol.~11, no.~1, pp. 37--57, 1985.

\bibitem{bankier1988power}
M.~D. Bankier, ``Power allocations: determining sample sizes for subnational
  areas,'' \emph{The American Statistician}, vol.~42, no.~3, pp. 174--177,
  1988.

\bibitem{berkhin2006survey}
P.~Berkhin, ``A survey of clustering data mining techniques,'' in
  \emph{Grouping multidimensional data}.\hskip 1em plus 0.5em minus 0.4em\relax
  Springer, 2006, pp. 25--71.

\bibitem{Medline}
``{MEDLINE Data},'' 2017,
  \url{https://www.nlm.nih.gov/databases/download/pubmed_medline.html/}.

\bibitem{Carpooling}
\BIBentryALTinterwordspacing
D.~Zhang, T.~He, F.~Zhang, M.~Lu, Y.~Liu, H.~Lee, and S.~H. Son, ``Carpooling
  service for large-scale taxicab networks,'' \emph{ACM Trans. Sen. Netw.},
  vol.~12, no.~3, pp. 18:1--18:35, Aug. 2016. [Online]. Available:
  \url{http://doi.acm.org/10.1145/2897517}
\BIBentrySTDinterwordspacing

\bibitem{Zhang:2014:EHM:2639108.2639116}
\BIBentryALTinterwordspacing
D.~Zhang, J.~Huang, Y.~Li, F.~Zhang, C.~Xu, and T.~He, ``Exploring human
  mobility with multi-source data at extremely large metropolitan scales,'' in
  \emph{Proceedings of the 20th Annual International Conference on Mobile
  Computing and Networking}, ser. MobiCom '14.\hskip 1em plus 0.5em minus
  0.4em\relax New York, NY, USA: ACM, 2014, pp. 201--212. [Online]. Available:
  \url{http://doi.acm.org/10.1145/2639108.2639116}
\BIBentrySTDinterwordspacing

\bibitem{liu2012sentiment}
B.~Liu, ``Sentiment analysis and opinion mining,'' \emph{Synthesis lectures on
  human language technologies}, vol.~5, no.~1, pp. 1--167, 2012.

\bibitem{tweets2011}
(2011) Tweets 2011. \url{http://trec.nist.gov/data/tweets/}.

\bibitem{stanfordcorenlp}
\BIBentryALTinterwordspacing
C.~D. Manning, M.~Surdeanu, J.~Bauer, J.~Finkel, S.~J. Bethard, and
  D.~McClosky, ``The {Stanford} {CoreNLP} natural language processing
  toolkit,'' in \emph{Association for Computational Linguistics (ACL) System
  Demonstrations}, 2014, pp. 55--60. [Online]. Available:
  \url{http://www.aclweb.org/anthology/P/P14/P14-5010}
\BIBentrySTDinterwordspacing

\bibitem{PageRank}
L.~Page, S.~Brin, R.~Motwani, and T.~Winograd, ``{The PageRank Citation
  Ranking: Bringing Order to the Web},'' Stanford InfoLab, Tech. Rep., 1999.

\bibitem{Wikipedia}
``Wikipedia database,'' \url{http://en.
  wikipedia.org/wiki/Wikipedia_database.}, 2016.

\bibitem{clickstream}
(2016) {Wikipedia clickstream}.
  \url{https://meta.wikimedia.org/wiki/Research:Wikipedia_clickstream/}.

\end{thebibliography}

\end{document}